\documentclass[journal]{IEEEtran}
\ifCLASSINFOpdf
\else
\fi

  \usepackage{amssymb}
  \usepackage{latexsym}
  \usepackage{amsfonts}
  \usepackage{amsthm}
  \usepackage{amsmath}
  \usepackage{graphics}
  \usepackage{setspace}
  \usepackage{graphicx}
  \usepackage{epstopdf}
  \usepackage{color}
  \usepackage{float}
  \usepackage{wrapfig}
  \usepackage{color}
  \usepackage{rotating}
  \usepackage{array} 
  \usepackage{dblfloatfix}
 \usepackage{breakurl}
  \usepackage[hyphens]{url}

  \usepackage[table]{xcolor}
  \usepackage{multirow}

\definecolor{litered}{RGB}{246,194,197}
\definecolor{liteyellow}{RGB}{250,227,199}
\definecolor{litegreen}{RGB}{206,235,183}

\newcommand{\llr}{\cellcolor{litered}} 
\newcommand{\lly}{\cellcolor{liteyellow}} 
\newcommand{\llg}{\cellcolor{litegreen}} 

\definecolor{red}{RGB}{0,0,0}
\definecolor{blue}{RGB}{0,0,255}
\newcommand{\gc}{\cellcolor[gray]{0.8}} 
\newcommand{\n}{--}

\definecolor{orange}{RGB}{246,0,0}

\definecolor{orange}{RGB}{0,0,0}
\definecolor{blue}{RGB}{0,0,0}

\newenvironment{noindlist}
 {
 \begin{itemize}}
 {\end{itemize}}
 
\newenvironment{noindlist2}
 {\begin{list}{\labelitemi}{\leftmargin=0.6em \itemindent=0em}}
 {\end{list}}

\newcommand{\comment}[1]{}

\begin{document}
%
\title{Context Aware Computing for \\The Internet of Things: A Survey}
%
%
%

\author{Charith~Perera,~\IEEEmembership{Student~Member,~IEEE,}
        Arkady~Zaslavsky,~\IEEEmembership{Member,~IEEE,} Peter~Christen,
        and~Dimitrios~Georgakopoulos,~\IEEEmembership{Member,~IEEE} \vspace{-0.8cm}

\thanks{Charith~Perera, Arkady~Zaslavsky and Dimitrios~Georgakopoulos are with the Information and Communication Centre, Commonwealth Scientific and Industrial Research Organisation,  Canberra, ACT, 2601, Australia (e-mail: firstname.lastname@csiro.au)}
\thanks{Peter~Christen is with the Research School of Computer Science, The Australian National University, Canberra, ACT 0200, Australia. (e-mail: peter.christen@anu.edu.au)}
\thanks{Manuscript received xxx xx, xxxx; revised xxx xx, xxxx.}}

%
%

\markboth{ IEEE Communications Surveys \& Tutorials,~Vol.~x, No.~x, xxxx~xxxx}%
{Shell \MakeLowercase{\textit{et al.}}: Bare Demo of IEEEtran.cls for Journals}
%



\maketitle

\begin{abstract}
As  we are moving towards the Internet of Things (IoT), the number of sensors deployed  around the world is growing  at a rapid pace. Market research has shown a significant growth of sensor deployments over the past decade and has predicted a significant increment of the growth rate in the future. These sensors continuously generate enormous amounts of data. However, in order to add value to raw sensor data we need to understand it. Collection, modelling, reasoning,  and distribution of context in relation to sensor  data plays critical role in this challenge. Context-aware computing has proven to be successful in understanding sensor data. In this paper, we survey context awareness from an IoT perspective.  We present the necessary background  by introducing the IoT paradigm and context-aware fundamentals at the beginning. Then we provide an in-depth analysis of context life cycle. We evaluate a subset of projects (50) which represent the majority of research and commercial solutions proposed in the field of context-aware computing conducted  over  the last decade (2001-2011)  based on our own taxonomy.  Finally, based on our evaluation, we highlight the lessons to be learnt from the past and some possible directions for future research. The survey addresses a broad range of techniques, methods, models, functionalities,  systems, applications, and middleware  solutions related to context awareness and IoT. Our goal is not only to analyse, compare and consolidate past research work but also to appreciate their findings and discuss their applicability towards the IoT.

\end{abstract}

\begin{IEEEkeywords}
Internet  of things,  context  awareness,  sensor networks,  sensor data, context life cycle,  context reasoning,  context modelling,  ubiquitous, pervasive, mobile, middleware.
\end{IEEEkeywords}

%
\IEEEpeerreviewmaketitle

\section{Introduction}
\label{Introduction}

%
%
%
%

\IEEEPARstart{C}{ontext} awareness, as a core feature of ubiquitous and pervasive computing systems, has existed  and been employed since the early 1990s.  The focus on context-aware computing evolved from desktop applications,  web applications, mobile computing, pervasive/ubiquitous computing to the Internet of Things (IoT) over the last decade. However,  context-aware computing  became more popular with the introduction  of the term  `\textit{ubiquitous  computing}' by Mark Weiser \cite{P506} in his ground-breaking paper \textit{The Computer for the 21st Century} in 1991. Then the term `\textit{context-aware}' was first used by Schilit and Theimer \cite{P173} in 1994.

Since then, research into context-awareness has been  established  as a  well known research area in computer science. Many researchers have proposed definitions  and explanations of different  aspects of context-aware  computing,  as we will discuss briefly in  Section \ref{chapter2:CAF}. The definitions for \textit{`context}' and `\textit{context-awareness}' that are widely accepted by the research community today were proposed by Abowd et al. \cite{P104} in 1999.

During the last two decades, researchers and engineers have developed a significant amount of prototypes, systems, and solutions using context-aware computing techniques. Even though the focus varied depending on each project, one aspect  remained  fairly  unchanged: that is the number of data sources (e.g. software and hardware sources). For example, most of the proposed solutions collect data  from a limited number of physical (hardware) and virtual (software) sensors.  In these situations, collecting and analysing sensor  data from all the sources  is possible  and feasible due to limited numbers.  In  contrast,  IoT  envisions an era where billions of sensors are connected  to the Internet, which means it is not feasible to process all the data collected  by those sensors. Therefore, context-awareness will play a critical role in deciding what data needs to be processed and much more.

Due to advances in sensor technology, sensors are getting more powerful, cheaper  and smaller in  size, which has stimulated  large scale deployments.  As a  result, today we have a large number of  sensors  already deployed  and it  is predicted that the numbers will grow rapidly over the next decade \cite{P029}. Ultimately, these  sensors  will  generate  \textit{big data} \cite{ZMP003}. The data we collect may not have any value unless we analyse, interpret, and understand it. Context-aware computing  has played an important role in tackling this challenge in previous paradigms, such as mobile  and pervasive, which lead us  to believe  that it  would continue  to be successful  in the IoT paradigm as  well.  Context-aware  computing allows us  to store context\footnote{The term \textit{`context}' implicitly provide the meaning of \textit{`information}' according to the widely accepted definition provided by \cite{P104}. Therefore, it is inaccurate to use the term `\textit{context information}' where \textit{`information}' is explicitly mentioned. However, research community and documents on the web frequently use the term `\textit{context information}'. Therefore, we also use both terms interchangeably.} information linked to sensor data so  the interpretation can be done easily and more meaningfully.  In addition, understanding  context makes  it  easier  to perform machine to machine communication  as it is a  core element in the IoT vision.

When large numbers  of  sensors  are deployed,  and start generating data, the traditional application based  approach (i.e. connect sensors directly to applications individually and manually) becomes infeasible. In order to address this inefficiency, significant amounts of middleware solutions are introduced by researchers. Each middleware solution focuses on different aspects  in the IoT, such as  device management,  interoperability, platform portability, context-awareness, security and privacy,  and many more. Even though, some solutions  address multiple aspects, an ideal middleware solution that addresses all the aspects required  by the IoT is yet to be designed.  In this survey, we consider identifying the context-aware computing related features and functionalities  that are required by an ideal IoT middleware solution  as a key task.

There have been several surveys conducted in relation to this field. We briefly introduce  these surveys  in chronological order. Chen and Kotz \cite{P431} (2000) have surveyed  context awareness, focusing  on applications, what context they use, and how contextual information is leveraged. In 2004, Strang and Linnhoff-Popien  \cite{P184} compared the most popular context modelling techniques in the field. Middleware  solutions for sensor networks  are surveyed by Molla and Ahamed \cite{P417} in 2006. Two separate surveys  were conducted by Kjaer \cite{P035} and Baldauf et al. \cite{P402} in 2007 on context-aware systems and middleware solutions using different taxonomies. Both surveys compared limited numbers, but different projects with very little overlap. c et al. \cite{P185} (2009) reviewed popular context representation  and reasoning  from a  pervasive computing  perspective. In 2010, Bettini et al. \cite{P216} also comprehensively  surveyed  context modelling and reasoning by focusing on techniques rather than projects. In the same year another survey was done by Saeed and Waheed  \cite{P359} focusing on architectures in the context-aware middleware domain. Bandyopadhyay et al. \cite{P118} have conducted a survey on existing popular Internet of Things middleware solutions in 2011. \textcolor{blue}{In 2012, Makris et al. \cite{P594} have conducted a survey on context-aware mobile and wireless networking (CAMoWiN) domain where they have identified all the possible components of a typical CAMoWiN architecture.} The latest survey is done by Bellavista et al. \cite{P291} (2013) which is focused on context distribution  for mobile ubiquitous systems.

\begin{figure*}[t]
\centering
\includegraphics[scale=.68]{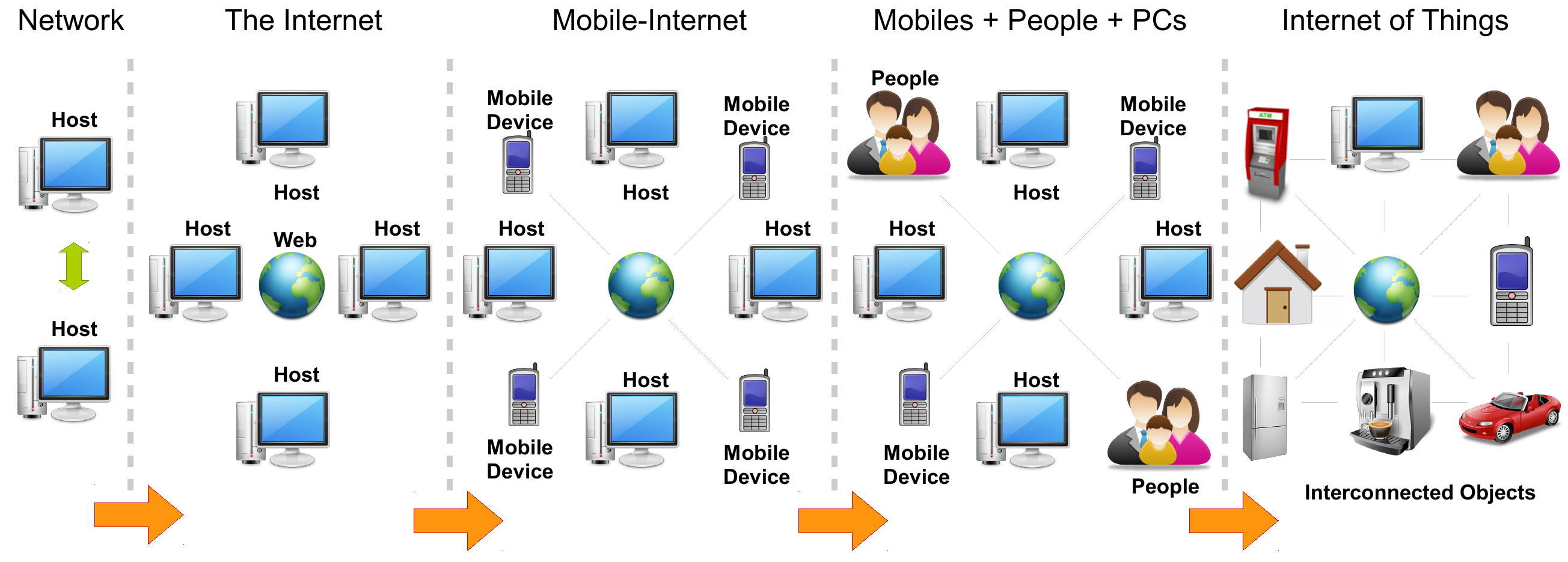}
\vspace{-10pt}
\caption{Evolution of the Internet in five phases. The evolution of Internet begins with connecting two computers together and then moved towards creating World Wide Web by connecting large number of computers together. The mobile-Internet emerged by connecting mobile devices to the Internet. Then, peoples' identities joined the Internet via social networks. Finally, it is moving towards Internet of Things by connecting every day objects
to the Internet.}
\label{Fig:Evolution_of_The_Internet}
\vspace{-12pt}
\end{figure*}

Our survey  differs from the previous literature surveys mentioned above in many ways. Most of the surveys evaluated a limited number of projects. In contrast,  we selected a  large number of  projects (50) covering a  decade,  based  on the unique criteria that will be  explained  at the end of this section. \textcolor{blue}{These projects are different in scale. Some are large scale projects and others corresponds to small scale contributions.} We took a much broader viewpoint  compared to some of the previous surveys, as they have  focused  on specific elements such as modelling, reasoning, etc. Finally and most importantly, our taxonomy formation and organisation  is  completely different. Rather than building  a theoretical taxonomy and then trying to classify existing research projects, prototypes and systems according to it, we use a practical  approach. We built our taxonomy based on past research projects by identifying the features, models, techniques, functionalities and  approaches they employed at higher levels (e.g. we do not consider implementation/code level differences between different solutions). We consolidated this information and analysed the capabilities of each solution or the project. We believe this approach allows us to highlight the areas where  researchers have mostly (priorities)  and rarely (non-priorities)  focused their attention and the reasons behind. Further, we have also used a non-taxonomical  project based evaluation, where we highlight how the different combinations of components are designed, developed and used in each project. This allows  to discuss their applicability from an IoT perspective.

Our objectives in revisiting the literature are threefold: 1) to learn how context-aware  computing techniques have helped  to develop solutions in the past, 2) how can we apply those techniques to solve problems in the future in different  paradigms such as the IoT, and 3) to highlight open challenges and to discuss future research directions.

This paper is organised into sections  as follows: Section  \ref{chapter2:IoTP} provides an introduction to the IoT. In this section, we briefly describe  the history and evolution  of the Internet. Then we explain what the IoT is, followed by a list of application  domains and statistics that show the significance of  the IoT. We  also  describe  the relationship between  sensor networks and the IoT. Comparisons of popular IoT middleware solutions are presented at the end of the section in order to highlight existing   research  gaps.  In  Section \ref{chapter2:CAF}, we present  context awareness fundamentals  such as context-aware  related definitions, context types and categorisation schemes, features and characteristics, and context awareness management design principles. In Section \ref{chapter2:CDLC}, we conduct our main discussion based on context  life cycle where we identify four stages: acquisition,  modelling,  reasoning, and distribution.  Section \ref{chapter2:PRE} briefly discusses the highlights of each project,  which we use for the comparison later. Finally, Section \ref{chapter2:LL} discusses  the lessons learn  from the literature and Section \ref{chapter2:LLFRD}  identifies future research directions and challenges. Conclusion remarks are presented in Section \ref{chapter2:Conclusions}.

For this literature review, we analyse, compare, classify a subset of both small scale and large scale projects (50) which represent the majority of research and commercial solutions proposed in the field of context-aware computing based on our own taxonomy. We selected the existing solutions to be reviewed based on different criteria. Mainly, we selected projects that were conducted over the last decade (2001-2011).  We also considered main focus, techniques used, popularity,  comprehensiveness, information availability,  and the year of publication, in order to make sure that our review provides  a balanced view on context-aware computing research.

\section{The Internet of Things Paradigm}
\label{chapter2:IoTP}

In  this section, we briefly introduce the IoT paradigm.  Our intention is not to survey the IoT, but to present some fundamental information (e.g. how Internet evolved, what is the IoT, statistics related to IoT, underline technologies, characteristics, and research gaps in IoT paradigm) that will help with understanding the historic movements and the direction into which technology  is moving today. The IoT paradigm  has  its own concepts and characteristics. It also shares significant amounts of concepts with other computer fields. The IoT bundles different technologies (e.g. sensor hardware/firmware, semantic, cloud, data modelling, storing, reasoning, processing, communication technologies) together to build its vision. We apply the existing technologies in different ways based on the characteristics and demands of the IoT. The IoT does not revolutionise our lives  or the field of computing. It is another step in the evolution of the Internet we already have.

\subsection{Evolution of Internet}
\label{chapter2:IoTP:Evolution of Internet}

Before we investigate the  IoT in depth, it  is worthwhile to look at the evolution  of the Internet. In the late 1960s,  communication between  two computers  was made  possible through a computer network \cite{P260}. In the early 1980s the TCP/IP stack was introduced.  Then, commercial   use of the Internet started in the late 1980s. Later, the World Wide Web (WWW) became available  in 1991 which made the Internet more popular and stimulate the rapid growth. Web of Things (WoT) \cite{P575}, which  based on WWW, is a part of IoT.

Later, mobile devices connected to the Internet and formed the mobile-Internet \cite{P018}. With the emergence of social networking, users started to become connected together over the Internet. The next step in the IoT is where objects around us will be able to connect to each other (e.g. machine to machine) and communicate via the Internet \cite{P006}. Figure \ref{Fig:Evolution_of_The_Internet} illustrates the five phases in the evolution of the Internet.

\subsection{What is the Internet of Things?}
\label{chapter2:IoTP:What is Internet of Things?}

\begin{figure}[b!]
 \centering
 \vspace{-8pt}
 \includegraphics[scale=1.1]{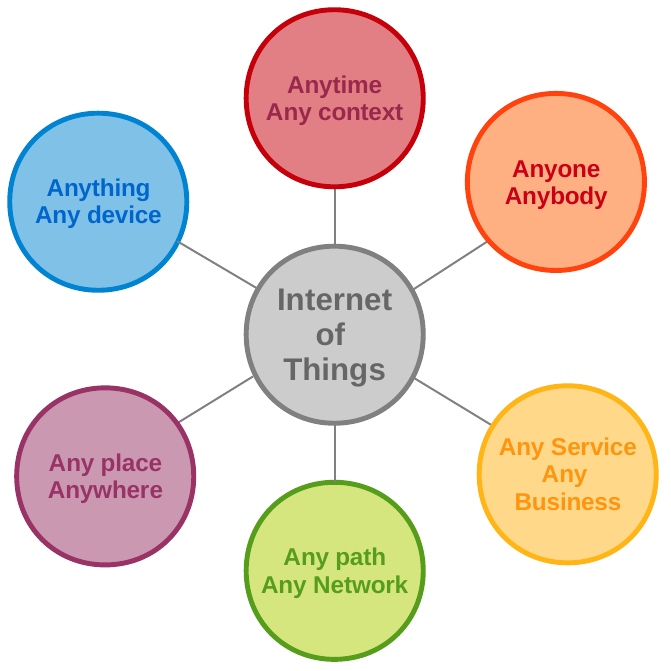}
 \vspace{-8pt}
 \caption{Definition of the Internet of Things: The Internet of Things allows people and things to be connected anytime, anyplace, with anything and anyone, ideally using any path/network and any service \cite{P019}.}
 \label{Fig:Definition_of_IoT}
 \vspace{-05pt}
\end{figure}

During the past decade, the IoT  has gained significant  attention in academia as well as industry.  The main reasons behind  this interest are the capabilities that the IoT  \cite{P007, P003} will  offer. It promises to create  a  world where all the objects (also called smart objects \cite{P041}) around us are connected to the Internet and communicate with each other with minimum human intervention \cite{P026}. The ultimate goal is to create `a better world for human beings', where objects around us know what we like, what we want, and what we need and act accordingly without explicit instructions \cite{P040}.

The term `Internet of Things' was firstly coined by Kevin Ashton \cite{P065} in a presentation in 1998. He has mentioned \textit{``The Internet of Things has the potential to change the world, just as the Internet did. Maybe even more so''}. Then, the MIT Auto-ID centre presented their IoT vision in 2001 \cite{P361}. Later, IoT was formally introduced by the International Telecommunication Union (ITU) by the \textit{ITU Internet report} in 2005 \cite{P020}.

The IoT encompasses a significant amount of technologies that drive its vision. In the document, \textit{Vision and challenges for realising the Internet of Things}, by CERP-IoT \cite{P029}, a comprehensive  set of technologies  was listed. IoT is a very broad vision. The research into the IoT is still in its infancy. Therefore,  there aren't any standard definitions  for IoT. The following definitions  were provided by different researchers.

\begin{noindlist2}
 \item Definition by \cite{P002}:  \textit{``Things have identities and virtual personalities operating in smart spaces using intelligent interfaces to connect and communicate within social, environment, and user contexts.''}  
 
  \item Definition by \cite{P006}:\textit{``The semantic origin of the expression is composed by two words and concepts: Internet and Thing, where Internet can be defined as the world-wide network of interconnected computer networks, based on a standard communication protocol, the Internet suite (TCP/IP), while Thing is an object not precisely identifiable Therefore, semantically, Internet of Things means a world-wide network of interconnected objects uniquely addressable, based on standard communication protocols.''}

   \item Definition by \cite{P019}: \textit{``The Internet of Things allows people and things\footnote{We use both terms, `\textit{objects}' and `\textit{things}' interchangeably to give the same meaning as they are frequently used in IoT related documentation. Some other terms used by the research community are `smart objects', `devices', `nodes'.} to be connected Anytime, Anyplace, with Anything and Anyone, ideally using Any path/network and Any service.''}
\end{noindlist2}

We accept the last definition provided by \cite{P019} for our research work, because we believe, this definition encapsulates the broader vision of IoT. Figure \ref{Fig:Definition_of_IoT} illustrates the definition more clearly. The broadness of IoT can be identified by evaluating the application domains presented in Section \ref{chapter2:IoTP:IoT Application Domains}.

\subsection{IoT Application Domains}
\label{chapter2:IoTP:IoT Application Domains}

The IoT, interconnection and communication  between everyday objects, enables many applications  in many domains. The application domain can be mainly divided in to three categories based on their focus \cite{P003, P029}: industry, environment, and society. The magnitude of the applications  can be seen in the statistics presented in Section \ref{chapter2:IoTP:IoT Related Statistics}.

Supply chain management \cite{P017}, transportation and logistics \cite{P005}, aerospace, aviation, and automotive are some of the industry focused applications of IoT. Telecommunication, medical technology \cite{P362}, healthcare, smart building, home \cite{P363} and office, media, entertainment, and ticketing are some of the society focused applications of IoT. Agriculture and breeding \cite{P244, P013}, recycling, disaster alerting, environmental monitoring are some of the environment focused applications.

Asin and Gascon \cite{P416} listed 54 application  domains under twelve  categories:  smart cities, smart environment, smart water, smart metering, security and emergencies, retail, logistics, industrial control, smart agriculture, smart animal farming, domestic and home automation, and eHealth.

\subsection{IoT Related Statistics}
\label{chapter2:IoTP:IoT Related Statistics}

The vision of the IoT is heavily  energised by statistics and predictions.  We present the statistics to justify our focus on the IoT and to show the magnitude of the challenges. It is estimated that there about 1.5 billion Internet-enabled PCs and over 1 billion Internet-enabled  mobile phones  today. These two categories  will  be joined with Internet-enabled devices (smart objects \cite{P041}))  in the future. By 2020, there will  be 50 to 100 billion devices connected to the Internet \cite{P029}.

According to BCC Research \cite{P255}, the global market for sensors was around \$56.3 billion in 2010. In 2011, it was around \$62.8 billion. Global market for sensors is expected to increase to \$91.5 billion by 2016, at a compound annual growth rate of 7.8\%.

\subsection{The Essential Component of IoT: Sensor Networks}
\label{chapter2:IoT:The Backbone of IoT: Sensor Networks}

We provide  a brief introduction to sensor networks  in this section  as it is the most essential component of the IoT. A sensor network  comprises one or more sensor nodes, which communicate between themselves using wired and wireless technologies. In sensor networks, sensors  can be homogeneous  or heterogeneous. Multiple sensor networks  can be connected together through different technologies  and protocols. One such approach  is through the Internet. The components and the layered structure of a typical sensor network  are discussed in Section  \ref{chapter2:IoT:Layers in Sensor Networks}.

We discuss how sensor networks  and the IoT work together in Section \ref{chapter2:IoT:Relationship Between Sensor Networks and IoT}. However, there are other technologies that can complement the sensing and communication infrastructure in IoT paradigm  such as traditional  ad-hoc networks. These are clearly a different technology from sensor networks  and have many weaknesses. The differences are comprehensively  discussed in  \cite{P009}.

There are three main architectures in sensor networks: flat architecture  (data transfers  from static sensor  nodes  to the sink node using a multi-hop  fashion), two-layer architecture (more static and mobile sink nodes are deployed to collect data from sensor nodes), and three-layer architecture (multiple  sensor networks are connected together over the Internet). Therefore, IoT follows a three-layer architecture.

Most of the sensors deployed  today are wireless.  There are several major wireless technologies used to build wireless sensor  networks: wireless personal   area  network (WPAN) (e.g. Bluetooth),  wireless local area  network (WLAN) (e.g. Wi-Fi), wireless metropolitan   area network (WMAN)  (e.g. WiMAX), wireless wide area network  (WWAN) (e.g. 2G and 3G networks),  and satellite network (e.g. GPS). Sensor networks  also use  two types of protocols  for communication: non-IP based (e.g: Zigbee and Sensor-Net) and IP-based protocols (NanoStack, PhyNet, and IPv6).

The sensor  network is not a  concept  that emerged  with
the IoT. The concept of a sensor network  and related research existed a long time before the IoT was introduced. However, sensor networks were used in limited domains to achieve specific purposes,  such as  environment    monitoring  \cite{P193}, agriculture \cite{P244}, medical care \cite{P158}, event detection \cite{P113}, structural health monitoring \cite{P067}, etc. Further, there are three categories of sensor networks that comprise the IoT \cite{P266}: body sensor networks (BSN), object sensor networks (OSN), and environment sensor networks (ESN).

Molla and Ahamed \cite{P417} identified  ten challenges that
need to be considered when developing  sensor network  middleware  solutions:  abstraction  support,  data fusion, resource constraints,  dynamic topology, application  knowledge,  programming paradigm, adaptability, scalability, security, and QoS support. A comparison of different sensor network  middleware solutions is also provided  based on the above parameters. Several selected projects  are also discussed in brief in order to discover the approaches they take to address various challenges associated with sensor networks.

Some of the major sensor network middleware approaches are IrisNet, JWebDust, Hourglass, HiFi, Cougar, Impala, SINA, Mate, TinyDB, Smart Object, Agilla, TinyCubus, TinyLime, EnviroTrack, Mires, Hood, and Smart Messages. \textcolor{blue}{Some of the above approaches are surveyed in \cite{P417, P086}.} A survey on web based wireless sensor architectures and applications is presented in \cite{P475}.

\subsection{Layers in Sensor Networks}
\label{chapter2:IoT:Layers in Sensor Networks}

\begin{figure}[h]
 \centering
    \vspace{-5pt}
 \includegraphics[scale=.56]{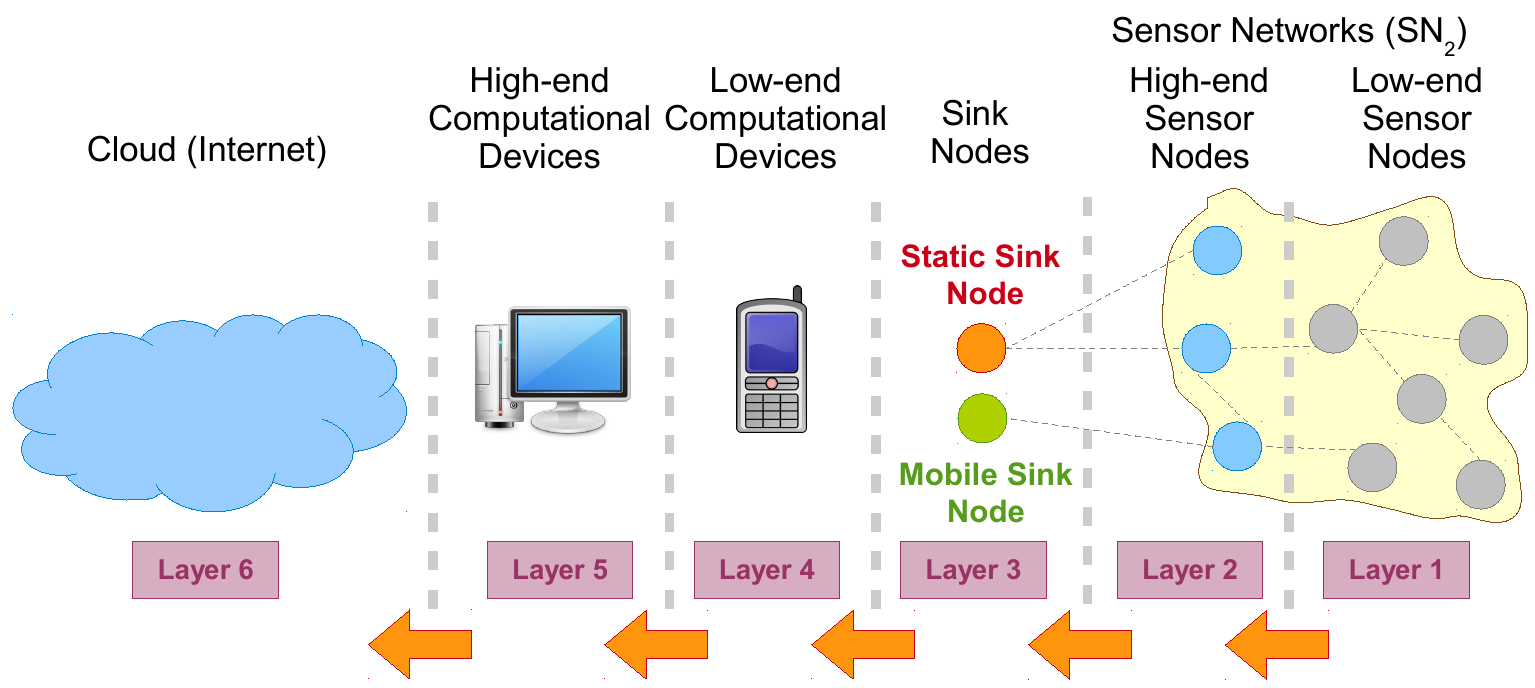}
 \caption{Layered structure of a sensor network: These layers are identified based on the capabilities posed by the devices. In IoT, this layered architecture may have additional number of sub layers as it is expected to comprises large verity of in sensing capabilities.}
 \label{Fig:Layered Structure on a Sensor Network}
   \vspace{-5pt}
\end{figure}

We have presented a typical structure of a sensor  network in Figure  \ref{Fig:Layered Structure on a Sensor Network}. It comprises the most common components in a sensor network.  As we have shown, with the orange coloured arrows, data flows from right to left. Data is generated by the low-end sensor nodes  and high-end  sensor nodes. Then, data is collected by mobile and static sink nodes.  The sink nodes send the data to low-end computational devices. These devices perform a certain amount of processing on the sensor data. Then, the data is sent to high-end  computational  devices to be processed further. Finally,  data reaches the cloud where it will be shared, stored, and processed significantly.

Based on the capabilities of the devices involved in a sensor network, we have identified  six layers. Information can be processed in any layer. Capability means the processing, memory, communication,  and energy capacity. Capabilities increase  from layer one to layer six. Based  on our identification of  layers, it  is evident  that an ideal system should understand the capability  differences, and perform data management accordingly.  It is all about efficiency and effectiveness. For example, perform processing in the first few layers could reduce data communication.  However,  devices in the first few layers do not have a sufficient amount of energy  and processing  power to do comprehensive  data processing \cite{P318}. \textcolor{blue}{IoT research needs to} find more efficient and effective  ways of  data management,  such as collecting, modelling, reasoning, distributing.

\subsection{Relationship Between Sensor Networks and IoT}
\label{chapter2:IoT:Relationship Between Sensor Networks and IoT}

\begin{figure}[h]
   \vspace{-8pt}
 \centering
 \includegraphics[scale=.40]{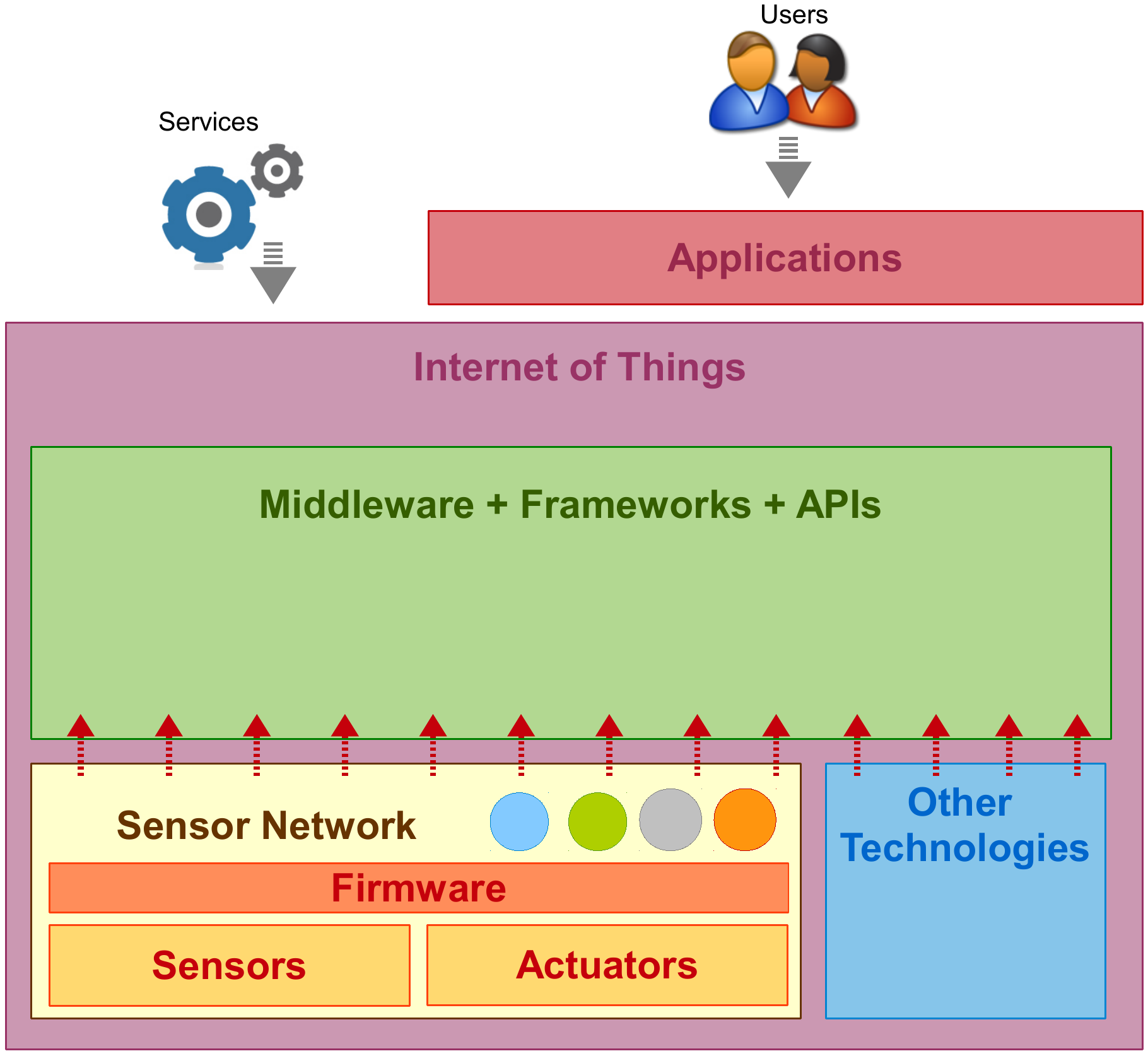}
  \vspace{-7pt}
 \caption{Relationship between sensor networks and IoT.}
 \label{Fig:Relationship Between Sensor Networks and IoT}
  \vspace{-7pt}
\end{figure}

In  earlier sections  we introduced both IoT  and sensor network  concepts. In this section we explain the relationship between the two concepts. Previously,  we argued that sensor networks are the most essential components of the IoT. Figure \ref{Fig:Relationship Between Sensor Networks and IoT} illustrates the big picture. The IoT comprises  sensors and actuators. The data is collected using sensors.  Then, it is processed and decisions are made. Finally, actuators perform the decided actions. This process is further discussed in Section \ref{chapter2:CDLC}. Further, integration  between wireless  sensor networks and the IoT are comprehensively  discussed in  \cite{P351}.  The difference  between sensor networks  (SN) and the IoT is largely unexplored  and blurred. We can elaborate  some of the characteristics  of both SN and IoT to identify the differences.

\begin{noindlist2}

 \item SN comprises of the sensor hardware  (sensors and actuators), firmware and a thin layer of software. The IoT comprises everything  that SN comprises and further it comprises a  thick layer of software  such as middleware   systems, frameworks, APIs and many more software components. The software layer is installed across computational  devices (both low and high-end) and the cloud.

 \item From their origin, SNs were designed, developed, and used for specific application purposes, for example, detecting bush fire \cite{P266}. In the early days, sensor networks were largely used  for monitoring purposes and not for actuation \cite{P277}. In contrast, IoT is not focused on specific applications. The IoT can be explained  as a general purpose sensor network  \cite{P285}. Therefore, the IoT should support many kinds of applications. During the stage  of deploying  sensors, the IoT would not be targeted to collect specific types of sensor data, rather it would deploy  sensors where they can be used for various application domains. For example, company may deploy sensors,  such as  pressure   sensors,  on a  newly built bridge to track its structural health. However, these sensors may be reused and connect with many other sensors in order to track traffic at a  later stage. Therefore, middleware solutions, frameworks, and APIs are designed to provide generic  services and functionalities such as intelligence, semantic interoperability,  context-awareness, etc. that are required to perform communication between sensors and actuators effectively.

 \item Sensor networks can exist without the IoT. However, the IoT cannot exist  without SN, because  SN  provides the majority of hardware (e.g. sensing and communicating) infrastructure support, through providing access to sensors and actuators. There are several other technologies  that can provide  access to sensor hardware,  such as wireless  ad-hoc  networks.  However, they are not scalable and cannot accommodate the needs of the IoT individually  \cite{P009}, though they can complement the IoT infrastructure. As is clearly depicted in Figure   \ref{Fig:Relationship Between Sensor Networks and IoT}, SN are a part of the IoT. However, the IoT is not a part of SN.

\end{noindlist2}

\subsection{Characteristics of the IoT}
\label{chapter2:IoT:Characteristics of IoT}

In Section \ref{chapter2:IoT:Relationship Between Sensor Networks and IoT}, we highlighted  the differences between sensor networks and the IoT. Further, we briefly explore the characteristics of the IoT from a research perspective. Based on previous research efforts \textcolor{blue}{we identified seven major characteristics} in the IoT \cite{P029}: \textit{intelligence},  \textit{architecture}, \textit{complex system}, \textit{size considerations}, \textit{time considerations}, \textit{space considerations}, and \textit{everything-as-a-service}. These characteristics need to be considered when developing IoT solutions throughout all the phases from design, development, implement and evaluation.

\vspace{-0.2cm}

\begin{noindlist2}

 \item \textbf{Intelligence:} This means the application of knowledge. First the knowledge  needs to be generated by collecting data and reasoning  it. Transforming the collected raw data into knowledge (high-level  information)  can be done by collecting, modelling, and reasoning the context. Context can be used  to fuse sensor  data together to infer new knowledge. Once we have knowledge,  it can be applied towards more intelligent interaction and communication.

 \item \textbf{Architecture:} IoT should be facilitated by a  hybrid architecture which comprises  many different architectures. Primarily there would be two architectures: event driven \cite{P038}  and time driven. Some  sensors produce  data when an event  occurs (e.g. door sensor);  the rest  produce  data continuously,   based on specified time frames (e.g. temperature  sensor). Mostly, the IoT and SN are event driven  \cite{P275}. Event-Condition-Action (ECA) rules are commonly  used in such systems.

 \item \textbf{Complex system:} The IoT comprises a large number of objects (sensors and actuators) that interact autonomously. New objects will  start communicating and existing  ones will  disappear. Currently, there are millions of sensors deployed  around the world  \cite{P069}. Interactions  may differ significantly  depending on the objects capabilities. Some objects  may have  very few capabilities, and as such store very limited  information and do no processing  at all. In contrast,  some objects may have  larger memory, processing, and reasoning capabilities,  which make them more intelligent.

 \item \textbf{Size considerations:} It is predicted that there will be 50-100 billion devices connected to the Internet by 2020 \cite{P029}. The IoT needs to facilitate the interaction  among these objects. The numbers will grow continuously and will never decrease. Similar to the number of objects, number of interactions may also increase significantly.

 \item \textbf{Time considerations:} The IoT could handle billions of parallel  and simultaneous events, due to the massive number of interactions. Real-time data processing is essential.

 \item \textbf{Space considerations:} The precise geographic location of a object will be critical \cite{P083} as location  plays a significant role in context-aware computing.  When the number of objects get larger, tracking becomes  a  key requirement. Interactions are highly dependent on their locations, their surroundings, and presence of other entities (e.g. objects and people).
 
 \item \textbf{Everything-as-a-service:} Due to the popularity of cloud computing \cite{P498}, consuming resources as a service \cite{P502} such as Platform-as-a-Service  (PaaS), Infrastructure-as-a-Service  (IaaS), Software-as-a-Service   (SaaS),  has become main stream. Everything-as-a-service   \cite{P533} model is highly efficient,  scalable, and easy to use. IoT demands significant amounts of infrastructure to be put in place in order to make its vision a reality, where it would follow a community  or crowd based approach. Therefore, sharing would be essential, where an everything-as-a-service model would suit mostly sensing-as-a-service \cite{ZMP003}.
\end{noindlist2}

\subsection{Middleware Support for IoT}
\label{chapter2:IoT:Middleware Support for IoT}

As we mentioned at the beginning, the IoT needs to be supported by middleware solutions. \textit{``Middleware is a software layer that stands between the networked operating system and the application and provides well known reusable solutions to frequently encountered problems like heterogeneity, interoperability, security, dependability \cite{P064}.''} The functionalities required by IoT middleware solutions are explained in detail in \cite{P029, P018, P006, P019, P020}. In addition, challenges in developing middleware solutions for the IoT are discussed in \cite{P028}. We present the summary of a survey conducted by Bandyopadhyay et al. \cite{P118}. They have selected the leading middleware solutions and analyse them based on their functionalities,  each one offers,  \textit{device management}, \textit{interoperation}, \textit{platform portability}, \textit{context-awareness}, and \textit{security and privacy}. Table \ref{Tbl:IoT Middleware Comparison} shows the survey results. By the time we were preparing this survey, some of the middleware solutions listed (i.e. GSN and ASPIRE) were in the processing of extending  towards next generation solutions (i.e. EU FP7 project OpenIoT (2012-2014)   \cite{P377}) by combining each other's strengths.

\begin{table}[h]
\centering
\footnotesize
\renewcommand{\arraystretch}{1.3}
\caption{IoT Middleware Comparison \cite{P118}}
\vspace{-0.3cm}

\begin{tabular}{l m{0.85cm} m{0.85cm} m{0.85cm} m{0.85cm} m{0.85cm}}\hline 

 Middleware  & DM    & I    &PP   &CA   &SP   \\ \hline \hline

Hydra \cite{P105}           &$\checkmark$      &$\checkmark$     &$\checkmark$       &$\checkmark$    &$\checkmark$ \\
ISMB \cite{P375}            &$\checkmark$      &$\times$         &$\checkmark$       &$\times$        &$\times$   \\ 
ASPIRE \cite{P366}          &$\checkmark$      &$\times$         &$\checkmark$       &$\times$        &$\times$  \\ 
UBIWARE \cite{P146}        &$\checkmark$      &$\times$         &$\checkmark$       &$\checkmark$    &$\times$   \\ 
UBISOAP \cite{P367}         &$\checkmark$      &$\checkmark$     &$\checkmark$       &$\times$        &$\times$   \\ 
UBIROAD \cite{P119}         &$\checkmark$      &$\checkmark$     &$\checkmark$       &$\checkmark$    &$\checkmark$   \\ 
GSN    \cite{P050}          &$\checkmark$      &$\times$         &$\checkmark$       &$\times$        &$\checkmark$   \\ 
SMEPP  \cite{P371}          &$\checkmark$      &$\times$         &$\checkmark$       &$\checkmark$    &$\checkmark$   \\ 
SOCRADES  \cite{P373}       &$\checkmark$      &$\checkmark$     &$\checkmark$       &$\times$        &$\checkmark$   \\ 
SIRENA   \cite{P368}       &$\checkmark$      &$\checkmark$     &$\checkmark$       &$\times$        &$\checkmark$   \\ 
WHEREX    \cite{P370}       &$\checkmark$      &$\checkmark$     &$\checkmark$       &$\times$        &$\times$   \\ \hline

\multicolumn{6}{p{0.42\textwidth}}{\textit{Legend}: Device Management (DM), Interoperation (I), Platform Portability (PP), Context Awareness (CA), Security \& Privacy (SP)}

\end{tabular}

\label{Tbl:IoT Middleware Comparison}
\vspace{-0.6cm}
\end{table}

\subsection{Research Gaps}
\label{chapter2:IoT:Research Gaps to Improve}

According to Table  \ref{Tbl:IoT Middleware Comparison}, it can be seen that the majority of  the IoT  middleware  solutions do not provide context-awareness functionality. In contrast, almost all the solutions are highly focused  on device management,  which involves connecting   sensors to the IoT  middleware.  In  the early days, context-awareness was strongly  bound to pervasive and ubiquitous computing. Even though there were some middleware solutions that provided an amount of context-aware functionality, they did not satisfy the requirements that the IoT demands. We  discuss the issues  and drawbacks  with existing solutions, in detail, in Section  \ref{chapter2:PRE}. We  discuss some of the research directions in Section \ref{chapter2:LLFRD}. 

In this section, we introduced the IoT paradigm and highlighted the importance of context-awareness for the IoT. We also learnt that context-awareness has not been addressed in existing IoT focused solutions,  which motivates  us to survey the solutions in other paradigms to evaluate the applicability of context-aware computing techniques toward IoT. In the next section we discuss context-aware  fundamentals that helps us understand the in-depth discussions in the later sections.

\section{Context Awareness Fundamentals}
\label{chapter2:CAF}

This section discusses definitions  of context and context awareness,  context-aware   features,  types of context and categorisation   schemes,  different levels and characteristics of context-awareness,  and finally, context management design principles in the IoT paradigm.

\subsection{Context-awareness Related Definitions}
\label{chapter2:CAF:Context-awareness Related Definitions}

\subsubsection{Definition of Context}
\label{chapter2:CAF:CARD:Definition of Context}

The term context has been defined by many researchers. Dey et al.  \cite{P143} evaluated and highlighted  the weaknesses of these definitions.  Dey claimed  that the definition provided by Schilit and Theimer  \cite{P173} was based on examples and cannot be used to identify new context. Further, Dey claimed that definitions provided by Brown  \cite{P175}, Franklin and Flachsbart \cite{P178}, Rodden et al. \cite{P181}, Hull et al. \cite{P179}, and Ward et al. \cite{P183} used synonyms  to refer to context,  such as environment   and situation.  Therefore,  these definitions  also cannot be used to identify new context. Abowd and Mynatt  \cite{P115} identified the five W's (Who, What, Where, When, Why) as the minimum  information that is necessary to understand context. Schilit et al.  \cite{P116} and Pascoe \cite{P180} have also defined the term context. Dey claimed that these definitions were too specific and cannot be used to identify context in a broader sense and provided a definition  for context  as follows:

\textit{``Context is any information that can be used to characterise the situation of an entity. An entity is a person, place, or object that is considered relevant to the interaction between a user and an application, including the user and applications themselves \cite{P104}.''} 

We accept the definition of context provided by Abowd et al. \cite{P104} to be used  in this research  work, because  this definition can be used to identify context from data in general.  If we consider  a data element, by using this definition, we can easily identify whether the data element is context   or not. A number of dictionaries  have also defined and explained the word context:

\begin{noindlist2}
 \item Synonyms \cite{P437}: \textit{``Circumstance, situation, phase, position, posture, attitude, place, point; terms; regime; footing, standing, status, occasion, surroundings, environment, location, dependence.''}
 \item Definition by FOLDOC \cite{P438}: \textit{``That which surrounds, and gives meaning to, something else.''}
 \item Definition by WordNet \cite{P439}: \textit{``Discourse that surrounds a language unit and helps to determine its interpretation''}
 \item Definition by Longman \cite{P440}: \textit{``The situation, events, or information that are related to something and that help you to understand it''}
\end{noindlist2}

In addition, Sanchez et al. \cite{P344} explained  the distinction between raw data and context information  as follows:
\begin{noindlist2}
 \item \textbf{Raw (sensor) data:} Is unprocessed and retrieved directly from the data source, such as sensors.
 \item \textbf{Context information:} Is generated by processing raw sensor  data. Further,  it is checked for consistency and meta data is added.
\end{noindlist2}
For example, the sensor readings produced by GPS sensors can be considered as raw sensor data. Once we put the GPS sensor readings in such a way that it represents a geographical location, we call it context information. Therefore in general, the raw data values produced by sensors can be considered as data. If this data can be used to generate context information, we identify these data as context. Therefore, mostly what we capture from sensors are data not the context information.

  Ahn and Kim  \cite{P278} define context (also called compound events)  as a set of interrelated  events with logical and timing relations among them. They also define an event  as an occurrence that triggers a condition in a target area. There are two categories of events: discrete events and continuous  events. If the sampling rate is \textit{p}:
\begin{noindlist2}
 \item \textbf{Discrete events:} An event that occurs at time t and t + \textit{p}, there are considered to have been two separate event instances. (e.g. a door open, lights on, etc.)
 
 \item \textbf{Continuous events:} An event instance lasting for at least time \textit{p}, where an event occurring at time t and t + \textit{p}, cannot be considered  as two separate events. (e.g. raining,  having a shower, driving a car, etc.)
\end{noindlist2}

\subsubsection{Definition of Context-awareness}
\label{chapter2:CAF:CARD:Definition of Context-awareness}

The term context awareness, also called sentient, was first introduced by Schilit and Theimer  \cite{P173}  in 1994. Later, it was defined  by Ryan et al.  \cite{P182}. In both cases, the focus was on computer applications and systems. As stated by Abowd et al.  \cite{P104}, those definitions  are too specific and cannot be used to identify whether a given system is a context-aware  system or not. Therefore, Dey has defined the term context-awareness  as follows:

\textit{``A system is context-aware if it uses context to provide relevant information and/or services to the user, where relevancy depends on the user's task. \cite{P104}''}

We accept the above definition  on context-awareness to be used in our research work, because we can use this definition to identify context-aware systems from the rest. If we consider a  system,   by using this definition we can easily identify whether this system is a context-aware  system or not. Context awareness frameworks   typically should support acquisition, representation, delivery, and reaction  \cite{P143}. In addition, there are three main approaches that we can follow to build context-aware applications   \cite{P339}.

\begin{noindlist2}
 \item \textbf{No application-level context model:} Applications perform all the actions, such as  context acquisition, pre-processing,  storing, and reasoning within  the application boundaries.
 
 \item \textbf{Implicit context model:} Applications uses libraries, frameworks, and toolkits to perform context acquisition, pre-processing, storing, and reasoning tasks. It provides a standard design to follow that makes it easier to build the applications quickly. However, still the context  is hard bound to the application.
 
 \item \textbf{Explicit context model:} Applications  uses a context management infrastructure  or middleware solution. Therefore, actions such as context acquisition,  pre-processing, storing, and reasoning lie outside the application boundaries. Context management  and application are clearly separated and can be developed and extend independently.
\end{noindlist2}

\subsubsection{Definition of Context Model and Context Attribute}
\label{chapter2:CAF:CARD:Definition_of_Context_Model_and_Context_Attribute}

We adopt the following interpretations of context model and context attributes provided by Henricksen \cite{P389} based on Abowd et al. \cite{P104} in our research work.

\textit{``A context model identifies a concrete subset of the context that is realistically attainable from sensors, applications and users and able to be exploited in the execution of the task. The context model that is employed by a given context-aware application is usually explicitly specified by the application developer, but may evolve over time \cite{P389}.''}

\textit{``A context attribute is an element of the context model describing the context. A context attribute has an identifier, a type and a value, and optionally a collection of properties describing specific characteristics \cite{P389}.''}

\subsubsection{Definition of Quality of Context}
\label{chapter2:CAF:CARD:Definition of Quality of Context}

There are number of definitions and parameters  that have  been proposed  in the literature regarding quality of context (QoC). A survey on QoC is presented in \cite{P291}. QoC is defined using a set of parameters that expresses the quality of requirements and properties of the context data. After evaluating a number of different parameter proposals in the literature, \cite{P291} has defined QoC based on three parameters: context data validity, context data precision, and context data up-to-dateness. QoC are being used to resolve context data conflicts. Further, they claim that QoC is depend on quality of the physical sensor, quality of the context data, and quality of the delivery process.

\subsection{Context-aware Features}
\label{chapter2:CAF:Context-aware Features}

After analysing and comparing the two previous efforts conducted  by Schilit et al.  \cite{P116} and Pascoe \cite{P180}, \textcolor{blue}{Abowd et al. \cite{P104} identified} three features that a context-aware  application  can support:  presentation,  execution, and tagging. Even though, the IoT vision was  not known at the time these  features  are identified,  they are highly applicable to the IoT paradigm  as well. We elaborate these features from an IoT perspective.

\begin{noindlist2}
 
 \item \textbf{Presentation:} Context  can be used to decide what information  and services need to be presented to the user. Let us consider a smart \cite{P007} environment  scenario. When a user enters a supermarket   and takes their smart phone out, what they want to see is their shopping list. Context-aware mobile  applications  need to connect to kitchen appliances  such as a smart refrigerator \cite{P352} in the home to retrieve  the shopping list and present it  to the user. This provides the idea of presenting information  based on context such as location,  time, etc. By definition, IoT promises to provide any service anytime, anyplace, with anything and anyone, ideally using any path/network.

 \item \textbf{Execution:} Automatic execution  of services  is also a critical feature in the IoT paradigm. Let us consider  a smart home  \cite{P007} environment. When a user  starts driving home from their office, the IoT application employed  in  the house should switch  on the air condition system and switch on the coffee machine  to be ready to use  by the time the user  steps into their house. These actions need to be taken automatically  based on the context. Machine-to-machine communication is a significant  part of the IoT.

 \item\textbf{Tagging:} In the IoT paradigm,  there will be a  large number of sensors attached to everyday objects. These objects will produce  large volumes  of sensor  data that has  to be collected, analysed,  fused and interpreted \cite{P109}. Sensor data produced by a  single sensor  will  not provide the necessary information that can be used to fully understand the situation.  Therefore,  sensor data collected  through multiple sensors  needs  to be fused together.  In order to accomplish the sensor data fusion task, context needs to be collected.  Context needs to be tagged together with the sensor data to be processed and understood later. Context annotation  plays a  significant role in context-aware computing research. We also call this \textit{tagging} operation as \textit{annotation} as well.

\end{noindlist2}

\subsection{Context Types and Categorisation Schemes}
\label{chapter2:CAF:context Types}

Different researchers  have  identified context types differently based of different perspectives. Abowd et al. \cite{P104} introduced  one of the leading  mechanisms of defining context types. They  identified location, identity, time, and activity as the primary context types. Further, they defined  secondary  context  as the context  that can be found using primary context. For example, given primary  context such as a person's identity,  we can acquire many pieces of related information such as phone numbers, addresses, email  addresses, etc.

However, using this definition we are unable to identify the type of a given  context. Let us consider two GPS sensors located in two different locations. We can retrieve \textcolor{blue}{GPS values} to identify the position of each sensor. However,  we can only find the distance between the two sensors by performing calculations based  on the raw values  generated by the two sensor.  The question is, `what is the category that \textit{distance} belongs to?' `is it primary or secondary?' The \textit{distance} is not just a value  that we sensed. We  computed  the \textit{distance} by fusing two pieces of context. The above definition does not represent this accurately.

\begin{figure}[h!]
 \centering
  \vspace{2pt}
 \includegraphics[scale=.44]{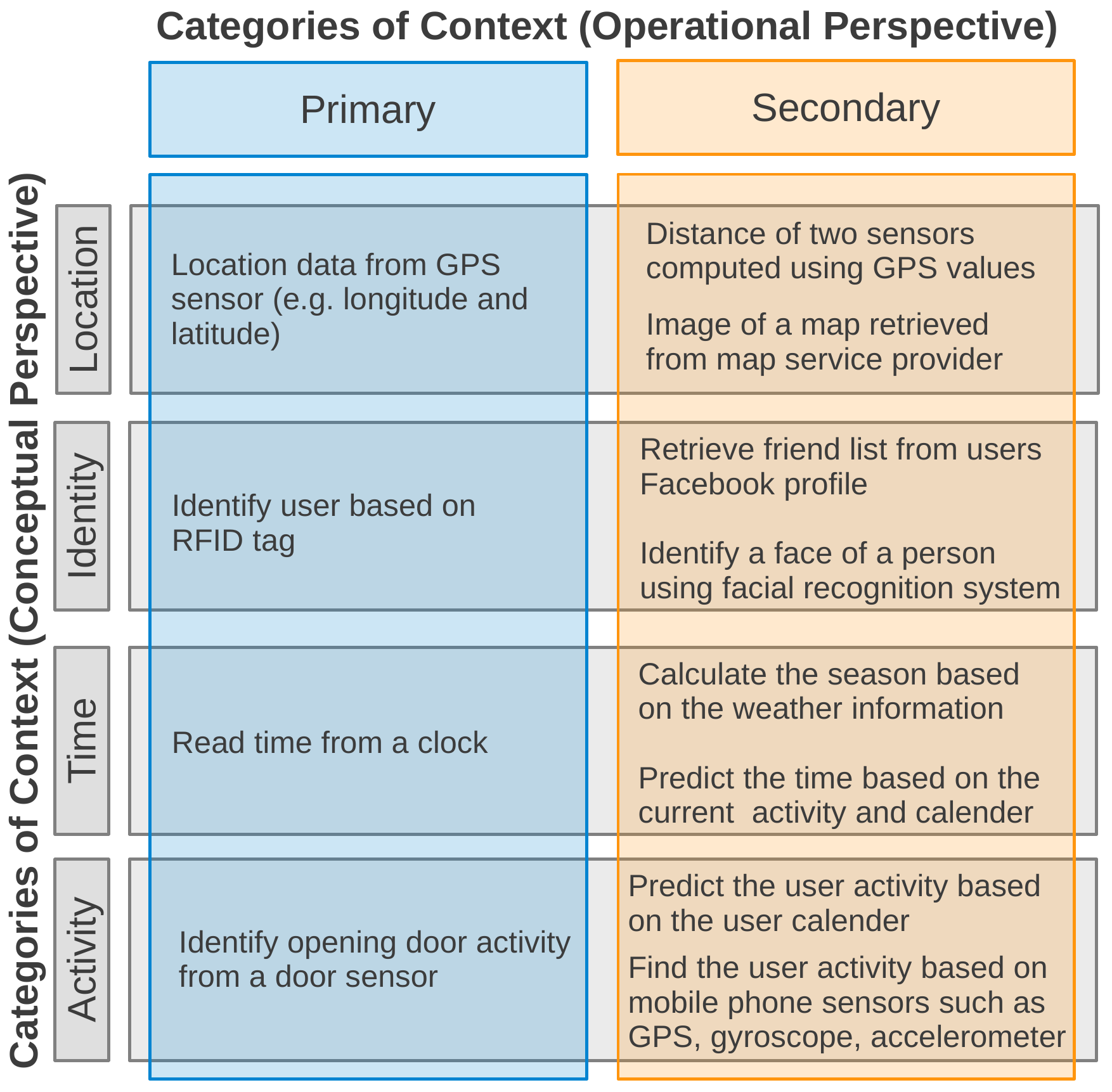}
 \vspace{-0.2cm}
 \caption{Context categorisation in two different perspectives: conceptual and operational. It shows why both operational and conceptual categorisation schemes are important in IoT paradigm as the capture different perspectives.}
 \label{Fig:Context Types and Categories of Context}
 \vspace{-20pt}
\end{figure}

Thus, we define  a context  categorisation  scheme (i.e. primary and secondary) that can be used to classify  a given data value (e.g. single data item such as current time) of context in terms of an operational perspective (i.e. how the data was acquired).  However,  the same data value  can be considered as primary  context in one scenario and secondary context in another. For example, if we collect the blood pressure level of a patient directly  from a sensor attached to the patient, it could be identified as primary context. However, if we derive the same information  from a patient's health record by connecting to the hospital  database, we call it secondary context. Therefore, the  same information can be acquired using different techniques. It is important to understand that the quality, validity, accuracy, cost and effort of acquisition, etc. may varied significantly  based on the techniques used. This would be more challenging in the IoT paradigm, because there would be a large amount of data sources that can be used to retrieve the same data value. To decide which source and technique to use would be a difficult task. We will revisit this challenge in Section VI. In addition, a similar type of context information  can be classified  as both primary and secondary. For example, location can be raw GPS data values or the name of the location (e.g. city, road, restaurant). Therefore, identifying a location  as primary context without examining how the data has been collected  is fairly inaccurate. Figure  \ref{Fig:Context Types and Categories of Context} depicts how the context can be identified  using our context type definitions.

\begin{table*}[t!]
\centering
\footnotesize

\renewcommand{\arraystretch}{1.2}
\caption{Different Context Categorisation Schemes and Their Scopes}
\vspace{-0.3cm}
\begin{tabular}{>{\footnotesize}l@{} p{0.35cm} p{0.40cm}p{0.40cm}p{0.40cm}p{0.40cm}p{0.40cm} c p{0.40cm}p{0.40cm}p{0.40cm}p{0.40cm}p{0.40cm} c p{0.38cm}p{0.38cm} c}
\hline
 Context Types 
 
 &  \begin{sideways}\scriptsize \parbox[b]{2.6cm}{(1994)\\ Schilit $\cite{P116}$ } \end{sideways}  
 &  \begin{sideways}\scriptsize \parbox[b]{2.6cm}{(1994)\\ Schilit  $\cite{P116}$} \end{sideways}  
 &  \begin{sideways}\scriptsize \parbox[b]{2.6cm}{(1997)\\ Ryan $\cite{P182}$} \end{sideways}  
 &  \begin{sideways} \scriptsize \parbox[b]{2.6cm}{(1999)\\ Abowd $\cite{P104}$} \end{sideways}  
 &  \begin{sideways}\scriptsize \parbox[b]{2.6cm}{(2000)\\ Chen and Kotz $\cite{P431}$}  \end{sideways}

&  \begin{sideways}\scriptsize \parbox[b]{2.6cm}{(2003)\\ Henricksen $ \cite{P389}$} \end{sideways}  
& \begin{sideways} \scriptsize \parbox[b]{2.6cm}{ (2003)\\ Prekop \& \\Burnett $\cite{P540}$,\\  Gustavsen $\cite{P541}$,\\ Hofer  $\cite{P542}$ } \end{sideways}   
&  \begin{sideways}  \scriptsize \parbox[b]{2.6cm}{(2005)\\ Van Bunningen  $\cite{P304}$}  \end{sideways}  
& \begin{sideways}\scriptsize  \parbox[b]{2.6cm}{(2006)\\ Miao and Yuan $\cite{P281}$ }\end{sideways}  
& \begin{sideways} \scriptsize \parbox[b]{2.6cm}{(2007)\\ Guan  $\cite{P331}$}  \end{sideways}  
&  \begin{sideways}\scriptsize \parbox[b]{2.6cm}{(2007)\\ Chong $\cite{P284}$}  \end{sideways}
&  \begin{sideways}\scriptsize \parbox[b]{2.6cm}{(2009)\\Zhong $\cite{P432}$ } \end{sideways}  
& \begin{sideways}\scriptsize \parbox[b]{2.6cm}{(2009) \\Mei \& \\ Easterbrook $\cite{P297}$} \end{sideways}  
& \begin{sideways} \scriptsize \parbox[b]{2.6cm}{(2010)\\ Rizou $\cite{P328}$ }\end{sideways}  
& \begin{sideways} \scriptsize \parbox[b]{2.6cm}{(2011)\\ Liu  $\cite{P211}$ } \end{sideways}  
& \begin{sideways} \scriptsize \parbox[b]{2.6cm}{(2011) \\ Yanwei $\cite{P271}$ } \end{sideways}

\\ \hline  \hline

User                                 &  &  $\checkmark$ &  &  & $\checkmark$ & & & &  & &  & $\checkmark$ &  &  & $\checkmark$ & $\checkmark$ \\ 
Computing (System)                   &  &  $\checkmark$ &  &  & $\checkmark$ & & & &  & & $\checkmark$ & $\checkmark$ &  & &  & $\checkmark$ \\ 
Physical (Environment)               &  &  $\checkmark$ & $\checkmark$ & & $\checkmark$ & & & & &  & $\checkmark$ & $\checkmark$ &  & & $\checkmark$ &  \\ 

Historical                           &  &  &  &  &  &  &  & &  &  & $\checkmark$ &  &  & &  &  \\ 
Social                               &  &  &  &  &  &  &  & &  &  &  & $\checkmark$ &  &  &  &  \\ 
Networking                           &  &  &  &  &  &  &  &  & &  &  &  &  &  & $\checkmark$ &  \\ 
Things                               &  &  &  &  &  &  &  &  &  & &  &  &  &  &  & $\checkmark$ \\ 
Sensor                               &  &  &  &  &  &  &  &  &  & & $\checkmark$ &  &  & &  &  \\ 

Who (Identity)                       &  $\checkmark$ &   & $\checkmark$ & $\checkmark$ & &  &  &  &  &  &  &  &  &  &  &  \\ 
Where (Location)                     &  $\checkmark$ &  & $\checkmark$ & $\checkmark$ & &  &  &  &  &  &  &  &  &  &  &  \\        
When (Time)                          &  &  & $\checkmark$ & $\checkmark$ & $\checkmark$ & & & &  &  & $\checkmark$ & $\checkmark$ &  & &  &  \\ 
What (Activity)                      &  $\checkmark$ &  & & $\checkmark$ & &  &  &  &  &  &  &  &  &  &  &  \\ 
Why                                  &  &  & & $\checkmark$ &  &  &  &  &  &  &  &  &  &  &  &  \\ 
Sensed                               &  &  &  &  &  & $\checkmark$ &  &  & $\checkmark$ &  & &  &  &  &  &  \\ 
Static                               &  &  &  &  &  & $\checkmark$ &  &  &  &  &  &  &  &  &  &  \\ 
Profiled                             &  &  &  &  &  & $\checkmark$ &  &  & $\checkmark$ &  & &  &  &  &  &  \\ 
Derived                              &  &  &  &  &  & $\checkmark$ &  &  & $\checkmark$ &  & &  &  &  &  &  \\ 
Operational                          &  &  &  &  &  &  &  & $\checkmark$ &  &  &  &  &  &  &  &  \\ 
Conceptual                           &  &  &  &  &  &  &  & $\checkmark$ &  &  &  &  &  &  &  &  \\ 
Objective                            &  &  &  &  &  &  &  &  &  &  &  & & $\checkmark$ &  &  &  \\ 
Cognitive                            &  &  &  &  &  &  &  &  &  &  &  & & $\checkmark$ &  &  &  \\ 
External (Physical)                  &  &  &  &  &  &  & $\checkmark$ &  &  &  &  &  &  &  &  & \\ 
Internal (Logical)                   &  &  &  &  &  &  & $\checkmark$ &  &  &  &  &  &  &  &  & \\ 
Low-level  (Observable)              &  &  &  &  &  &  &  &  &  & $\checkmark$ &  &  &  & $\checkmark$ & &  \\ 
High-level  (Non-Observable)         &  &  &  &  &  &  &  &  &  & $\checkmark$ &  &  &  & $\checkmark$ & &  \\ 

\hline
\end{tabular}
\label{Tbl:Different_Context_Categorization_Schemes}
\vspace{-0.50cm}
\end{table*}


\vspace{0.2cm}

\begin{noindlist2}
 \item \textbf{Primary context:} Any information retrieved without using existing  context and without performing any kind of sensor  data fusion operations (e.g. GPS sensor readings as location  information).

 \item \textbf{Secondary context:} Any information that can be computed using primary context. The secondary context can be computed by using sensor data fusion operations or data retrieval operations such as web service calls (e.g. identify the distance between two sensors by applying sensor data fusion operations on two raw GPS sensor values). Further, retrieved context such as phone numbers, addresses, email addresses, birthdays, list of friends from a contact  information provider based on a personal identity as the primary  context can also be identified  as secondary context.

\end{noindlist2}

\begin{table*}[t!]
\centering
\footnotesize
\renewcommand{\arraystretch}{1.2}
\caption{Relationship Between Different Context Categories}
\vspace{-0.3cm}
\begin{minipage}{16cm}

\begin{tabular}{>{\footnotesize}m{4cm}@{}
 p{0.07cm} 
p{0.07cm}
p{0.07cm}
p{0.07cm}
p{0.07cm}
p{0.07cm}
p{0.07cm}
p{0.07cm}
p{0.07cm}
p{0.07cm}
p{0.07cm}
p{0.07cm}
p{0.07cm}
p{0.07cm}
p{0.07cm}
p{0.07cm}
p{0.07cm}
p{0.07cm}
p{0.07cm}
p{0.07cm}
p{0.07cm}
p{0.07cm}
p{0.07cm}
p{0.07cm}
c}
\hline

&  \begin{sideways}  \scriptsize User  \end{sideways}  
&  \begin{sideways}\scriptsize Computing (System)    \end{sideways}  
&  \begin{sideways} \scriptsize Physical (Environment)  \end{sideways}  

&  \begin{sideways}\scriptsize Historical \end{sideways}  

&  \begin{sideways}\scriptsize Social    \end{sideways}  
&  \begin{sideways}\scriptsize Networking   \end{sideways}  
&  \begin{sideways}\scriptsize Things  \end{sideways}  
& \begin{sideways} \scriptsize Sensor    \end{sideways}

&  \begin{sideways}\scriptsize Who (Identity)  \end{sideways}  
& \begin{sideways} \scriptsize Where (Location)  \end{sideways}  
& \begin{sideways} \scriptsize When (Time)   \end{sideways}  
& \begin{sideways}\scriptsize What (Activity)   \end{sideways}  
& \begin{sideways}\scriptsize Why \end{sideways}  
& \begin{sideways} \scriptsize Sensed  \end{sideways}  
&  \begin{sideways}\scriptsize Static  \end{sideways}  
& \begin{sideways} \scriptsize Profiled  \end{sideways}
& \begin{sideways} \scriptsize Derived  \end{sideways}  
& \begin{sideways} \scriptsize Operational    \end{sideways}   
& \begin{sideways} \scriptsize Conceptual    \end{sideways}  
& \begin{sideways} \scriptsize Objective    \end{sideways}  
& \begin{sideways} \scriptsize Cognitive    \end{sideways}  
& \begin{sideways} \scriptsize External (Physical)    \end{sideways}  
& \begin{sideways} \scriptsize Internal (Logical)     \end{sideways}  
& \begin{sideways} \scriptsize Low-level  (Observable)    \end{sideways}  
& \begin{sideways} \scriptsize High-level  (Non-Observable)    \end{sideways}  
 
\\ \hline  \hline

User                        &\gc  &\gc  &\gc  &\gc  &\gc  &\gc  &\gc  &\gc  &\gc  &\gc  &\gc  &\gc  &\gc  &\gc  &\gc  &\gc  &\gc  &\gc  &\gc  &\gc  &\gc  &\gc  &\gc  &\gc  &\gc  \\ 
Computing (System)          & 3\llr\llr\llg   &\gc  &\gc  &\gc  &\gc  &\gc  &\gc  &\gc  &\gc  &\gc  &\gc  &\gc  &\gc  &\gc  &\gc  &\gc  &\gc  &\gc  &\gc  &\gc  &\gc  &\gc  &\gc  &\gc  &\gc  \\ 
Physical (Environment)      & 3\llr\llr   & 3\llr\llr   &\gc  &\gc  &\gc  &\gc  &\gc  &\gc  &\gc  &\gc  &\gc  &\gc  &\gc  &\gc  &\gc  &\gc  &\gc  &\gc  &\gc  &\gc  &\gc  &\gc  &\gc  &\gc  &\gc  \\
Historical                  & 3\llr\llr   & 2\lly   & 2\lly   &\gc  &\gc  &\gc  &\gc  &\gc  &\gc  &\gc  &\gc  &\gc  &\gc  &\gc  &\gc  &\gc  &\gc  &\gc  &\gc  &\gc  &\gc  &\gc  &\gc  &\gc  &\gc  \\ 
Social                      & 3\llr\llr   & 2\lly   & 2\lly   &2\lly    &\gc  &\gc  &\gc  &\gc  &\gc  &\gc  &\gc  &\gc  &\gc  &\gc  &\gc  &\gc  &\gc  &\gc  &\gc  &\gc  &\gc  &\gc  &\gc  &\gc  &\gc  \\ 
Networking                  & 3\llr\llr   & 2\lly   & 3\llr\llr   &2\lly    & 2\lly   &\gc  &\gc  &\gc  &\gc  &\gc  &\gc  &\gc  &\gc  &\gc  &\gc  &\gc  &\gc  &\gc  &\gc  &\gc  &\gc  &\gc  &\gc  &\gc  &\gc  \\ 
Things                      & 3\llr\llr   & 2\lly   & 2\lly   &2\lly    & 2\lly   & 2\lly   &\gc  &\gc  &\gc  &\gc  &\gc  &\gc  &\gc  &\gc  &\gc  &\gc  &\gc  &\gc  &\gc  &\gc  &\gc  &\gc  &\gc  &\gc  &\gc  \\ 
Sensor                      & 3\llr\llr   & 2\lly   & 1\llg   &2\lly    & 2\lly   & 2\lly   & 2\lly   & \gc &\gc  &\gc  &\gc  &\gc  &\gc  &\gc  &\gc  &\gc  &\gc  &\gc  &\gc  &\gc  &\gc  &\gc  &\gc  &\gc  &\gc  \\ 
Who (Identity)              & 2\lly   & 2\lly   & 2\lly   &2\lly    & 2\lly   & 2\lly   & 2\lly   & 2\lly   &\gc  &\gc  &\gc  &\gc  &\gc  &\gc  &\gc  &\gc  &\gc  &\gc  &\gc  &\gc  &\gc  &\gc  &\gc  &\gc  &\gc  \\ 
Where (Location)            & 3\llr\llr   & 3\llr\llr   & 2\lly   &2\lly    & 2\lly   & 2\lly   & 2\lly   & 3\llr\llr   & 3\llr\llr   &\gc  &\gc  &\gc  &\gc  &\gc  &\gc  &\gc  &\gc  &\gc  &\gc  &\gc  &\gc  &\gc  &\gc  &\gc  &\gc  \\ 
When (Time)                 & 3\llr\llr   & 3\llr\llr   & 3\llr\llr   &2\lly    & 3\llr\llr   & 3\llr\llr   & 3\llr\llr   & 3\llr\llr   & 3\llr\llr   & 3\llr\llr   &\gc  &\gc  &\gc  &\gc  &\gc  &\gc  &\gc  &\gc  &\gc  &\gc  &\gc  &\gc  &\gc  &\gc  &\gc  \\ 
What (Activity)             & 3\llr\llr   & 2\lly   & 2\lly   &2\lly    & 2\lly   & 2\lly   & 2\lly   & 2\lly   & 3\llr\llr   & 3\llr\llr   & 3\llr\llr   &\gc  &\gc  &\gc  &\gc  &\gc  &\gc  &\gc  &\gc  &\gc  &\gc  &\gc  &\gc  &\gc  &\gc  \\ 
Why                         & 3\llr\llr   & 3\llr\llr   & 3\llr\llr   &2\lly    & 3\llr\llr   & 3\llr\llr   & 3\llr\llr   & 3\llr\llr   & 3\llr\llr   & 3\llr\llr   & 3\llr\llr   & 3\llr\llr   &\gc  &\gc  &\gc  &\gc  &\gc  &\gc  &\gc  &\gc  &\gc  &\gc  &\gc  &\gc  &\gc  \\ 
Sensed                      & 1\llg   & 1\llg   & 1\llg   &2\lly    & 1\llg   & 1\llg   & 1\llg   & 1\llg   & 1\llg   & 1\llg   & 1\llg   & 1\llg   & 1\llg   &\gc  &\gc  &\gc  &\gc  &\gc  &\gc  &\gc  &\gc  &\gc  &\gc  &\gc  &\gc  \\ 
Static                      & 2\lly   & 3\llr\llr   & 3\llr\llr   &2\lly    & 3\llr\llr   & 3\llr\llr   & 3\llr\llr   & 3\llr\llr   & 3\llr\llr   & 3\llr\llr   & 3\llr\llr   & 3\llr\llr   & 3\llr\llr   & 3\llr\llr   &\gc  &\gc  &\gc  &\gc  &\gc  &\gc  &\gc  &\gc  &\gc  &\gc  &\gc  \\ 
Profiled                    & 2\lly   & 2\lly   & 2\lly   &2\lly    & 2\lly   & 2\lly   & 2\lly   & 2\lly   & 2\lly   & 2\lly   & 2\lly   & 2\lly   & 2\lly   & 3\llr\llr   & 3\llr\llr   &\gc  &\gc  &\gc  &\gc  &\gc  &\gc  &\gc  &\gc  &\gc  &\gc  \\ 
Derived                     & 2\lly   & 2\lly   & 2\lly   &2\lly    & 2\lly   & 2\lly   & 2\lly   & 2\lly   & 2\lly   & 2\lly   & 2\lly   & 2\lly   & 2\lly   & 3\llr\llr   & 3\llr\llr   & 3\llr\llr   &\gc  &\gc  &\gc  &\gc  &\gc  &\gc  &\gc  &\gc  &\gc  \\ 
Operational                 & 3\llr\llr   & 3\llr\llr   & 3\llr\llr   &2\lly    & 3\llr\llr   & 3\llr\llr   & 3\llr\llr   & 3\llr\llr   & 3\llr\llr   & 3\llr\llr   & 3\llr\llr   & 3\llr\llr   & 3\llr\llr   & 2\lly   & 2\lly   & 2\lly   & 2\lly   &\gc  &\gc  &\gc  &\gc  &\gc  &\gc  &\gc  &\gc  \\ 
Conceptual                  & 1\llg   & 1\llg   & 1\llg   &2\lly    & 1\llg   & 1\llg   & 1\llg   & 1\llg   & 1\llg   & 1\llg   & 1\llg   & 1\llg   & 1\llg   & 2\lly   & 2\lly   & 2\lly   & 2\lly   & 2\lly   &\gc  &\gc  &\gc  &\gc  &\gc  &\gc  &\gc  \\ 
Objective                   & 2\lly   & 2\lly   & 2\lly   &2\lly    & 2\lly   & 2\lly   & 2\lly   & 2\lly   & 1\llg   & 1\llg   & 1\llg   & 1\llg   & 1\llg   & 2\lly   & 2\lly   & 2\lly   & 2\lly   & 3\llr\llr   & 2\lly   &\gc  &\gc  &\gc  &\gc  &\gc  &\gc  \\ 
Cognitive                   & 1\llg   & 3\llr\llr   & 3\llr\llr   &2\lly    & 3\llr\llr   & 3\llr\llr   & 3\llr\llr   & 3\llr\llr   & 3\llr\llr   & 3\llr\llr   & 3\llr\llr   & 3\llr\llr   & 1\llg   & 3\llr\llr   & 2\lly   & 1\llg   & 1\llg   & 3\llr\llr   & 2\lly   & 3\llr\llr   &\gc  &\gc  &\gc  &\gc  &\gc  \\ 
External (Physical)         & 2\lly   & 2\lly   & 2\lly   &2\lly    & 2\lly   & 2\lly   & 2\lly   & 2\lly   & 2\lly   & 2\lly   & 2\lly   & 2\lly   & 2\lly   & 1\llg   & 2\lly   & 3\llr\llr   & 3\llr\llr   & 2\lly   & 2\lly   & 2\lly   & 3\llr\llr   &\gc  &\gc  &\gc  &\gc  \\ 
Internal (Logical)          & 2\lly   & 2\lly   & 2\lly   &2\lly    & 2\lly   & 2\lly   & 2\lly   & 2\lly   & 2\lly   & 2\lly   & 2\lly   & 2\lly   & 2\lly   & 3\llr\llr   & 2\lly   & 1\llg   & 1\llg   & 2\lly   & 2\lly   & 2\lly   & 1\llg   & 3\llr\llr   &\gc  &\gc  &\gc  \\ 
Low-level (Observable)      & 2\lly   & 2\lly   & 2\lly   &2\lly    & 2\lly   & 2\lly   & 2\lly   & 2\lly   & 2\lly   & 2\lly   & 2\lly   & 2\lly   & 3\llr\llr   & 1\llg   & 2\lly   & 3\llr\llr   & 3\llr\llr   & 2\lly   & 2\lly   & 2\lly   & 3\llr\llr   & 1\llg   & 3\llr\llr   &\gc  &\gc  \\ 
High-level (Non-Observable) & 2\lly   & 2\lly   & 2\lly   &2\lly    & 2\lly   & 2\lly   & 2\lly   & 2\lly   & 2\lly   & 2\lly   & 2\lly   & 2\lly   & 2\lly   & 3\llr   & 2\lly   & 1\llg   & 1\llg   & 2\lly   & 2\lly   & 2\lly   & 1\llg   & \llr   & 1\llg   & 3\llr   &\gc  \\ 

\hline 

\multicolumn{26}{p{1\textwidth}}{\textit{Notes}: We denote row labels as (\textit{P}) and column labels as (\textit{Q}). 1 means (\textit{P}) $\cap$ (\textit{Q}) $\approx$ very high; 2 means (\textit{P}) $\cap$ (\textit{Q}) $\approx$ moderate; 3 means (\textit{P}) $\cap$ (\textit{Q}) $\approx$ very low.}

\end{tabular}
\end{minipage}

\label{Tbl:Relationship_Between_Different_Context_Categories}
\vspace{-0.6cm}
\end{table*}


We acknowledge  location, identity, time, and activity as important context information. The IoT paradigm  needs to consider more comprehensive categorisation  schemes in a hierarchical  manner,  such as major categories, sub categories and so on. Operational  categorisation  schemes allow us to understand the issues and challenges in data acquisition techniques, as well as quality  and cost factors related to context. In contrast, conceptual categorisation allows an understanding of the conceptual relationships between context. We have to integrate perspective in order to model context precisely. We compare different  context categorisation schemes in Table  \ref{Tbl:Comparison of Context Categorization Schemes}. In addition to the two categorisation schemes we discussed earlier there are several other schemes introduced  by different researchers  focusing on different perspectives.  Further, we highlight relationships  between different context categories (also called context types) in different perspectives in Table \ref{Tbl:Different_Context_Categorization_Schemes} and in Table \ref{Tbl:Relationship_Between_Different_Context_Categories}. These context categories  are not completely different from each other. Each category shares common characteristics with the others. The similarities and difference among categories are clearly presented in Table \ref{Tbl:Relationship_Between_Different_Context_Categories}. Further, we have listed and briefly explained three major context categorisation schemes and their categories proposed by previous researchers. In Table \ref{Tbl:Different_Context_Categorization_Schemes}, we present each categorisation  effort in chronological order from left to right.

\begin{noindlist2}

\item Schilit et al. \cite{P116} (1994): They categorised  context into three categories using a conceptual categorisation based technique on three common questions that can be used to determine the context.
\begin{enumerate}
 \item Where you are: This includes all location related information such  as GPS coordinates, common names  (e.g. coffee shop, university, police), specific  names (e.g. Canberra city police), specific addresses,  user preferences  (e.g. user's  favourite coffee shop).
 
 \item Who you are with: The information  about the people present around the user.
 
 \item What resources are nearby: This includes information about resources available in the area where the user is located, such as machinery,  smart objects, and utilities.
\end{enumerate}

\end{noindlist2}


\begin{table*}[ht!]

\centering
\footnotesize

\caption{Comparison of Context Categorisation Schemes}
\renewcommand{\arraystretch}{-4}
\vspace{-0.3cm}
\begin{tabular}{ |p{0.2cm}| p{2.9cm} p{6.7cm} p{6cm}| }\hline

\begin{center}   \end{center}       &    
\begin{center} Categorisation Schemes \end{center} & 
\begin{center} Pros \end{center} & 
\begin{center} Cons \end{center}

\\ \hline \hline

\multirow{3}{*}{ \begin{sideways}Conceptual \hspace{2.5cm}     \end{sideways}} &

\begin{center} Where, when, who, what, objective  
\end{center}         
&   
\begin{noindlist}
		\item Provide a broader guide that helps to identify the related context
		\item Less comprehensive 
 		
\end{noindlist}  
&

\begin{noindlist}

 		\item Do not provide information about operational aspects such as cost, time, complexity, techniques, and effort of data acquisition
 		\item Do not provide information about frequency of update required
\end{noindlist} 

\\ 

& \begin{center} User, computing, physical, environmental, time, social, networking, things, sensors contexts 
\end{center}         
&   
\begin{noindlist}
	\item More clear and structured method to organise context
 	\item More extensible and flexible
 	\item More comprehensive
\end{noindlist}  
&

\begin{noindlist}
 		\item Do not provide information about operational aspects such as cost, time, complexity, techniques, and effort of data acquisition
 		\item Do not provide information about frequency of update required
\end{noindlist} 

\\ 

& \begin{center}  Why, cognitive 
\end{center}         
&   
\begin{noindlist}
		\item Allow to model mental reasoning behind context
 		
\end{noindlist} 
&
\begin{noindlist}	
 		\item Do not provide information about core context, relationships between context or operational aspects such as cost, time, complexity, techniques, and effort of data acquisition
\end{noindlist}

\cr\cline{0-0}

\\ 
\multirow{2}{*}{\begin{sideways}Operational \hspace{2cm}   \end{sideways}} &
\begin{center} Sensed, static, profiled, derived
\end{center}         
&
\begin{noindlist}
		\item Provide information about programming and coding level 
		\item Provide information about context source and computational complexity
 		\item Allow to track information such as frequency of update required, validation, quality, etc.
 		\item Provide information about cost and effort of data acquisition
\end{noindlist}  
 &
\begin{noindlist}
		\item Weak in representing the relationship among context
		\item Difficult to classify context information due to ambiguity. Same piece of data can belong to different categories depending to the situation (e.g. location can be derived as well as sensed)
\end{noindlist}

\\ 

& \begin{center} Internal (physical), internal (logical), low-level (observable), high-level (non-observable)
\end{center}         
&  
\begin{noindlist}
		\item Provide information about context sources and the process of accessing data (e.g. whether more reasoning is required or not)
 		\item Provide information about cost and effort of data acquisition
 		\item Provide information about computational complexity
\end{noindlist} 
&
\begin{noindlist}
		\item Weak in representing the relationship among context
 		\item Difficult to classify context information due to ambiguity. Same piece of data can belong to different categories depending to the situation (e.g. temperature can be physical or virtual sensor)
\end{noindlist} 
\\ \hline

\end{tabular}

\label{Tbl:Comparison of Context Categorization Schemes}
\vspace{-0.6cm}
\end{table*}


\begin{noindlist2}

\item Henricksen \cite{P389} (2003): Categorised context into four categories  based on an operational categorisation technique.
\begin{enumerate}
 \item Sensed: Sensor data directly sensed from  the sensors, such as  temperature   measured  by a  temperature sensor. Values will be changed over time with a high frequency.
 
 \item Static: Static information which will  not change over time, such  as manufacturer   of the sensor, capabilities of the sensor, range  of the sensor measurements.
 
 \item Profiled: Information  that changes over time with a low frequency,  such as once per month (e.g. location of sensor, sensor ID).
 
 \item Derived: The information  computed using primary context   such as distance of two sensors calculated using two GPS sensors.
\end{enumerate}

\item Van Bunningen et al. \cite{P304} (2005): Instead of categorising context, they classified the context categorisation schemes into two broader categories: operational and conceptual.
\begin{enumerate}
\item Operational categorisation: Categorise context based  on how they were acquired,  modelled, and treated.
\item Conceptual categorisation:  Categorise context based on the meaning and conceptual relationships between the context.
\end{enumerate}

\end{noindlist2}

Based on the evaluation of context categorisation, it is evident that no single categorisation scheme can accommodate all the demands  in the IoT paradigm.  We  presented a comparison between conceptual and operational categorisation  schemes in Table  \ref{Tbl:Comparison of Context Categorization Schemes}. To build an ideal context-aware middleware solution for the IoT, different  categorisation schemes need to be combined together  in order to complement their  strengths  and mitigate their weaknesses.

\vspace{-0.3cm}

\subsection{Levels of Context Awareness and characteristics}
\label{chapter2:CAF:Levels of Context Awareness}

Context  awareness can be identified in three levels based on the user interaction \cite{P430}.

\begin{noindlist2}
 \item \textbf{Personalisation}: It allows the users to set their preferences, likes, and expectation to the system manually. For example,  users may set  the preferred  temperature in a smart home environment where the heating system of the home can maintain the specified  temperature  across all rooms.

 \item \textbf{Passive context-awareness}: The system constantly  monitors the environment and offers the appropriate options to the users so they can take actions. For example, when a user enters a super market, the mobile  phone alerts the user with a list of discounted products to be considered.

 \item \textbf{Active context-awareness}: The system continuously and autonomously  monitors the situation and acts autonomously.  For example,  if  the smoke  detectors  and temperature  sensors detect a fire in a room in a smart home environment,  the system will  automatically  notify the fire brigade  as well as the owner of the house via appropriate methods  such as phone calls.
\end{noindlist2}

In addition, Van Bunningen et al. \cite{P304} has identified comprehensively, and discussed, eight characteristics of context: context 1) is sensed though  sensors or sensor networks,  2) is sensed by small and constrained devices, 3) originates from distributed sources, 4) is continuously  changing, 5) comes from mobile objects 6) has a temporal  character  7) has a spatial character, 8) is imperfect  and uncertain.

\subsection{Context Awareness Management Design Principles}
\label{chapter2:CAF:Context Awareness Management Design Principles}

Martin et al. \cite{P294} have identified  and comprehensively discussed six design principles related to context-aware  management frameworks (middleware). Further, Ramparany et al. \cite{P340} and Bernardos et al. \cite{P302} have  also identified several design requirements.  We  summarise the findings below with brief explanations. This list is not intended to be exhaustive. Only the most important  design aspects are considered.

\begin{noindlist2}
 \item \textbf{Architecture layers and components}: The functionalities need to be divided into layers and components in a meaningful  manner. Each component should perform a very limited amount of the task and should be able to perform independently up to a large  extent.
 
 \item \textbf{Scalability and extensibility}: The component  should be able to added or removed  dynamically. For example. new functionalities (i.e. components) should be able to be add without altering the existing components (e.g. Open Services  Gateway  initiative). The component  needs to be developed according  to   standards across the solutions, which improves  scalability  and extensibility (e.g. plug-in architectures).

 \item \textbf{Application programming interface (API)}: All  the functionalities should be available  to be accessed  via a  comprehensive   easy  to learn and easy  to use  API. This allows the incorporation of  different solutions very easily. Further, API can be used  to bind context management frameworks to applications. Interoperability among different IoT solutions heavily depends on API and their usability.

 \item \textbf{Debugging mechanisms and tools}: Debugging is  a critical task in any software development process. In the IoT paradigm, debugging would be difficult due to the exponential number of possible alternative interactions. In order to win the trust of the consumers, the IoT should prove its trustworthiness.  Integrated  debug mechanisms inbuilt into the framework will  help to achieve this challenge. For example, the justifications behind the results produced by the reasoners should be available to be evaluated to find possible inaccuracies so further development can be carried out. Some initial work in this area is presented in the Intelligibility Toolkit  \cite{P384}.

 \item \textbf{Automatic context life cycle management}: Context-aware frameworks should be able to be understand by the available  context sources  (i.e. physical and virtual sensors), their data structure, and automatically  built internal data models to facilitate them. Further, raw context needs to be retrieved and transformed into appropriate context representation  models  correctly with minimum human intervention.

 \item \textbf{context model in-dependency}: Context needs to be modelled and stored separately from context-aware framework  related code and data structures, which allows both parts to be altered independently.

 \item \textbf{Extended, rich, and comprehensive  modelling}: Context models should be able to extend  easily. The IoT will  need to deal with enormous  amount  of devices,  and will be required to handle vast amounts of domain specific context. It also needs to support complex relationships, constrains, etc. In an ideal context-aware framework for the IoT, multiple different context representation models should be incorporated  together to improve their efficiency and effectiveness.

 \item \textbf{Multi-model reasoning}: No single reasoning model can accommodate the demands of the IoT. We  will discuss reasoning in Section  \ref{chapter2:CAF:Context Reasoning Decision Models}. Each reasoning model has its own strengths  and weaknesses.  An ideal framework should incorporate  multiple reasoning  models together to complement  each others'  strengths  and mitigate  their weaknesses.

 \item \textbf{Mobility support}: In the IoT, most devices  would be mobile, where each one has a different set  of hardware and software capabilities. Therefore, context-aware frameworks  should be developed  in  multiple flavours (i.e. versions), which can run on different hardware and software configurations (e.g. more capabilities for server level software and less capabilities  for mobile phones).

 \item \textbf{Share information (real-time and historic)}: In the IoT, there is  no single point of  control. The architecture would  be distributed. Therefore, context sharing  should happen at  different levels:  framework-to-framework and framework-to-application.  Context model in-dependency has been discussed earlier and is crucial in sharing.

 \item \textbf{Resource optimisation}: Due to the scale (e.g. 50 billion devices),   a  small improvement  in data structures  or processing can make a huge impact in storage and energy consumption. This stays true for any type of resource used in the IoT.

 \item \textbf{Monitoring and detect event}: Events  play a significant role in the IoT, which is complement  by monitoring. Detecting an event triggers an action autonomously in the IoT paradigm. This is how the IoT will help humans carry out their day-to-day  work easily and efficiently. Detecting events in real time is a major challenge for context-aware frameworks in the IoT paradigm.
 
\end{noindlist2}

\section{Context Life Cycle}
\label{chapter2:CDLC}

A data life cycle shows how data moves  from phase to phase  in software  systems (e.g. application, middleware). Specifically, it explains where the data is generated and where the data is consumed. In this section we consider movement of context  in context-aware  systems. Context-awareness is no longer limited to desktop, web, or mobile applications. It has already become a service: Context-as-a-Service (CXaaS)  \cite{P024}. In other terms, context  management has become an essential  functionality in software  systems. This trend will grow in the IoT paradigm.

There	are	web-based	context	management	services (WCXMS) that provide context information management throughout the context's life  cycle. Hynes et al.  \cite{P024} have classified data life cycles into two categories: Enterprise Lifecycle  Approaches (ELA)   and  Context  Lifecycle Approaches (CLA).

ELA are focused on context. However, these life cycles
are robust and well-established,   based on industry standard strategies for data management in general. In contrast, CLA are specialised in context management. However, they are not tested or standardised strategies as much as ELA. We have selected ten popular  data life cycles to analyse in this survey. In the following list, 1-5 belong to ELA category and 6-10 belong to CLA category. Three dots (...) denotes reconnecting to the first phase by completing the cycle. The right arrow ($\rightarrow$) denotes data transfer form one phase to another.

\begin{enumerate}
 \item \textit{Information Lifecycle Management (ILM)} \cite{P516}: creation and receipt $\rightarrow$ distribution $\rightarrow$ use $\rightarrow$ maintenance $\rightarrow$ disposition $\rightarrow$ ...
 
 \item \textit{Enterprise Content Management (ECM)} \cite{P517}: capture $\rightarrow$ manage $\rightarrow$ store $\rightarrow$ preserve $\rightarrow$ deliver $\rightarrow$ ...

 \item \textit{Hayden's Data Lifecycle} \cite{P515}: collection $\rightarrow$ relevance $\rightarrow$ classification $\rightarrow$ handling and storage $\rightarrow$ transmission and transportation $\rightarrow$ manipulate, conversion and alteration $\rightarrow$ release $\rightarrow$ backup $\rightarrow$ retention destruction $\rightarrow$ ...

  \item \textit{Intelligence Cycle}  \cite{P170}: collection $\rightarrow$ processing $\rightarrow$ analysis$\rightarrow$ publication $\rightarrow$ feedback $\rightarrow$ ...
  
  \item \textit{Boyd Control Loop} (also called OODA loop) \cite{P171}: observe $\rightarrow$ orient $\rightarrow$ decide $\rightarrow$ act $\rightarrow$ ...

 \item \textit{Chantzara and Anagnostou Lifecycle} \cite{P114}: sense (context provider) $\rightarrow$ process (context broker) $\rightarrow$ disseminate (context broker) $\rightarrow$ use (service provider) $\rightarrow$ ...
 
 \item \textit{Ferscha et al. Lifecycle} \cite{P518}: sensing $\rightarrow$ transformation $\rightarrow$ representation $\rightarrow$ rule base $\rightarrow$ actuation $\rightarrow$ ...
 
 \item \textit{MOSQUITO} \cite{P519}: context information discovery $\rightarrow$ context information acquisition $\rightarrow$ context information reasoning $\rightarrow$ ...

 \item \textit{WCXMS Lifecycle} \cite{P024}:  (context sensing $\rightarrow$ context transmission $\rightarrow$ context acquisition $\rightarrow$ ... ) $\rightarrow$ context classification $\rightarrow$ context handling $\rightarrow$ (context dissemination  $\rightarrow$ context usage $\rightarrow$ context deletion $\rightarrow$ context request $\rightarrow$... ) $\rightarrow$ context maintenance $\rightarrow$ context disposition $\rightarrow$...
 
 \item \textit{Baldauf et al.} \cite{P402}: sensors $\rightarrow$ raw data retrieval $\rightarrow$ reprocessing $\rightarrow$ storage $\rightarrow$ application.
 
\end{enumerate}

\begin{figure}[h]
 \vspace{-0.2cm}
  \begin{center}
    \includegraphics[scale=1]{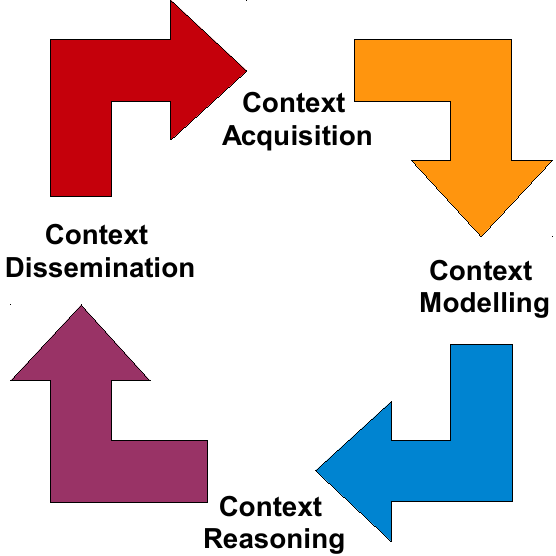}
  \end{center}
  \vspace{-15pt}
  \caption{This is the simplest form of a context life cycle. These four steps are essential in context management systems and middleware solutions. All the other functions that may offer by systems are value added services.}
  \label{Fig:Context_Data_Life_Cycle}
\end{figure}

In addition to the life cycles, Bernardos et al. \cite{P302} identified three  phases in a typical context  management system: context acquisition, information  processing, and reasoning and decision. After reviewing the above  life cycles,  we derived an appropriate (i.e. minimum number of phases but includes all essential) context life cycle  as depicted in Figure  \ref{Fig:Context_Data_Life_Cycle}.

This context  life cycle consists of four phases. First, context   needs to be acquired from various  sources. The sources could be physical  sensors or virtual sensors (context acquisition).  Second, the collected  data needs to be modelled and represent  according  to a  meaningful manner (context modelling). Third, modelled data needs to be processed to derive high-level context information from low-level raw sensor data (context reasoning).  Finally, both high-level and low-level context  needs to be distributed  to the consumers  who are interested  in context (context dissemination).  The following discussion is based on these four phases.

\begin{table*}[ht!]
\caption[Comparison of Context Acquisition Methods based on Responsibility]{Comparison of Context Acquisition Methods based on Responsibility (Push, Pull)}
\centering
\footnotesize
\renewcommand{\arraystretch}{-4}
\vspace{-0.3cm}
\begin{tabular}{ p{1.5cm} p{7.6cm} p{7.4cm} }\hline
 
\begin{center} Criteria \end{center}       &    
\begin{center} Push \end{center} & 
\begin{center} Pull \end{center}

\\ \hline \hline

\begin{center} Pros  \end{center}      
&   
\begin{noindlist}
		 \item Sensor hardware make the major decisions on sensing and communication
 		 \item Can be both instant or interval sensing and communication
\end{noindlist}  
 &
 
 \begin{noindlist}
 		 \item Software of the sensor data consumer makes the major decisions on sensing and communication
  		 \item Decision on when to collect data is based on reasoning significant amount of data in software level
  		 \item Can be both instant or interval sensing and communication
 \end{noindlist} 

\\ 
\begin{center} Cons  \end{center}      
&   
\begin{noindlist}
		 \item Decision on when to send data based on reasoning less amount of data
 		 \item Sensors are required to program when the requirements are changed
 		 
\end{noindlist} 
 &
\begin{noindlist}
 		 \item More communication bandwidth is required where software level has to send data requests to the sensors all the time
 		
\end{noindlist} 
\\ 

\begin{center} Applicability \end{center}      
&   
Can be used when sensors know about when to send the data and have enough processing power and knowledge to reason locally. (e.g. event detection where one or small number of sensors can reason and evaluate the conditions by their own without software level complex data processing and reasoning.)

&
Can be used when sensors do not have knowledge on when to send the data to the consumer. (e.g. event detection where large amount of data need to be collected, processed, and reasoned in order to recognize the event.)
\\ \hline

\end{tabular}

\label{Tbl:Comparison_of_Context_Acquisition_Methods_based_on_Responsibility}
\vspace{-0.1cm}
\end{table*}

\begin{table*}[ht!]
\centering
\footnotesize
\caption[Comparison of Context Acquisition Methods based on Frequency]{Comparison of Context Acquisition Methods based on Frequency (Instant, Interval)}

\renewcommand{\arraystretch}{-4}
\vspace{-0.3cm}
\begin{tabular}{ p{1.5cm} p{7.5cm} p{7.5cm} } \hline
 
\begin{center} Criteria \end{center}       &    
\begin{center} Instant \end{center} & 
\begin{center} Interval \end{center}

\\ \hline \hline

\begin{center} Pros  \end{center}      
&  
\begin{noindlist}
		 \item Save energy due to no redundant network communications are involved
 		 \item More accurate data can be gather as the network transmission would be triggered as soon as the conditions are met
\end{noindlist}  
 &
\begin{noindlist}
		 \item Either sensors can be configured to sense and communicate with data consumers in a predefined frequency or the sensor data consumers can retrieve data explicitly from the sensors in a predefined frequency
 		 \item Sensors do not need to be intelligent/knowledge or have significant processing and reasoning capabilities
 		 \item Allows to understand the trends or behaviour by collecting sensor data over time
\end{noindlist}  

\\ 
\begin{center} Cons  \end{center}      
&   
\begin{noindlist}
		 \item More knowledge is required to identify the conditions and the satisfaction of the conditions
 		 \item Hardware level (i.e. sensor) or software level should know exactly what to look for
 		 \item Difficult to detect events which require different types of data from number of different sensors
 		 \item Comparatively consume more energy for data processing
\end{noindlist}  
 &
 
 \begin{noindlist}
 		 \item May waste energy due to redundant data communication
  		 \item Less accurate as the sensor readings can be change over the interval between two data communications
  		 \item Reasoning need to be done in software level by the data consumer which will miss some occurrence of events due to above inaccuracy
 \end{noindlist}  

\\ 

\begin{center}Applicability\end{center}      
& 
Can be used to detect frost events or heat events in agricultural domain. In smart home domain, this method can be used to detect some one entering to a room via door sensors. Ideally, applicable for the situations where expected outcome is well-known by either hardware level (i.e. sensors) or software level
 &
Can be used to collect data from temperature sensors for controlling air condition or measure air pollution where actions are not event oriented but monitoring oriented. Ideally, applicable for the situations where expected outcome is not known by either hardware level (i.e. sensors) or software level

\\ \hline

\end{tabular}

\label{Tbl:Comparison_of_Context_Acquisition_Methods_based_on_Frequency}
\vspace{-0.1cm}
\end{table*}

\begin{table*}[ht!]

\centering
\footnotesize
\caption[Comparison of Context Acquisition Methods based on Source]{Comparison of Context Acquisition Methods based on Source (Direct Sensors, Middleware, Context Servers)}
\renewcommand{\arraystretch}{-4}
\vspace{-0.3cm}
\begin{tabular}{ p{1.5cm} p{5cm} p{5cm} p{5cm}  }\hline 
 
\begin{center} Criteria \end{center}       &    
\begin{center} Direct Sensor Access \end{center} & 
\begin{center} Through Middleware \end{center} & 
\begin{center} Through Context Server \end{center}

\\  \hline \hline

\begin{center} Pros  \end{center}      
&   
\begin{noindlist}
\vspace{2pt}
		 \item Efficient as it allows direct communication with the sensors
		 \item Have more control over sensor configuration and data retrieval process
 	
\end{noindlist}  
 & 
  \begin{noindlist}
 		 \item Easy to manage and retrieve context as most of the management tasks are facilitated by the middleware.
  		 \item Can retrieve data faster with less effort and technical knowledge
 \end{noindlist}  
 &
\begin{noindlist}
		 \item Less resources required
 		 \item Can retrieve data faster with less effort and technical knowledge
\end{noindlist}   

\\ 
\begin{center} Cons  \end{center}      
&   
\begin{noindlist}
		 \item Significant technical knowledge is required including hardware level embedded device programming and configuring
 		 \item Significant amount of time, effort, cost involved
 		 \item Updating is very difficult due to tight bound between sensor hardware and consumer application
\end{noindlist}   
 &
\begin{noindlist}
		 \item Require more resources (e.g. processing, memory, storage) as middleware solutions need to be employed
 		 \item Less control over sensor configuration
 		 \item Moderately efficient as data need to be retrieve through middleware
\end{noindlist}   
 &
\begin{noindlist}
		 \item No control over sensor configuration
 		 \item Less efficient as the context need to be pulled from server over the network
\end{noindlist}   

\\ 

\begin{center} Applicability \end{center}      
&   
 Can be used for small scale scientific experiments. Can also be used for situation where limited number of sensors are involved
&
IoT application will use this methods in most cases. Can be used in situations where large number of heterogeneous sensors are involved
&
Can be used in situations where significant amount of context are required but have only limited resources (i.e. cannot employ context middleware solutions due to resource limitations) that allows run the consumer application
\\ \hline

\end{tabular}

\label{Tbl:Comparison_of_Context_Acquisition_Methods_based_on_Source}
\vspace{-0.6cm}
\end{table*}

\begin{table*}[ht!]

\centering
\footnotesize
\caption[Comparison of Context Acquisition Methods based on Sensor Types]{Comparison of Context Acquisition Methods based on Sensor Types (Physical, Virtual, Logical)}

\renewcommand{\arraystretch}{-4}
\vspace{-0.3cm}
\begin{tabular}{ p{1.5cm}  p{5cm} p{5cm} p{5cm}   }\hline
 
\begin{center} Criteria \end{center}       &    
\begin{center} Physical Sensors \end{center} & 
\begin{center} Virtual Sensors \end{center} & 
\begin{center} Logical  Sensors \end{center}

\\ \hline \hline

\begin{center} Pros  \end{center}      
&   
\begin{noindlist}
 		 \item Error detection is possible and relatively easy
 		 \item Missing value identification is also relatively easy
 		 \item Have access to low-level sensor configuration therefore can be more efficient
\end{noindlist}  
 & 
\begin{noindlist}
		 \item Provide moderately meaningful data
 		 \item Provide high-level context information
 		 \item Provided data are less processed
 		 \item Do not need to deal with hardware level tasks
\end{noindlist}   
 &
  \begin{noindlist}
  		 \item Provide highly meaningful data
   		 \item Provide high-level context information
   		 \item Usually more accurate
   		 \item Do not need to deal with hardware level tasks
  \end{noindlist}

\\ 
\begin{center} Cons  \end{center}      
&
\begin{noindlist}
		 \item Hardware deployment and maintenance is costly
 		 \item Have to deal with sensor and hardware level programming, design, development, test, debug
 		 \item Provide less meaningful and low-level raw sensor data
\end{noindlist}  
 &
 \begin{noindlist}
 		 \item Difficult to find errors in data
  		 \item Filling missing values is not easy as they are mostly non-numerical and unpredictable
 \end{noindlist}  
 &
 
  \begin{noindlist}
  		 \item Difficult to find error in data
  		 \item Filling missing values is not easy as they are mostly non-numerical
  		 \item Do not have control over data production process
  		 \item License fees and other restrictions may apply
  \end{noindlist}  

\\ 

\begin{center} Applicability  \end{center}      
&   

Can be used to collect physically observable phenomenon such as light, temperature, humidity, gas, etc.

 &

Can be used to collect information that cannot be measure physically such as calendar details, email, chat, maps, contact details, social networking related data, user preferences, user behaviour, etc.

 &
 
 Can be used to collect information that are costly and impossible to collect directly through single physical sensor where advance processing and fusing data from multiple sensors are required (e.g. weather information, activity recognition, location recognition, etc.).

\\ \hline

\end{tabular}

\label{Tbl:Comparison_of_Context_Acquisition_Methods_based_on_Sensor_Types}
\vspace{-0.6cm}
\end{table*}

\subsection{Context Acquisition}
\label{chapter2:CDLC:Context Acquisition}

In this section we discuss five factors that need to be considered when developing context-aware middleware solutions in the IoT paradigm. The techniques  used to acquire context can be varied based on responsibility,  frequency, context source, sensor type, and acquisition process.

\subsubsection{Based on Responsibility}

Context  (e.g. sensor data) acquisition  can be primarily accomplished using two methods  \cite{P334}: push and pull. A comparison is presented in Table \ref{Tbl:Comparison_of_Context_Acquisition_Methods_based_on_Responsibility}.

\begin{noindlist2}

 \item Pull: The software  component which is responsible for acquiring  sensor data from sensors  make  a request  (e.g. query) from the sensor hardware  periodically (i.e. after certain intervals) or instantly to acquire data.
 
 \item Push:  The physical or virtual sensor  pushes data to the software component which is responsible to acquiring sensor data periodically  or instantly. Periodical or instant pushing can be employed to facilitate a publish and subscribe model.

\end{noindlist2}

\subsubsection{Based on Frequency}

Further, in the IoT paradigm, context can be generated based on two different event types: instant events and interval  events  \ref{Tbl:Comparison_of_Context_Acquisition_Methods_based_on_Frequency}.

\begin{noindlist2}
 \item Instant (also known as threshold violation): These events occur instantly. The events  do not span across certain amounts  of time. Open a  door, switch on a  light, or animal enters experimental   crop field are some types of instant events. In order to detect this type of event, sensor data needs to be acquired when the event occurs. Both push and pull methods can be employed.
 
 \item Interval (also known as periodically):  These events span a  certain period of time. Raining, animal eating  a  plant, or winter are some interval events. In order to detect this type of event, sensor data needs to be acquired periodically (e.g. sense  and send  data to the software every 20 seconds). Both push and pull methods can be employed.
\end{noindlist2}

\subsubsection{Based on Source}

In addition, context acquisition methods can be categorised into three categories  \cite{P419} based on where the context came from. A comparison is presented in Table  \ref{Tbl:Comparison_of_Context_Acquisition_Methods_based_on_Source}.

\begin{noindlist2}
 \item Acquire directly from sensor hardware:  In this method, context is directly acquired  from the sensor  by communicating with the sensor hardware  and related APIs. Software drivers and libraries need to be installed locally. This method is typically used to retrieve data from sensors attached  locally. Most devices  and sensors  today require some amount of driver support and can be connected  via USB, COM, or serial ports. However, wireless technologies are becoming popular in the sensor community,  which allows data transmission without driver installations. In the IoT paradigm most objects will communicate with each other via a wireless means.

 \item Acquire through a  middleware infrastructure:  In  this method, sensor (context)  data is acquired by middleware solutions  such as GSN. The applications  can retrieve sensor  data from the middleware and  not from the sensor hardware directly.  For example, some GSN instances will directly access sensor hardware and rest of the GSN instances will communicate with other GSN instances to retrieve data.

 \item Acquire from context servers:  In this method,  context is acquired from several other context storages (e.g. databases, RSS (Really Simple Syndication)  feeds, web services) via different mechanisms such as  web service calls. This mechanism  is useful when the hosting device of the context-aware application has limited computing  resources. Resource-rich context servers can be used to acquire and process context.
\end{noindlist2}

\subsubsection{Based on Sensor Types}
\label{chapter2:CDLC:CA:Based_on_Sensor_Types}

There are different types of sensors  that can be employed  to acquire context. In general usage, the term `sensor' is used to refer to tangible sensor hardware devices. However, among the technical community, sensors are refer to as any data source that provides relevant context. Therefore,  sensors can be divided into three categories  \cite{P543}: physical, virtual, and logical. A comparison is presented in Table  \ref{Tbl:Comparison_of_Context_Acquisition_Methods_based_on_Sensor_Types}.

\begin{noindlist2}
 \item Physical sensors: These are the most commonly  used type of sensors  and they are tangible. These  sensors generate sensor data by themselves. Most of the devices we use today are equipped with a variety of sensor (e.g. temperature, humidity,  microphone, touch). A discussion on commonly used sensor  data types and sensors  is presented in  \cite{P544}. The data retrieved from physical  sensors is called low-level context. They are less meaningful, trivial, and vulnerable  to small changes. IoT solutions needs to understand the physical world  using imperfect, conflicting  and imprecise data.

 \item Virtual sensors: These sensors do not necessarily generate sensor data by themselves. Virtual sensors retrieve  data from many sources and publish it  as sensor data (e.g. calendar, contact number directory, twitter statuses, email and chat applications).  These sensors do not have a physical presence. They commonly  use web services technology to send and receive data.

\item Logical sensors  (also called software sensors):  They combine  physical  sensors and virtual sensors in order to produce  more meaningful  information. A  web service dedicated to providing weather information can be called  a  logical sensor. Weather  stations  use thousands  of physical sensors to collect weather information. They also collect information from virtual sensors  such as  maps, calendars, and historic data. Finally, weather information is produced by combing both physical and virtual sensors. In addition, the android mobile operating  system  consists of a number  of software  sensors such as  gravity, linear accelerometer, rotation vector, and orientation sensors.

\end{noindlist2}

\subsubsection{Based on Acquisition Process}

There are three ways to acquire  context:  sense, derive,  and manually provided.

\begin{noindlist2}
 \item Sense:  The data is sensed  through sensors,  including the sensed  data stored  in databases  (e.g. retrieve  temperature from a sensor, retrieve  appointments  details  from a calendar).
 
 \item Derive: The information is generated by performing computational  operations on sensor data. These operations could be as simple as web service calls or  as complex  as mathematical functions  run over sensed  data (e.g. calculate distance  between  two sensors  using GPS coordinates). The necessary  data should be available  to apply any numerical or logical reasoning technique.
 
 \item Manually provided: Users  provide context information manually via predefined settings options such as  preferences  (e.g. understand  that user doesn't like  to receive event notifications  between 10pm to 6.00am). This method can be use to retrieve any type of information.
\end{noindlist2}

\subsection{Context Modelling}
\label{chapter2:CAF:Context Modelling}

We discuss the basic definition of context modelling in Section \ref{chapter2:CAF:CARD:Definition_of_Context_Model_and_Context_Attribute}. Context modelling is also widely refereed to as context representation. There are several popular context modelling \textcolor{blue}{techniques \cite{P402, P269} used}  in context-aware computing. Before we present the discussion on context modelling  techniques, let's briefly introduce context modelling fundamentals. Context models can be static or dynamic. Static models have a predefined set of context information that will be collected and stored \cite{P271}. The requirements  that need  to be taken into consideration when modelling context information are identified and explained in  \cite{P216}  as heterogeneity  and mobility, relationships  and dependencies, timeliness  (also called freshness), imperfection,  reasoning, usability of modelling formalisms, and efficient context provisioning.  Typically, there are two steps in representing context according to a model:

\begin{noindlist2}
 \item Context modelling process:  In the first step, new context information needs to be defined in terms of attributes, characteristics, relationships with previously specified context, quality-of context attributes and the queries for synchronous context requests.

 \item Organize context according to the model: In the second step, the result of the context modelling step needs  to be validated. Then the new context information  needs to be merged and added to the existing context information repository. Finally, the new context information is made available to be used when required.
\end{noindlist2} 

The first step performs the actual modelling of context. However, the factors and parameters that are considered for the modelling context are very subjective. It varies from one solution to another.  We  use two examples to demonstrate the variance. Currently,  there is no standard  to specify what type of information  needs to be considered in context modelling. We discussed context  categories proposed by the researcher in Section \ref{chapter2:CAF:context Types}. Even though these categories provide high-level  guidelines towards choosing relevant context, choosing specific context attributes is a subjective  decision.

\textit{Example 1:} MoCA \cite{P277} has used an object oriented  approach to model context using XML. There are three sections in the proposed context model: structural information  (e.g. attributes and dependencies among context types), behavioural information (e.g. whether the  context  attribute  has a constant or variable value), and context-specific abstractions (e.g. contextual events and queries).

\textit{Example 2:} W4 Diary \cite{P287} uses a  W4 (who, what, where, when) based  context model to structure data in  order to extract high-level information  from location  data. For example, W4 represents context   as  tuples (e.g. Who: John, What: walking:4km/h,  Where: ANU, Canberra, When: 2013-01-05:9.30am).

 In the IoT paradigm,  context information has  six states   \cite{P335}: ready, running,  suspended, resumed, expired,  and terminated. These states  are also similar to the process  states  in  an operating system. They align context to an event. An example scenario from the smart agriculture  domain can be used to explain the state transition  of context.
  
\begin{noindlist2}
 
 \item Ready: Every context is in the ready state at the initial stage (e.g. possible event can be `an animal eating crop').
 
 \item Suspended: When the context   seems to be invalid temporally  (e.g. sensors detect that animal stops eating crop temporarily).
 
 \item Resumed:  When the context becomes  valid from being suspended (e.g. sensors detect animal  starts to eat crop again).
 
 \item Expired: When the context   has expired and further information is not available (e.g. sensor data has  not been received  by the system  for the last 60 seconds where all sensor data is considered to be expired (based on policy) within 20 seconds  from the time it is collected).
 
 \item Terminated: When the context is no longer valid (i.e. inferred something  else) and further information is not available (e.g. sensors detects that animal moves away from the crops).
\end{noindlist2}


\begin{table*}[ht!]
\centering
\footnotesize
\renewcommand{\arraystretch}{-4}
\caption{Comparison of Semantic Web Ontology Languages (RDF(S), OWL(2))}
\vspace{-0.3cm} 
\begin{tabular}{ m{0.05cm} m{6.5cm} m{10cm}  }

\hline  
   \multicolumn{1}{r}{}   &    
\begin{center} RDF(S) \end{center} & 
\begin{center} OWL(2) \end{center}

\\ \hline \hline


\begin{sideways}Pros\end{sideways}     
&   \begin{noindlist}
 		 \item Provide basic elements to describe and organize knowledge. Further, OWL is build on top of RDFS
		 \item Relatively simple
 		 \item Faster processing and reasoning

    \end{noindlist}  
 &  \begin{noindlist}
  		 \item Improved version of RDFS. Therefore adaptability from RDF(S) to OWL is high
 		 \item Increasing number of tools are supported
  		 \item More expressive (e.g. larger vocabulary/constraints, rules, more meaningful)
  		 \item Higher machine interoperability (e.g. strong syntax)
  		 \item W3C approved standard for semantics (since 2004)
  	     \item Comes in three versions (i.e. OWL light, OWL DL, OWL Full) where each one has more expressive and reasoning power that previous
     \end{noindlist}  

\\ 
\begin{sideways} Cons  \end{sideways}      
&   \begin{noindlist}
		 \item Lack of inconsistency checking and reasoning
 		 \item Limited expressiveness (e.g. no cardinality support)
\end{noindlist}  
 &
  \begin{noindlist}
 		 \item Relatively Complex
  		 \item Low performance (e.g. require more computation power and time)
 \end{noindlist}  

\\ \hline

\end{tabular}

\label{Tbl:Comparison of Semantic Technologies}
\vspace{-0.6cm}
\end{table*}


The most popular context modelling techniques  are surveyed in  \cite{P431, P184}. These  surveys  discuss a  number  of systems  that have  been developed  based  on the following techniques.  Each of the following techniques  has its own strengths and weaknesses. We discuss context modelling techniques at a high-level. The actual implementations  of these  techniques  can vary widely depending on application domain (e.g. implementation details may differ from embedded environments to mobile environments to cloud based environments). Therefore, our focus is on conceptual  perspective of each modelling technique no on specific implementation.  Our discussion  is  based  on the six most popular context modelling techniques:  \textit{key-value, markup schemes, graphical, object based, logic based,} and \textit{ontology based modelling}. A comparison of these models is presented in Table \ref{Tbl:Comparison_of_Context_Modelling_and_Representation_Techniques}.

\subsubsection{Key-Value Modelling}
It models context information as key-value   pairs in different formats such as  text files and binary files. This is the simplest form of context representation  among  all the other techniques. They are easy to manage  when they have  smaller amounts  of data. However, key-value modelling is not scalable and not suitable to store complex data structures.  Further, hierarchical  structures  or relationships cannot be modelled using key-value pairs. Therefore, lack of data structuring  capability makes it difficult to retrieve  modelled information efficiently. Further, attaching meta information is not possible. The key-value  technique is an application oriented and application  bounded technique that suits  the purpose  of  temporary  storage  such as  less complex application configurations and user preferences.

\subsubsection{Markup Scheme Modelling (Tagged Encoding)} It models data using tags. Therefore,  context   is stored within tags. This technique is an improvement over the key-value modelling technique.  The advantage  of using markup tags is that it allows efficient data retrieval. Further, validation  is supported through schema definitions.  Sophisticated validation tools are available  for popular markup techniques  such as XML.  Range checking  is also possible  up to some  degree for  numerical values. Markup schemas  such as  XML  are widely used  in almost all application  domains to store data temporarily,  transfer data among applications, and transfer data among application  components. In contrast, markup languages do not provide advanced expressive capabilities which allow reasoning. Further, due to lack of design specifications, context modelling, retrieval, interoperability,  and re-usability over different markup schemes  can be difficult. A common application of markup based modelling  is modelling profiles. Profiles are commonly developed  using languages  such as XML.  However, the concept  of markup languages  are not restricted only to XML. Any language or mechanism (e.g. JSON) that supports tag based storage allows markup scheme modelling. An example of popular markup scheme modelling is Composite  Capabilities/Preference  Profiles (CC/PP) \cite{P529}. There are a significant number of similar emerging applications such as  ContextML \cite{P423} in context-aware computing. Tuples are also used to model context \cite{P271}.

\subsubsection{Graphical Modelling} It models context with relationships.  Some examples of this modelling technique are Unified Modelling Language (UML) \cite{P530} and Object Role Modelling (ORM) \cite{P531}. In terms of expressive richness, graphical modelling is better than markup and key-value modelling as  it allows relationships  to be captured  into the context model. Actual low-level  representation of the graphical modelling technique could be varied. For example, it could be a SQL database, noSQL  database, XML, etc. Many other extensions have also been proposed and implemented  using this technique  \cite{P389}. Further, as we are familiar with databases, graphical  modelling is a  well known, easy  to learn, and easy  to use technique. Databases can hold massive amounts of data and provide simple data retrieval operations, which can be performed relatively quickly. In contrast, the number of different implementations (i.e. different databases and other solutions)  makes it difficult with regards to interoperability.  Further, there are limitations  on data retrieval mechanisms such as  SQL. In addition, sophisticated context retrieval  requirements may demand very complex SQL queries to be employed. The queries can be difficult to create, use, and manage even with the sophisticated tools that exist today. Adding context information  and changing the data structure is also difficult in later stages. However,  some of the recent trends and solutions  in the noSQL  \cite{P556} movement allows these structure alteration  issues to be overcome. Therefore, graphical modelling techniques  can be used as  persistent   storage  of context.

\subsubsection{Object Based Modelling} Object based (or object oriented) concepts are used to model data using class hierarchies and relationships. Object oriented paradigm promotes encapsulation and re-usability.  As most of the high-level programming languages support object oriented concepts, modelling  can integrated into context-aware systems easily. Therefore, object based  modelling is  suitable to  be used as an internal, non-shared,  code based,  run-time context modelling, manipulation, and storage mechanism. However,  it does not provide inbuilt reasoning  capabilities.  Validation of  object oriented designs is also difficult due to the lack of standards and specifications.

\subsubsection{Logic Based Modelling} Facts,  expressions,  and rules are used  to represent  information about the context. Rules are used by other modelling  techniques, such as ontologies,  as well. Rules are primarily used to express policies, constraints, and preferences. It provides much more expressive richness compared to the other models discussed previously. Therefore, reasoning is possible up to a certain  level. The specific structures and languages that can be used to model context using rules are varied. However, lack of standardisation reduces the re-usability and applicability.  Furthermore, highly sophisticated and interactive  graphical  techniques can be employed to develop logic based or rule based representations.  As a  result, even non-technical  users can add rules and logic to the systems during run time. Logic based  modelling allows new high-level context information to be extracted using low-level context. Therefore, it has the capability  to enhance other context modelling techniques by acting  as a supplement.

\subsubsection{Ontology Based Modelling} The context   is organised into ontologies using semantic technologies. A number of different standards (RDF, RDFS, OWL) and reasoning capabilities are available to be used depending on the requirement. A wide range of development tools and reasoning engines are also available. However,  context retrieval can be computationally intensive and time consuming when the amount of data is increased. According  to many surveys, in context-aware computing  and sensor data management, ontologies  are the preferred mechanism of managing and modelling  context despite its weaknesses. Due to its popularity  and wider adaptation during the last five years in both academia and industry we present   a  brief discussion  on semantic  modelling and reasoning. However,  our intention is not to survey semantic technologies but to highlight the applicability  of semantics in a context-aware domain from an IoT perspective. Comprehensive and extensive amounts of information on semantic technology are available in  \cite{P557, P558, P378}.

\label{chapter2:CAF:CM:Ontology_Based_Modelling}


\begin{table*}[t]

\centering
\footnotesize
\caption{Comparison of Context Modelling and Representation Techniques}
\vspace{-0.3cm}
\renewcommand{\arraystretch}{-4}
\begin{tabular}{ m{1.2cm} m{4.8cm} m{4.8cm} m{5.5cm}  }\hline
 
\begin{center} Techniques \end{center}       &    
\begin{center} Pros \end{center} & 
\begin{center} Cons \end{center} & 
\begin{center} Applicability \end{center}

\\ \hline \hline

\begin{center} Key-Value  \end{center}      
&   
\begin{noindlist}
		 \item Simple
 		 \item Flexible
 		 \item Easy to manage when small in size
\end{noindlist}  
 & 
 \begin{noindlist}
 		 \item Strongly coupled with applications
  		 \item Not scalable
  		 \item No structure or schema
  		 \item Hard to retrieve information
  		 \item No way to represent relationships
  		 \item No validation support
  		 \item No standard processing tools are available
 \end{noindlist} 
 &
Can be used to model limited amount of data such as user preferences and application configurations. Mostly independent and non-related pieces of information. This is also suitable for limited data transferring and any other less complex temporary modelling requirements.

\\ 
\begin{center} Markup Scheme Tagged Encoding (e.g. xml)  \end{center}      
&   
\begin{noindlist}
		 \item Flexible
 		 \item More structured
 		 \item Validation possible through schemas
 		 \item Processing tools are available 		 
\end{noindlist}  
 &
 \begin{noindlist}
  		 \item Application depended as there are no standards for structures
  		 \item Can be complex when many levels of information are involved
  		 \item Moderately difficult to retrieve information
  		 
 \end{noindlist}  
 &
Can be used as intermediate data organisation format as well as mode of data transfer over network. Can be used to decouple data structures used by two components in a system. (e.g. SensorML \cite{P256} for store sensor descriptions, JSON as a format to data transfer over network)
\\ 

\begin{center} Graphical (e.g. databases)   \end{center}      
&   
\begin{noindlist}
		 \item \textcolor{blue}{Allows} relationships modelling
 		 \item Information retrieval is moderately easier
 		 \item Different standards and implementations are available.
 		 \item Validation possible through constraints 
\end{noindlist}  
 & 
 \begin{noindlist}
 		 \item Querying can be complex
  		 \item Configuration may be required
  		 \item Interoperability among different implementation is difficult
  		 \item No standards but governed by design principles
 \end{noindlist}  
 &
Can be used for long term and large volume of permanent data archival. Historic context can be store in databases.
\\ 

\begin{center} Object Based  \end{center}      
&   
\begin{noindlist}
		 \item \textcolor{blue}{Allows} relationships modelling
 		 \item Can be well integrated using programming languages
 		 \item Processing tools are available
\end{noindlist}  
 & 
\begin{noindlist}
		 \item Hard to retrieve information
 		 \item No standards but govern by design principles
 		 \item Lack of validation
\end{noindlist}   
 &
Can be used to represent context in programming code level. \textcolor{blue}{Allows} context runtime manipulation. Very short term, temporary, and mostly stored in computer memory. Also support data transfer over network.
\\ 

\begin{center}Logic Based  \end{center}      
&   
\begin{noindlist}
		 \item \textcolor{blue}{Allows} to generate high-level context using low-level context
 		 \item Simple to model and use
 		 \item support logical reasoning
 		 \item Processing tools are available
\end{noindlist}  
 & 
 \begin{noindlist}
 		 \item No standards
  		 \item Lack of validation
  		 \item Strongly coupled with applications
 \end{noindlist} 
 &
Can be used to generate high-level context using low-level context (i.e. generate new knowledge), model events and actions (i.e. event detection), and define constrains and restrictions.

\\ 

\begin{center} Ontology Based  \end{center}      
&   
\begin{noindlist}
		 \item Support semantic reasoning
 		 \item \textcolor{blue}{Allows} more expressive representation of context
 		 \item Strong validation
 		 \item Application independent and \textcolor{blue}{allows} sharing
 		 \item Strong support by standardisations
 		 \item Fairly sophisticated tools available 		 
\end{noindlist}  
 & 
 \begin{noindlist}
    \item Representation can be complex
    \item Information retrieval can be complex and resource intensive
 \end{noindlist}
 
 &
Can be used to model domain knowledge and structure context based on the relationships defined by the ontology. Rather than storing data on ontologies, data can be stored in appropriate data sources (i.e. databases) while structure is provided by ontologies. 
\\ \hline

\end{tabular}

\label{Tbl:Comparison_of_Context_Modelling_and_Representation_Techniques}
\vspace{-0.6cm}
\end{table*}


Khoo \cite{P057} has explained the evolution  of the web in four stages: basic  Internet  as Web 1.0, social media and user generated content  as web 2.0, semantic web as  web 3.0 and IoT as  web 4.0. In this identification,  semantic web has  been given a separate  phase  to show its importance  and the significant  changes that semantic technologies can bring to the web in general.

Ontology is the main component in semantic technology that allows it to model data. Based on the previous approaches and survey \cite{P184}, one of the most appropriate formats to manage  context is ontologies.  Ontologies  offer an expressive language to represent the relationships and context. IT also provides  comprehensive  reasoning  mechanisms as well. Ontologies also allow knowledge sharing and they decouple the knowledge from the application  and program codes \cite{P419}.

There are several reasons to develop and use ontologies  in contrast  to other modelling techniques.  The most common reasons  are to  \cite{P191, P447} share a common  understanding  of the structure  of information among  people  or software  agents, analyse domain knowledge,  separate domain knowledge  from operational knowledge, enable reuse of domain knowledge, high-level knowledge inferring, and make domain assumptions explicit. Due to the dynamic nature, the IoT middleware solutions should support applications which are not even known at the middleware design-time. Ontologies allow the integration of knowledge on different domains into applications when necessary.

Studer et al. \cite{P546} defined the concept of ontology as follows. \textit{``An ontology is a formal, explicit specification of a shared conceptualisation. A conceptualisation refers to an abstract model of some phenomenon in the world by having identified the relevant concepts of that phenomenon. Explicit means that the type of concepts used, and the constraints on their use are explicitly defined. For example, in medical domains, the concepts are diseases and symptoms, the relations between them are causal and a constraint is that a disease cannot cause itself. Formal refers to the fact that the ontology should be machine readable, which excludes natural language. Shared reflects the notion that an ontology captures consensual knowledge, that is, it is not private to some individual, but accepted by a group.''} Another  acceptable definition  has been presented by Noy and McGuinness \cite{P447}. Further ontologies  are discussed extensively as principles, methods, and applications in perspective \cite{P445}.

Some of the requirements and objectives behind designing an ontology are simplicity, flexibility and extensibility, generality, and expressiveness  \cite{P545}. In addition, some of the general requirements in context modelling  and representation are unique identification, validation, reuse,  handling uncertainty, and incomplete information  \cite{P185}. A further eight principles for developing ontologies are identified by Korpipaa and Mantyjarvi  \cite{P034} as:  domain, simplicity, practical access,  flexibility  and expandability, facilitate inference, genericity,  efficiency, and expressiveness.

Ontologies consists of several common key components  \cite{P197,P332} such as individuals,  classes, attributes, relations, function terms, restrictions,  rules, axioms, and events. Furthermore, there are two steps in developing ontologies. First, the domain and scope need to be clearly defined. Then existing ontologies need to be reviewed to find the possibilities of leverage existing  in ontologies.  One of the main goals of ontologies is the reusability of shared knowledge. By the time this survey was prepared, there were several popular domains that design, develop, and use ontologies. Sensor domain is one of them. A survey of the semantic specification of sensors is presented in  \cite{P103}. They have evaluated and compared a number of ontologies and their capabilities.

There are several popular semantic web ontology languages that can be used to develop ontologies:  RDF \cite{P252}, RDFS \cite{P559}, OWL \cite{P148}. The current recommendation is OWL 2 which is an extended  version of OWL. A significant amount  of OWL usage  has  been  noticed in the context modelling ad reasoning domain  \cite{P185}. It further emphasises the requirement of having the modelling  language, reasoning engines, and mechanism to define rules as a bundle, rather than choosing different  available options arbitrarily, to get the real power of semantic technologies. SWRL is one of the available  solutions  to add rules in OWL  \cite{P216}. SWRL is not a hybrid  approach as it is fully integrated into ontological  reasoning. In contrast, when the amount of data becomes larger and structure becomes complex,  ontologies can becomes exceedingly  complex  causing the reasoning process to be resource intensive and slow. However,  some of the main reasons to choose OWL as the context modelling  mechanism are  \cite{P419, P332}.

\begin{noindlist2}
 \item W3C  strongly supports the standardisation   of  OWL.
 Therefore, a variety of development tools are available for integrating  and managing OWL ontologies, which makes it easier to develop and share.

 \item OWL allows interoperability  among other context-aware systems. These features, such as classes, properties and constraints, and individuals  are important for supporting ontology  reuse, mapping and interoperability.
 
 \item OWL supports a high-level of inference / reasoning support.

 \item OWL is more expressive. For example, it provides cardinality constraints,  which enables  imposing additional restrictions on the classes.
\end{noindlist2}

We compare the two most popular web ontology languages, RDF(S) and OWL(2) in
Table \ref{Tbl:Comparison of Semantic Technologies}, to highlight the fundamental differences.

\begin{figure*}[t]
 \centering

 \includegraphics[scale=.62]{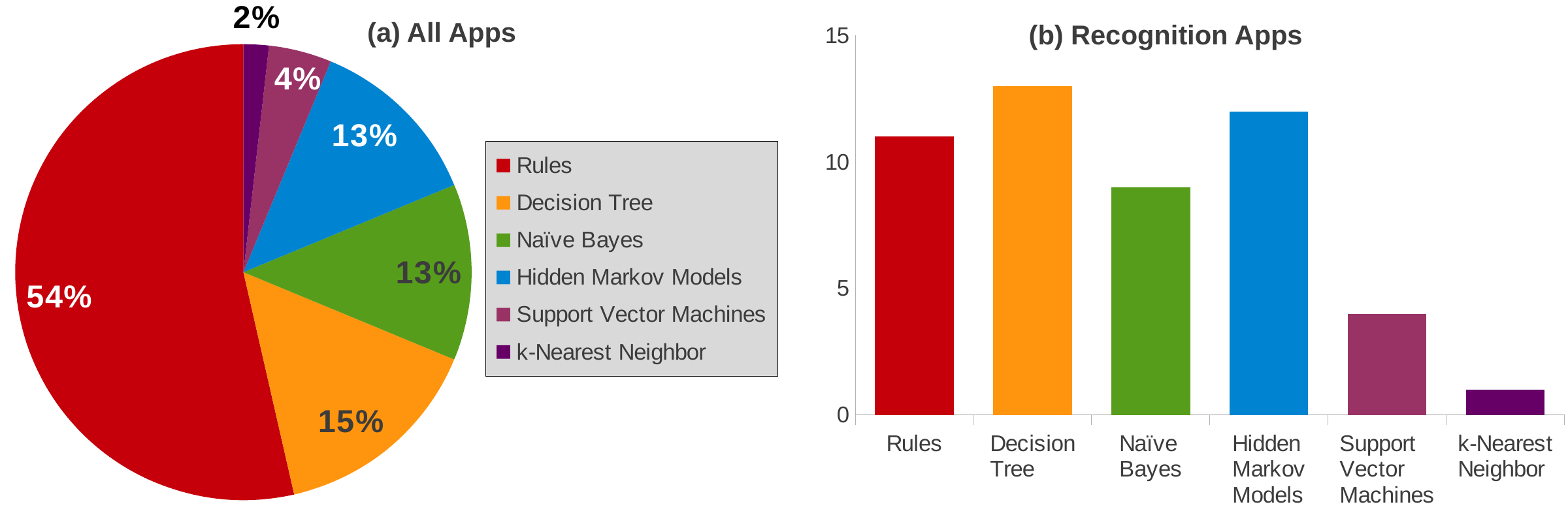}
\vspace{-10pt} 
 \caption{(a) Counts of model types used in 109 of 114 reviewed context-aware applications. (b) Counts for 50 recognition applications; classifiers are used most often for applications that do recognition \cite{P384}.}
 
 \label{Fig:Survey_on_Context_Reasoning_Techniques}
 \vspace{-14pt} 
\end{figure*}


After evaluating  several context modelling techniques,  it was revealed that incorporating multiple modelling techniques is the best way to produce efficient and effective results, which will  mitigate each other's weaknesses. Therefore,  no single modelling  technique is ideal to be used in a standalone fashion. There is a strong  relationship  between context  modelling and reasoning. For example, some reasoning techniques prefer some modelling  techniques. However, it should not limit the employability of different context reasoning  and modelling techniques together. In the next section we discuss reasoning context-aware computing.

\subsection{Context Reasoning Decision Models}
\label{chapter2:CAF:Context Reasoning Decision Models}

Context  reasoning can be defined  as a method  of deducing new knowledge,  and understanding  better,  based on the available  \textcolor{blue}{context \cite{P215}. It  can} also be explained   as  a process of giving high-level  context deductions from a set of contexts  \cite{P331}. The requirement of reasoning also emerged due to two characteristics of raw context: imperfection  (i.e. unknown,  ambiguous, imprecise, or erroneous) and uncertainty. Reasoning  performance  can be measured  using efficiency, soundness, completeness,  and interoperability  \cite{P185}. Reasoning is also called inferencing. Contest reasoning comprises several steps. Broadly  we can divide them into three phases  \cite{P214}. 

\begin{noindlist2}
 \item Context pre-processing: This phase cleans the collected sensor  data. Due to inefficiencies  in sensor hardware and network communication, collected data may be not accurate  or missing.  Therefore, data needs to be cleaned by filling missing values, removing  outliers, validating context via multiple sources, and many more. These tasks have been extensively  researched by database, data mining,  and sensor network  research communities over many years.
 
 \item Sensor data fusion: It is a method  of combining  sensor data from multiple sensors  to produce  more accurate, more complete, and more dependable  information that could not be achieve through  a single  sensor \cite{P248}. In the IoT, fusion is extremely  important,  because there will be billions of sensors available. As a result, a large number of alternative sources will exist to provide the same information.
 
 \item Context inference: Generation of high-level  context information using lower-level context. The inferencing can be done in a single interaction or in multiple interactions. Revisiting an example from a different  perspective, W4 Diary  \cite{P287} represented context   as tuples \textcolor{blue}{(e.g. Who: }John, What: walking:4km/h,  Where: ANU,Canberra, When: 2013-01-05:9.30am). This low-level  context can be inferred through a number of reasoning mechanisms to generate  the final results.  For example,  in the first iteration, longitude and latitude values of a GPS  sensor may be inferred  as  \textit{PurplePickle cafe in canberra}. In the next iteration \textit{PurplePickle cafe in canberra} may be inferred as \textit{John's favourite cafe}. Each iteration gives more accurate and meaningful information.
 
\end{noindlist2}

There are a large number of  different context reasoning decision  models, such as decision   tree, naive Bayes, hidden Markov models, support vector machines, k-nearest neighbour, artificial neural networks,  Dempster-Shafer, ontology-based, rule-based,  fuzzy reasoning  and many more. Most of  the models originated  and are employed in the fields of artificial intelligence  and machine learning. Therefore, these models are not specific to context-reasoning but commonly  used across many different fields in computing and engineering.

We present the results of a survey  conducted  by Lim and Dey \cite{P384} in Figure \ref{Fig:Survey_on_Context_Reasoning_Techniques}. They have investigated the popularity  of  context reasoning decision models. The survey is based on literature from three major conferences over five years: Computer-Human Interaction (CHI) 2003-2009, Ubiquitous Computing (Ubicomp)  2004-2009, and Pervasive 2004-2009.

In the IoT paradigm, there are many sensors  that sense  and produce context information.  The amount of information that will be collected by over 50 billion sensors is enormous. Therefore, using all this context for reasoning in not feasible for many reasons, such as processing time, power, storage, etc. Furthermore, Guan et al.  \cite{P331} has proved  that using more context will  not necessarily improve  the accuracy  of the inference in a considerable  manner.  They have used two reasoning models in their research: back-propagation neural networks and k-nearest neighbours. According  to the results, 93\% accuracy  has been achieved  by using ten raw context. Adding 30 more raw context to the reasoning model has increased the accuracy by only 1.63\%. Therefore, selecting  the appropriate  raw context for reasoning  is critical to infer high-level context with high accuracy.

Context reasoning  has been  researched  over many years. The most popular context reasoning techniques (also called decision models) are surveyed in \cite{P185, P216, P215}. Our intention in this paper is not to survey context reasoning techniques but to briefly introduce them so  it will  help to understand and appreciate the role of context reasoning in the IoT paradigm.  We classify context reasoning techniques broadly into six categories: \textit{supervised learning, unsupervised learning, rules, fuzzy logic, ontological reasoning} and \textit{probabilistic reasoning}. A comparison of these techniques is presented in Table \ref{Tbl:Context_Reasoning_Decision_Models}

\begin{table*}[htbp]
\centering
\footnotesize
\caption{Comparison of Context Reasoning Decision Modelling Techniques}

\renewcommand{\arraystretch}{-4}
\begin{tabular}{ m{2.1cm} m{4.8cm} m{4.5cm} m{5cm}   }\hline
 
\begin{center}  Techniques \end{center}       &    
\begin{center} Pros \end{center} & 
\begin{center} Cons \end{center} & 
\begin{center} Applicability \end{center}

\\ \hline \hline

\begin{center} Supervised Learning (Artificial neural network, Bayesian Networks, Case-based reasoning, Decision tree learning, Support vector machines)  \end{center}      
&   
\begin{noindlist}
		 \item Fairly accurate
 		 \item Number of alternative models are available
 		 \item Have mathematical and statistical foundation
\end{noindlist}  
 & 
 \begin{noindlist}
 		 \item Require significant amount of data
  		 \item Every data element need to be converted in to numerical values
  		 \item Selecting feature set could be challenging
  		 \item Can be more resource intensive (processing, storage, time)
  		 \item less semantic so less meaningful
  		 \item Training data required
  		 \item Models can be complex
  		 \item Difficult to capture existing knowledge
  		 
 \end{noindlist}  
 &
For situation where the feature set is easily identifiable, possible out comes are known, and large data sets (for training as well) are available in numerical terms. (For example: activity recognition, missing value identification)

\\ 
\begin{center} Unsupervised Learning (Clustering, k-Nearest Neighbour)  \end{center}      
&   
\begin{noindlist}
		 \item No training data required
 		 \item No need to know the possible outcome
\end{noindlist}  
 & 
 \begin{noindlist}
 		 \item Models can be complex
  		 \item Less semantic so less meaningful
  		 \item Difficult to validate
  		 \item Outcome is not predictable
  		 \item Can be more resource intensive (processing, storage, time)
 \end{noindlist} 
 &
 For situations where possible out comes are not known (For example: unusual behaviour detection, analysing agricultural fields to identify appropriate location to plant a specific type of crop)
 
\\ 

\begin{center} Rules  \end{center}      
&   
\begin{noindlist}
		 \item Simple to define
		 \item Easy to extend
 		 \item Less resource (e.g. processing, storage) intensive
\end{noindlist}  
 &
\begin{noindlist}
 		 \item Should define manually
 		 \item Can be error prone due to manual work
 		 \item No validation or quality checking
\end{noindlist}  
  &
For situations where raw data elements need to be converted in to high level context information. Suitable to be used to define events.
\\ 

\begin{center} Fuzzy Logic \end{center}      
&   \begin{noindlist}
		 \item Allow more natural representation
 		 \item Simple to define
 		 \item Easy to extend
 		 \item Less resource (e.g. processing, storage) intensive
 		 \item Can handle uncertainty
\end{noindlist}  
 & 
\begin{noindlist}
 		 \item Should define manually
 		 \item Can be error prone due to manual work
 		 \item No validation or quality checking
 		 \item May reduce the quality (e.g. precision) of the results due to natural representation
\end{noindlist}  
&
For situation where  low-level context need to be converted in to high-level more natural context information. This type of simplification will make it easy to process further. For example, control automated irrigation system where water will be released when the system detect the soil is `dry'
\\ 

\begin{center} Ontology based (First-Order Predicate Logic)  \end{center}      
&  
\begin{noindlist}
		 \item Allow complex reasoning
 		 \item Allow complex representation
 		 \item More meaningful results
 		 \item Validation and quality checking is possible
 		 \item Can reason both numerical and textual data
\end{noindlist}  
 &
 \begin{noindlist}
 		 \item Data need to be modelled in a compatible format (e.g. OWL, RDF)
 		 \item Limited numerical reasoning

  		 \item Low performance (e.g. require more computation power and time)
 \end{noindlist} 
 &
For situations where knowledge is critical. For example, store and reason domain knowledge about agricultural domain. It allows the context information to be store according to the ontology structure and automatically reason later when required
\\ 

\begin{center} Probabilistic logic (Dempster-Shafer, hidden Markov Models, naive Bayes)  \end{center}      
&   
\begin{noindlist}
		 \item Allows to combine evidence
 		 \item Can handle unseen situations
 		 \item Alternative models are available
 		 \item Can handle uncertainty
 		 \item provide moderately meaningful results
\end{noindlist}  
 & 
 \begin{noindlist}
 		 \item Should know the probabilities
 		 \item Reason numerical values only
 \end{noindlist} 
 &
For situations where probabilities are known and combing evidence from different sources are essential. For example, evidence produced from a camera, infra-red sensors, acoustics  sensor, and motion detector can be combined to detect a wind animal  infiltrate to a agricultural field 
\\ \hline

\end{tabular}

\label{Tbl:Context_Reasoning_Decision_Models}
\vspace{-0.6cm}
\end{table*}

\subsubsection{Supervised learning} In this category of techniques, we first collect training examples. Then we label them according to the results we expect.  Then we derive   a  function that can generate  the expected  results  using the training data. This technique is widely used in mobile phone  sensing \cite{P217} and activity recognition \cite{P187}. \textit{Decision tree} is a supervised learning technique where it builds a  tree from a dataset  that can be used  to classify data.  This technique  has been used  to develop a  student  assessment  system  in  \cite{P561}. \textit{Bayesian Networks} is a technique  based on probabilistic reasoning  concepts.  It uses directed  acyclic  graphs to represent events and relationships among  them. It is a widely used technique  in statistical reasoning.  Example applications  are presented  in   \cite{P197, P289}. Bayesian networks are commonly  used in combining uncertain information from a  large number  of sources  and deducing higher-level contexts. \textit{Artificial neural networks} is a technique that attempts to mimic the biological  neuron system. They are typically used to model complex relationships between inputs and outputs or to find patterns in data. Body sensor networks domain  has employed  this technique for pervasive healthcare monitoring in  \cite{P267}. \textit{Support vector machines} are widely used for pattern recognition in context-aware computing. It has been used to detect activity recognition of patients in the healthcare domain  \cite{P562} and to learn situations in a smart home environment \cite{P209}.

\subsubsection{Unsupervised learning} This category of  techniques can find hidden structures  in unlabelled  data. Due to the use of no training data, there is no error or reward signal to evaluate a potential  solution.  Clustering  techniques such as \textit{K-Nearest Neighbour} is popularly  used in context-aware reasoning. Specifically,  clustering is used in low-level (sensor hardware  level) sensor  network operations  such as  routing and high level tasks such  as indoor and outdoor positioning and location  \cite{P565}. Unsupervised neural network techniques such as Kohonen  Self-Organizing  Map (KSOM) are used to classify incoming sensor data in a real-time  fashion  \cite{P566}. Noise detection and outlier detection are other applications in  context-aware  computing. Applications of  unsupervised learning  techniques in relation to body sensor networks  are surveyed in  \cite{P267}. The unsupervised clustering  method has been employed to capturing user contexts by dynamic profiling in  \cite{P268}.

\subsubsection{Rules} This is the simplest and most straightforward methods of reasoning out of all of them. Rules are usually structure in an IF-THEN-ELSE format. This is the most popular method of reasoning according  to Figure \ref{Fig:Survey_on_Context_Reasoning_Techniques}. It allows the generation of high level context information using low level context. Recently, rules have been heavily used when combined with ontological reasoning  \cite{P243, P420, P421}. MiRE \cite{P298} is a minimal rule engine for context-aware mobile devices. Most of the user preferences are encoded using rules. Rules are also used in event detection  \cite{P136, P128}. Rules are expected  to play a significant role in the IoT, where they are the easiest and simplest  way to model human thinking and reasoning in machines. PRIAMOS  \cite{P139}  has used semantic  rules to annotate sensor data with context information. Application of rule based reasoning  is clearly explained in relation to context-aware I/O control in \cite{P567}.

\subsubsection{Fuzzy logic} This allows approximate reasoning instead of fixed and crisp reasoning. Fuzzy logic is similar to probabilistic reasoning but confidence  values represent degrees of membership rather than probability \cite{P444}. In traditional logic theory, acceptable truth values are  0 or 1. In fuzzy logic partial truth values are acceptable. It allows real world scenarios to be represented more naturally;  as most real world facts are not crisp. It further allows the use of natural  language (e.g. temperature: slightly warm, fairly cold) definitions rather than exact numerical  values (e.g. temperature:  10 degrees Celsius). In other words it allows imprecise  notions such as  tall,  short, dark, trustworthy and confidence to be captured, which is critical in context information  processing. In most cases, fuzzy reasoning cannot be used as a standalone reasoning technique. It is usually  used to complement  another techniques  such as rules based, probabilistic or ontological reasoning. Gaia  \cite{P568} has used fuzzy logic in context providers to handle uncertainty. Several examples of applying fuzzy logic to represent context information are presented in  \cite{P547, P548}.

\subsubsection{Ontology based}

: It is based on description logic, which is a family of logic based knowledge  representations of formalisms. Ontological  reasoning is mainly supported by two  common representations  of  semantic  web languages: RDF(S) \cite{P252} and OWL(2) \cite{P148}. We discussed ontology based modelling in Section \ref{chapter2:CAF:CM:Ontology_Based_Modelling}. Semantic web languages are also complemented by several semantic query languages: RDQL, RQL, TRIPLE and number of reasoning engines: FACT \cite{P253}, RACER, Pellet \cite{P150}. Rules such as SWRL \cite{P243}  are increasingly popular in ontological reasoning. The advantage of ontological reasoning is that it integrates well with ontology modelling. In contrast,  a  disadvantage   is that ontological reasoning  is not capable  of finding missing values  or ambiguous  information where statistical reasoning  techniques  are good at that. Rules can be used to minimise this weakness by generating new context information based  on low-level context. Missing values can also be tackled by having rules that enable  missing values to be replaced with suitable  predefined  values. However, these  mechanism  will  not perform accurately  in highly dynamic and uncertain domains. Ontological  reasoning is heavily  used in a wide range of applications,  such as  activity recognition \cite{P187}, hybrid reasoning \cite{P187}, and event detection \cite{P128}. A survey on semantic based reasoning is presented in  \cite{P215}. It also compares a number of context aware frameworks  based on modelling  technique, reasoning techniques, and architectures used in their systems. Comprehensive and extensive amounts of information on semantic technology  are available in  \cite{P557, P558, P378}. In addition,  a semantic based architecture  for sensor data fusion is presented in  \cite{P072, P073, P071}.

\subsubsection{Probabilistic logic} This category of techniques allows  decisions to be made based on probabilities  attached to the facts related to the problem. It can be used to combine  sensor data from two different  sources. Further, it can be used to identify resolutions to conflicts  among context. Most often these techniques are used to understand  occurrence  of events. Probabilistic logic has been used in  \cite{P444} to encode  access control policies. \textit{Dempster-Shafer}, which is based on probabilistic logic, allows different evidence to be combined to calculate  the probability of an event.  Dempster-Shafer  is commonly used  in sensor data fusion for activity recognition. In \cite{P548, P238}, it has been used to understand whether  there is a meeting  in the room. Other example applications  are presented in  \cite{P236, P235}. \textit{hidden Markov Models \cite{P553}} are also a probabilistic  technique that allows state to be represented using  observable evidence without directly reading the state. For example, it provides a method to bridge the gap between raw GPS sensor measurements and high level information such as a user destination,  mode of transportation, calendar  based observable  evidence  such as  user calendar, weather,  etc. hidden Markov Models are commonly  used in activity recognition in context-aware domains. For example, it has been used to learn situation models in a smart home \cite{P209}.


Up to now, we have presented and discussed a number of context modelling and reasoning  techniques.  However, it  is clear that each technique  has its own strengths and weakness. No single technique can be used to accomplish  perfect results. Therefore, the best method  to tackle the problem of context awareness it to combine multiple models in such  a way that, as  a  whole, they reduce weaknesses by complementing each other. For example, Alternative  Context Construction Trees (ACCT)  \cite{P326} is an approach that enables the concurrent evaluation  and consolidation  of different reasoning models such  as logic rules, Bayesian networks and CoCoGraphs  \cite{P560}. There are two reasons that context information can become uncertain,   as discussed  in  \ref{chapter2:CDLC:ESRF:QCRV}. Therefore,  employing or incorporating strategies  that can reason  under uncertainty such as Bayesian networks,  Dempster-Shafer or fuzzy logic is essential in such situations. The process of how the multiple techniques can be combined together is presented in  \cite{P216, P463}. We briefly explain the hybrid context modelling  and reasoning approach  as follows.

At the lowest level, statistical techniques can be used to fuse sensor data.  Then, fuzzy logic can be employed to convert fixed data in to more natural terms. In the future, Dempster-Shafer can be used to combine sensor data from different  sources. In addition,  machine learning  techniques, such as support vector machines and artificial  neural networks, can be used for further reasoning. After completing statistical reasoning, the high level data can be modelled using semantic technologies such as ontologies.  Ontological  reasoning can be applied to infer additional context information using domain knowledge at the higher level. A similar process is explained  in detail in \cite{P463}.

\subsection{Context Distribution}
\label{chapter2:CAF:Context Distribution}

Context distribution is a fairly straightforward  task. It provides methods to  deliver context  to  the consumers. From the consumer  perspective  this task can be called  context acquisition,  where the discussion we presented in Section \ref{chapter2:CDLC:Context Acquisition} is completely applicable. Therefore all the factors we discussed under context acquisition  need to be considered for context distribution  as well. Other than that there are two other methods to that are used commonly  in context distribution:

\begin{noindlist2}
 \item Query:  Context  consumer  makes a request in terms of a query, so  the context  management system can use  that query to produce results.

 \item Subscription  (also called publish / subscribe):  Context consumer  can be allowed to subscribe  with a  context management system by describing the requirements. The system will then return the results periodically or when an event  occurs (threshold  violation). In other terms, consumers can subscribe for a specific  sensor or to an event.  However, in underline  implementations,  queries may also use to define subscriptions. Further, this method is typically use in real time processing.
\end{noindlist2}

\section{Existing Research Prototypes and Systems}
\label{chapter2:PRE}

In this section, first we present our evaluation framework and then we briefly discuss some of the most significant projects and  highlight their significance. Later, we identify the lessons  we can learn from them towards context-aware development in the IoT paradigm in Section \ref{chapter2:LL}. The projects are discussed in the same order  as in Table  \ref{Tbl:Evaluation_of_Previous_Research_Efforts}. Our taxonomy is summarized in Table \ref{Tbl:Summarized taxonmy}.

\vspace{-0.2cm}

\subsection{Evaluation Framework}
\label{chapter2:CDLC:Evaluation of Surveyed Research Efforts}

\begin{figure*}[hbtp]
\centering
\includegraphics[scale=.80]{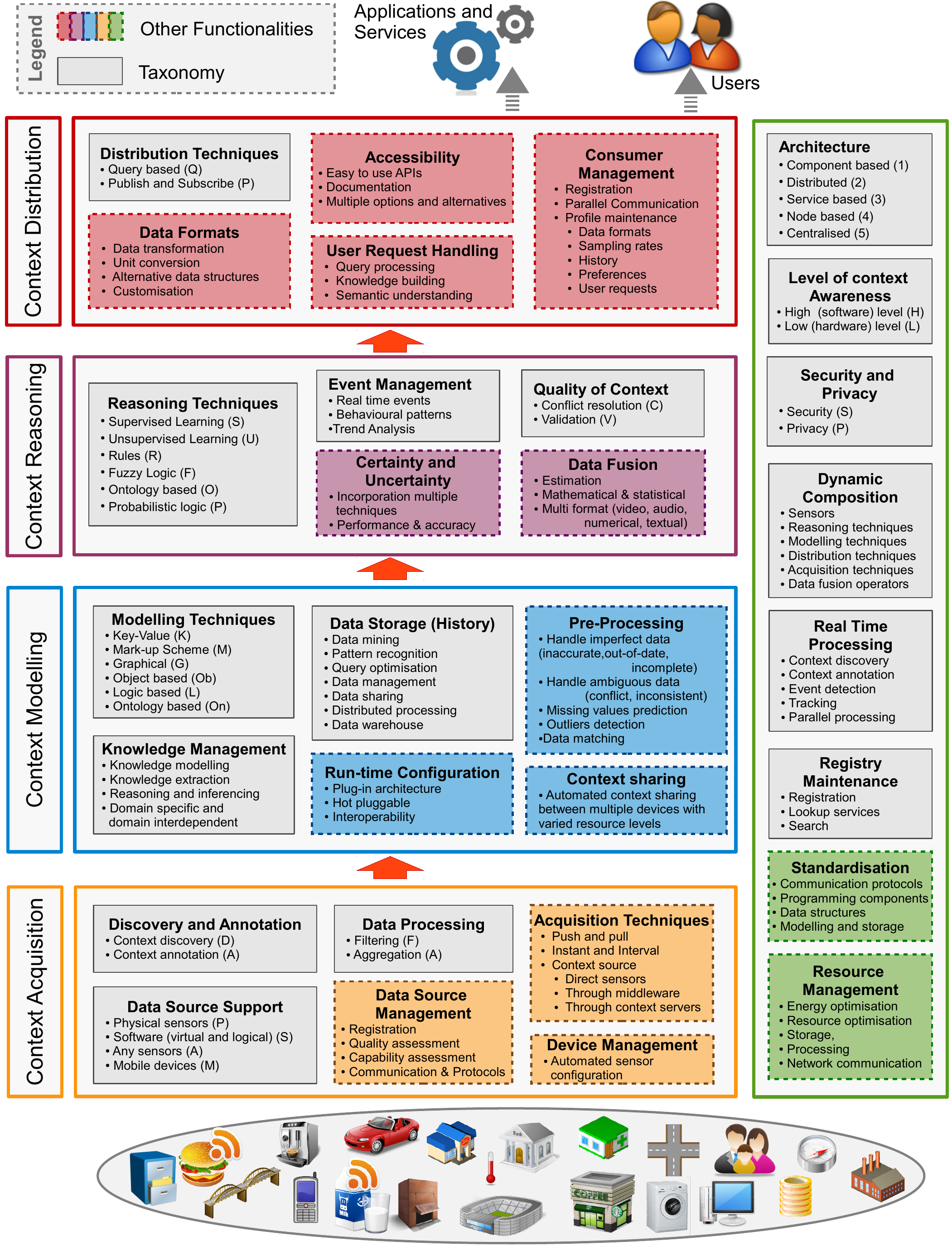}
\caption{Taxonomy (functionalities commonly supported in existing research prototypes and systems); Conceptual Framework (value added features that need to be supported by ideal context-aware IoT middleware solution)}
\label{Fig:Taxonomy_and_Conceptual_Framework}
\end{figure*}

We used abbreviations  as much as possible   to make  sure that the structure allowed  all 50 projects to be presented in a single  page, which enables the readers to analyse and identify positive  and negative patterns that we have not explicitly  discussed.  In Table  \ref{Tbl:Evaluation_of_Previous_Research_Efforts}, we use a dash (\n) symbol across  all columns to denote  that the functionality is either missing or not mentioned in related publications  that are available.  In order to increase  the readability,  we have numbered the columns of the Table   \ref{Tbl:Evaluation_of_Previous_Research_Efforts} corresponding  to the taxonomy numbered below. Our taxonomy  and several other features that will provide additional value in IoT solutions are visually illustrated in Figure  \ref{Fig:Taxonomy_and_Conceptual_Framework}.

\subsubsection{\textbf{Project Name}} This is the name given to the project by the authors of the related publications. Most of the project names are abbreviations  that are used to refer to the project.  However, some project do not have an explicit project name, here we used a dash (\n) symbol.

\subsubsection{\textbf{Citation}} We provide only one citation due to space limitations. Other citations are listed under  each project's descriptions and highlights  in Section \ref{chapter2:PRE}.

\subsubsection{\textbf{Year}} Table \ref{Tbl:Evaluation_of_Previous_Research_Efforts} is ordered according  to chronological order (i.e. from oldest to newest) based  on the year of publication.

\subsubsection{\textbf{Project Focus}} Based on our evaluation,  each project has  its own focus on whether to build a  system,   a  toolkit, or a  middleware   solution. The following abbreviations  are used to denote the focus: system (S), toolkit (T), and middleware (M). Systems  focus on developing an end-to-end solution where it involves hardware, software and application  layer. Systems  cannot  be used as  middleware.   It is designed to provide one or a few tasks. Building different functionalities on top of the system is not an option.  Systems are designed and developed for a use by the end users. Toolkits are not designed  to be used  by the end users.  They are employed by system,  application,  and middleware developers.  They provide very specific functionalities. Toolkits  are usually designed according to well-known  design principles and standards  and always released  with  proper documentation that shows how to use them  at programming code level. Middleware \cite{P064} can be explained   as a software layer that lies between the hardware and application  layers. It provides reusable functionalities  that are required by the application to meet complex customer requirements. They are usually built to address common  issues in application development such as heterogeneity, interoperability,  security, and dependability. A goal of middleware is to provide  a set of programming  abstractions to help software  development where  heterogeneous components  need to be connected and communicate  together. Middleware  is designed to be used by application developers, where the middleware solution handles most of the common functionalities  leaving more time and effort for the application  developers to deal with application functionalities.

\subsubsection{\textbf{Modelling}} This has been discussed in detail in Section \ref{chapter2:CAF:Context Modelling}. We  use the following abbreviations  to denote the context modelling techniques  employed by  the project: key-value modelling (K), markup Schemes (M), graphical modelling (G), object oriented modelling (Ob), logic-based modelling (L), and ontology-based modelling (On).

\subsubsection{\textbf{Reasoning}} This has been discussed in detail in Section \ref{chapter2:CAF:Context Reasoning Decision Models}. We  use the following abbreviations  to denote the context reasoning  techniques  employed  by the project: supervised learning (S), un-supervised learning (U), rules (R), fuzzy logic (F), ontology-based (O), and probabilistic  reasoning (P).  The symbol ($\checkmark$) is used where reasoning functionality is provided but the specific technique is not mentioned.

\subsubsection{\textbf{Distribution}}  This has been discussed in detail in Section \ref{chapter2:CAF:Context Distribution}. We  use the following  abbreviations  to denote the context distribution  techniques employed by the project: publish/subscribe (P) and query (Q).

\subsubsection{\textbf{Architecture}}
This varied widely from one solution to another. Architecture  can be classified into different categories based on different perspectives. Therefore, there is no common classification  scheme that can be used for all situations. We consider  the most significant architectural  characteristics to classify the solution. Different architectural  styles are numbered  as  follows. (1) Component  based architecture  where the entire solution is based  on loosely coupled major components, which interact  each other.  For example, Context Toolkit \cite{P143} has three major components which perform the most critical functionalities of the system.
(2) Distributed architecture enables peer-to-peer interaction in a distributed fashion, such as in Solar  \cite{P569}.
(3) Service based architecture  where the entire solution consists of several services working together. However, individual access to each service may not be provided in solutions such as  Gaia \cite{P444}.
(4) Node based architecture allows to deployment of  pieces of software with similar or different capabilities, which communicate and collectively  process data in sensor networks \cite{P344}.
(5) Centralised  architecture which  acts as a complete  stack  (e.g. middleware) and  provides applications  to be developed on top of that, but provides no communication between different instances of the solution.
(6) Client-server architecture  separates sensing and processing from each other, such as in CaSP  \cite{P317}.

\subsubsection{\textbf{History and Storage}} Storing context history is critical \cite{P290} in  both traditional context-aware  computing and the IoT. Historic data allows sensor data to be better understood.  Even though most of the IoT solutions and applications are focused on real time interaction, historic data has its own role to play. Specifically, it allows user behaviours, preferences, patterns,  trends,  needs, and many more to be understood. In contrast, due to the scale of the IoT, storing all the context  for the long term may not feasible. However,  storage devices are getting more and more powerful and cheap. Therefore,  it would be a tradeoff between cost and understanding.   The symbol ($\checkmark$) is used denote that context history functionality is facilitated and employed by the project.

\subsubsection{\textbf{Knowledge Management}} This functionality is broader than any others. Most of the tasks that are performed  by IoT middleware solutions require knowledge in different perspectives, such as knowledge on sensors, domains,  users, activities, and many more. One of  the most popular techniques  to represent knowledge in context-aware computing is using ontologies. However,  several other techniques are also available such as rules. Knowledge  can be used for tasks such as automated configuration  of sensors to IoT middleware, automatic sensor  data annotation,  reasoning, and event detection.  The symbol ($\checkmark$) is used to denote that knowledge management functionality is facilitated and employed by the project in some perspective.

\subsubsection{\textbf{Event Detection}}
\label{chapter2:CDLC:ESRF:Event_Detection} This is one of  the most important functionalities in  IoT solutions. IoT envisions machine-to-machine  (M2M)  and machine-to-person communication. Most of these interactions  are likely to occur based on an event. Events can referred to many  things,  such as an observable occurrence, phenomenon, or an extraordinary  occurrence. We define one or more conditions and identify it as an occurrence of an event once all the defined conditions  are satisfied. In the IoT, sensors collect data and compare it with conditions to decide whether the data satisfies the conditions. An occurrence event is also called a \textit{event trigger}. Once an event has been triggered,  a notification or action may be executed. For example, detecting current activity of a person or detecting  a meeting  status in a room, can be considered as events. Mostly, event detection needs to be done in real-time.  However,  events such as trends may be detected using historic data.  The symbol ($\checkmark$) is used to denote that event detection functionality is facilitated and employed by the project in some perspective.

\subsubsection{\textbf{Context Discovery and Annotation}}

We use the following abbreviations to denote context discovery and annotation facilitated and employed by the project: context discovery (D) and context annotation (A). Context annotation allows context related information and raw sensors data to be attached, modelled, and stored. Some of the most common and basic information that needs to be captured in relation to context are context type, context value, time stamp,  source, and confidence. Context-aware  geographical  information retrieval approach \cite{P421} has proposed  a mechanism  to map raw sensor data to semantic ontologies using SWRL. This is critical in all types of systems. Even though, statistical reasoning systems can use raw sensor data directly, semantic mapping before the reasoning allows more information to be extracted.  Context information  only becomes meaningful when it is interpreted with respect to the user. This can be achieved by knowledge  base integration  and reasoning using ontologies. Another application is discussed in \cite{P420}. Ontologies and other context modelling techniques  allow structure data to be more meaningful which express relationships among data.

End-users in the IoT paradigm are more interested in high-level information  compared to low-level raw sensor data \cite{P285}. The following examples explain  the difference  between high-level information and low-level raw sensor  data. It  is raining (high-level  information)  can be derived from humidity is 80\% (low-level sensor data). Further, high-level  sensor data can be explained as semantic information  as it provides more meaning to the end users. Challenges of semantic sensor webs are identified  and discussed in  \cite{P031}. This is the most common form of discovery.

\subsubsection{\textbf{Level of Context Awareness}} Context-awareness can be employed at two levels: low (hardware) level and high (software) level. At the hardware level, context-awareness is used to facilitate tasks such as efficient  routing, modelling,  reasoning, storage and event detection (considering energy consumption and availability)   \cite{P288}. At the hardware level, data and knowledge available  for decision making is less. Further, sensors  are resource constraint  devices, so complex  processing cannot be performed at the hardware level. However, applying context-aware technologies in the hardware level allows resources to be saved, such as network communication costs by preliminary filtering. The software level has access to a broader range of data and knowledge   as well as more resources, which  enables more complex  reasoning to be performed.  We  use the following abbreviations  to denote the level of context awareness facilitated and employed by the project: high level (H) and low level (L).

\subsubsection{\textbf{Security and Privacy}}
This is a  major concern  in context-aware  computing in  all  paradigms.  However, the IoT paradigm will intensify  the challenges in security and privacy. In the IoT, sensors are expected to collect more information about users (i.e. people) in all aspects. This includes both physical and conceptual data, such as  location, preferences, calendar data, and medical information  to name  a few. As a result, utmost care needs to be taken when collecting,  modelling,  reasoning, and with persistent storage. Security and privacy need to be handled at different levels in the IoT. At the lowest level, the hardware layer should ensure security and privacy  during collecting and temporary storage within the device. Secure protocols  need to ensure communication is well protected. Once the data is received, application level protection  needs to be in placed to monitor and control who can  see or use context   and so on. Different  projects use different  techniques such as policies, rules, and profiles to provide security and privacy. The symbol ($\checkmark$) denoted the presence of security and privacy related functionality in the project, in some form.

\subsubsection{\textbf{Data Source Support}} There are different  sources that are capable of providing context.  Broadly we call them \textit{sensors}. We discussed different  types of sensors in Section  \ref{chapter2:CAF}. Based on the popularity of the data sources supported by each solution, we selected the following classification. (P) denotes that the solution supports only physical sensors. Software sensors (S) denotes that the solution supports either virtual sensors, logical sensors or both. (A) denotes that the solution supports all kinds of data sources (i.e. physical, virtual, and logical). (M) denotes that the solution supports mobile devices.

\subsubsection{\textbf{Quality of Context}}
\label{chapter2:CDLC:ESRF:QCRV}
We denote the presence of conflict resolution functionality  using (C) and context validation  functionality using (V). Conflict resolution is critical in the context management  domain  \cite{P310}. There has  to be a  consistency   in collecting , aggregating,  modelling, and reasoning.  In the IoT paradigm, context may not be accurate.  There are two reasons for context information not to be certain. First is that the sensor technology  is not capable of producing 100\% accurate  sensor data due to various technical and environmental challenges. Secondly,  even with sensors that produce 100\% accurate sensor data, reasoning models are not 100\% accurate. In summary, problems in sensor technology  and problems  in reasoning techniques  contribute  to context conflicts. There are two types of context conflicts that can occurred and they are defined in  \cite{P310}:

 \begin{noindlist2}
  \item Internal   context   conflict:   Fusing two or more context elements  that characterises  the situation from different dimensions of the same observed  entity in a given moment may lead to internal   context   conflict. (e.g. motion sensor detects that a user is in the kitchen  and calendar shows that the user is supposed to be in a meeting.  Therefore, it is unable to correctly deduce  the current location by fusing two data sources: calendar and motion sensor.)
 
  \item External   context   conflicts:   The   context   conflict/inconsistency that  may  occur  between two  or more bits of context that describe  the situation of an observed entity from the same point of view. (e.g. two motion  sensors located in the same area provide  two completely different readings, where one sensor  detects  a person and other sensor detects three people.)
 \end{noindlist2}
 
Context validation ensures  that collected data is correct and meaningful.  Possible validations are checks for range, limit, logic, data type, cross-system consistency, uniqueness, cardinality, consistency, data source quality, security, and privacy.
 
\begin{table*}[htbp]
\centering
\footnotesize
\renewcommand{\arraystretch}{1.15}
\vspace{-0.3cm}
\caption{Summarized taxonomy used in Table \ref{Tbl:Evaluation_of_Previous_Research_Efforts}}
\begin{tabular}{ c l m{11cm} }
\hline
 & Taxonomy & Description \\ \hline \hline
5 & Modelling & Key-value modelling (K), Markup schemes (M), Graphical modelling (G), Object oriented modelling (Ob), Logic-based modelling (L), and Ontology-based modelling (On) \\ 
6 & Reasoning & Supervised learning (S), Un-supervised learning (U), rules (R), Fuzzy logic (F), Ontology-based (O), and Probabilistic reasoning (P) \\ 
7 & Distribution & Publish/subscribe (P) and Query (Q) \\ 
8 & Architecture & Component based architecture (1) , Distributed architecture  (2), Service based architecture (3), Node based architecture (4) ,  Centralised architecture (5), Client-server architecture (6)  \\ 
9 & History and Storage & Available ($\checkmark$) \\ 
10 & Knowledge Management & Available ($\checkmark$) \\ 
11 & Event Detection & Available ($\checkmark$) \\ 
12 & Context Discovery and Annotation & context Discovery (D) and context Annotation (A) \\ 
13 & Level of Context Awareness & High level (H) and Low level (L). \\ 
14 & Security and Privacy & Available ($\checkmark$) \\ 
15 & Data Source Support & Physical sensors (P), Software sensors (S), Mobile devices (M), Any type of sensor (A) \\ 
16 & Quality of Context & Conflict resolution (C), context Validation (V) \\ 
17 & Data Processing & Aggregate (A), Filter (F) \\ 
18 & Dynamic Composition & Available ($\checkmark$) \\ 
19 & Real Time Processing & Available ($\checkmark$) \\ 
20 & Registry Maintenance & Available ($\checkmark$) \\ \hline
\end{tabular}
\label{Tbl:Summarized taxonmy}
\vspace{-0.6cm}
\end{table*}

\subsubsection{\textbf{Data Processing}} 
We denote the presence of context aggregation functionality  using (A) and context filter functionality using (F). Aggregation can be explained in different ways; for example, Context Toolkit  \cite{P143} has a dedicated component called context aggregator  to collect data related  to a specific entity (e.g. person)  from different context sources  and act as  a  proxy to context applications.  They do not perform any complex operations; just collect similar information together. This is one of the simplest forms of aggregation of context.

Context filter functionality makes sure the reasoning engine processes only important  data. Specially in IoT, processing all  the data collected by all  the sensors  is not possible due to scale. Therefore,  IoT solutions should process  only selected  amounts of data that allows it to understand  context accurately. Filtering functionality  can be presented in different solutions in different forms: filter data, filter context sources, or filter events. Filtering  helps both at the low (hardware) level and software level. At the hardware level, it helps to reduce the network communication cost by transmitting only important data. At the high-level, filtering can save process energy by only processing important data.

Context  processing can be classified into three categories (also called layers) \cite{P185}. Typical methods and techniques used in each layer  are also presented as follows:

\begin{noindlist2}

 \item \textit{Activity and context recognition layer}: Feature extraction, classification, clustering, fuzzy rules
 
 \item \textit{Context and representation  layer}: Conceptual  models, logic programming,  ontology based representation  and reasoning,  databases and query languages,  rule based representation and reasoning, cased based representation and reasoning, representing uncertainty,  procedural programming
 
 \item \textit{Application and adaptation layer}: Rules, query languages, procedural programming
\end{noindlist2}

Data fusion, which is also considered a data processing technique,  is critical in understanding  sensor data. In order to lay a  solid foundation  to our discussion,  we adopt the definition provided by Hall and Llinas  \cite{P248} on sensor data fusion. \textit{``Sensor data fusion is a method of combining sensor data from multiple sensors to produce more accurate, more complete, and more dependable information that could not be possible to achieve through a single sensor \cite{P248}.''} For example, in positioning,  GPS does not work indoors. In contrast, there are a variety of other indoor positioning  schemes that can be used. Therefore, in order to continuously track the positioning regardless of indoor or outdoor,  sensor data fusion is essential  \cite{P115}. Data fusion methods, models, and classification techniques in the wireless sensor networks domain are comprehensively  surveyed in  \cite{P130}.

In order to identify context, it is possible to combine data
from different  data sources. For example, consider  a situation where we want to identify the location of a user. The possible sources  that can be used  to collect evidence regarding  the location are GPS sensors,  motion sensor,  calendar,  email, social networking services, chat clients, ambient sound (sound level, pattern), users nearby, camera sensors, etc. This long list shows the possible alternatives. It is always a tradeoff between required  resource (e.g. processing power, response time) and accuracy. Processing and combining  all the above sensor readings would produce a more accurate result; however,  it would require more resources and time. There is a significant  gap between low-level  sensor readings
and high-level  `situation-awareness' \cite{P287}. Collecting low-level sensor data is becoming significantly  easier and cheaper than ever due to advances in sensing technology.  As a result, enormous  amounts of  sensor  data  (e.g. big data \cite{ZMP003})  is available. In order to understand big data, a variety of different reasoning  techniques  need to employed   as we discussed  in Section  \ref{chapter2:CAF:Context Reasoning Decision Models}.

\subsubsection{\textbf{Dynamic Composition}} As explained in Solar \cite{P569}, IoT solutions must have a programming  model that allows dynamic composition without requiring the developer or user to identify specific sensors and devices. Dynamic  organising  is critical in environments like the IoT, because it is impossible to identify or plan possible interaction  at the development stage. Software solutions should be able to understand the requirements and demands  on each situation, then organise  and structure  its internal components according to them.  Components  such as reasoning  models,  data fusion operators,  knowledge   bases, and context discovery  components can be dynamically  composed according to the needs. The symbol ($\checkmark$) denoted the presence of dynamic composition functionality in the project in some form.
\label{chapter2:CDLC:ESRE:Dynamic_Composition}

\begin{table*} [t!]
\centering
\scriptsize
\renewcommand{\arraystretch}{1.15}
\caption{Evaluation of Surveyed Research Prototypes, Systems, and Approaches}
\label{Tbl:Evaluation_of_Previous_Research_Efforts}

\begin{tabular}{
m{1.9cm} 
c 
m{0.25cm} 
c
c
c
c
c
c
c
c
c
c
c
c
c
c
c
c
c } \hline

Project Name     & 

\begin{sideways}Citations \end{sideways}   & 
Year & 
\begin{sideways}Project Focus \end{sideways} &
\begin{sideways}Modelling \end{sideways} &
\begin{sideways}Reasoning  \end{sideways} &
\begin{sideways}Distribution \end{sideways} & 

\begin{sideways}Architecture \end{sideways} &
\begin{sideways}History and Storage \end{sideways} & 
\begin{sideways}\begin{minipage}[b]{2cm}Knowledge Management\end{minipage} \end{sideways} & 
\begin{sideways}Event Detection \end{sideways} &

\begin{sideways} \begin{minipage}[b]{2cm}Context Discovery and Annotation \end{minipage} \end{sideways} & 
\begin{sideways} \begin{minipage}[b]{2cm}Level of Context Awareness \end{minipage} \end{sideways} & 

\begin{sideways}Security and Privacy\end{sideways} &
\begin{sideways}Data Source Support \end{sideways} &
\begin{sideways}\begin{minipage}[b]{2cm}Quality of Context \end{minipage} \end{sideways} &

\begin{sideways}Data Processing \end{sideways} & 
\begin{sideways}Dynamic Composition \end{sideways} &
\begin{sideways}Real Time Processing\end{sideways} &
\begin{sideways}Registry Maintenance\end{sideways} 
\\  \hline \hline

 (1)                                 & (2)          & (3)  & (4)             &(5)        & (6)& (7)& (8)& (9)      &  (10)   &  (11)   & (12)   & (13)   &  (14)  & (15)   & (16)    & (17)    & (18)   &  (19)   & (20)
 \\ \\

\comment{01}Context Toolkit           &  \cite{P143}   & 2001 &  T    & K           & $\checkmark$      & Q & 1,5 & $\checkmark$      &  \n    &  \n    &  \n    & H &  \n    & A &  \n    & A &  \n    &  \n    &   \n   \comment{01}\\ \\   
\comment{02}Solar                     &  \cite{P569}   & 2002 & M & K,M,Ob           & R & P & 2 &  \n    &  \n    & $\checkmark$      & D & H & $\checkmark$      & P & $\checkmark$      & A & $\checkmark$      &  \n    &   \n   \comment{02}\\ 
\comment{03}Aura                      &  \cite{P555}   & 2002 &  M                & M           & R & P & 2 &  \n    &  \n    & $\checkmark$      & D & H &  \n    & A &  \n   &  \n   &  \n   &  \n    &  $\checkmark$    \comment{03}\\ \\ 
\comment{04}CoOL                      &  \cite{P190}   & 2003 & T & On           & R,O & Q & 1 &  \n    & $\checkmark$      & $\checkmark$      & D & H &  \n    & S &  \n    &  \n    &  \n   &  \n    &  $\checkmark$   \comment{04}\\ 
\comment{05}CARISMA                   &  \cite{P386}   & 2003 &  M                & M           & R & Q & 2 &  \n    &  \n    &  \n    &  \n    & H &  \n    & M & C &  \n    &  \n   &  \n    &  \n    \comment{05}\\ \\ 
\comment{06}CoBrA                     &  \cite{P419}   & 2004 &  M                & On           & R,O & Q & 1 & $\checkmark$      & $\checkmark$      & $\checkmark$      &  \n    & H & $\checkmark$      & A &  \n    &  \n    &  \n   &  \n    &  \n    \comment{06}\\ 
\comment{07}Gaia                      &  \cite{P444}   & 2004 &  M                & F,On           & S,P, F & Q & 2,3 & $\checkmark$      & $\checkmark$      & $\checkmark$      & D & H & $\checkmark$      & A &  \n    &  \n    & $\checkmark$      &  \n    & $\checkmark$       \comment{07}\\ 
\comment{08}SOCAM                     &  \cite{P570}   & 2004 &  M                & On           & R,O & Q,P & 3 & $\checkmark$      & $\checkmark$      & $\checkmark$      & D & H &  \n    & A &  \n    & A &  \n   &  \n    &  $\checkmark$      \comment{08}\\ \\ 
\comment{09}CARS                      &  \cite{P311}   & 2005 &  S                & K           & U &  \n    &  \n    &  \n    &  \n    & $\checkmark$      & A & H &  \n    & P &  \n    &  \n    &  \n   &  \n    &  \n    \comment{09}\\ 
\comment{10}CASN                      &  \cite{P288}   & 2005 &  M                & F,On           & F,O & P & 2 &  \n    & $\checkmark$      &  \n    & D & L &  \n    & P &  \n    &  \n    &  \n   &  \n    &   \n   \comment{10}\\ 
\comment{11}SCK                       &  \cite{P332}   & 2005 &  M                & M,On         & R,O & Q & 1 & $\checkmark$      & $\checkmark$      & $\checkmark$      & A,D & H &  \n    & A & V & \n  &  \n   &  \n    &  $\checkmark$     \comment{11}\\ 
\comment{12}TRAILBLAZER               &  \cite{P305}   & 2005 &  S    & K           & R & Q & 2 &  \n    &  \n    &  \n    & D & L &  \n    & P &  \n    &  \n    &  \n   &  \n    &  \n   \comment{12}\\ \\ 
\comment{13}BIONETS                   &  \cite{P316}   & 2006 &  M    & On           & R,O & Q & 1 &  \n    & $\checkmark$      &  \n    & A & H &  \n    & A &  \n    & \n  &  \n   &  \n    &  \n    \comment{13}\\ 
\comment{14}PROCON                    &  \cite{P278}   & 2006 &  S                & K           & R & Q & 2 &  \n    &  \n    & $\checkmark$      & D & L &  \n    & P &  \n    & A,F &  \n   &  \n    &  \n    \comment{14}\\ 
\comment{15}CMF (MAGNET)              &  \cite{P344}   & 2006 &  M                & M           & R & P,Q & 2,4 & $\checkmark$      &  \n    &  \n    & D & H &  \n    & A & C &  \n    & $\checkmark$      &  \n    &  \n    \comment{15}\\ 
\comment{16}e-SENSE                   &  \cite{P266}   & 2006 &  M                &  \n    & R & Q & {2,4} &  \n    & $\checkmark$      &  \n    & D & H & $\checkmark$      & P &  \n    & F &  \n   &  \n    &  \n    \comment{16}\\ \\ 
\comment{17}HCoM                      &  \cite{P336}   & 2007 &  M                & G,On         & R,O & Q & 5 & $\checkmark$      & $\checkmark$      &  \n    & D & H &  \n    & S & V & F &  \n   &  \n    &  $\checkmark$      \comment{17}\\ 
\comment{18}CMS                       &  \cite{P340}   & 2007 &  M                & On           & O & P,Q & {1,2} & $\checkmark$      &  \n    & $\checkmark$      & S & H &  \n    & A &  \n    & A &  \n   &  \n    &  $\checkmark$     \comment{18}\\ 
\comment{19}MoCA                      &  \cite{P338}   & 2007 &  M                & M,Ob     & O & P,Q & {4,5} &  \n    &  \n    & $\checkmark$      & D & H & $\checkmark$      & A & V & \n  &  \n   & $\checkmark$      &  $\checkmark$     \comment{19}\\ 
\comment{20}CaSP                      &  \cite{P317}   & 2007 &  M                & M,On         & O & P,Q & 6 & $\checkmark$      &  \n    &  \n    & D & H &  \n    & A &  \n    &  \n    &  \n   &  \n    &   $\checkmark$       \comment{20}\\ 
\comment{21}SIM                       &  \cite{P349}   & 2007 &  M                & K,G         & R &  \n    & 2 & $\checkmark$      &  \n    &  \n    &  \n    & H &  \n    & P & C & A &  \n   &  \n    &  \n    \comment{21}\\ 
\comment{22}---                       &  \cite{P335}   & 2007 &  M                & On           & O & Q & {    } &  \n    &  \n    & $\checkmark$      & D & H &  \n    & P & V & A &  \n   &  \n    &  \n    \comment{22}\\ \\ 
\comment{23}COSMOS                    &  \cite{P403}   & 2008 &  M    & Ob          & R & Q & {2,4} &  \n    &  \n    & $\checkmark$      &  \n    & H &  \n    & P &  \n    & A & $\checkmark$      &  \n    & $\checkmark$     \comment{23}\\ 
\comment{24}DMS-CA                    &  \cite{P308}   & 2008 &  S                & M           & R & Q & 5 &  \n    &  \n    & $\checkmark$      &  \n    & H &  \n    & A &  \n    &  \n    &  \n   &  \n    &  \n    \comment{24}\\ 
\comment{25}CDMS                      &  \cite{P293}   & 2008 &  M    & K,M           & R & Q & 2 & $\checkmark$      &  \n    & $\checkmark$      & D & H &  \n    & A &  \n    & A,F &  \n   &  \n    &  $\checkmark$     \comment{25}\\ 
\comment{26}---                       &  \cite{P197}   & 2008 &  M                & On           & O,P & Q & 5 &  \n    & $\checkmark$      &  \n    & D & H &  \n    &  \n    & V &  \n    &  \n   &  \n    &  \n    \comment{26}\\ 
\comment{27}---                       &  \cite{P333}   & 2008 &  M                & On           & R,O & P,Q & 5 &  \n    &  \n    & $\checkmark$      & D & H &  \n    & P &  \n    & A &  \n   &  \n    &  \n    \comment{27}\\ 
\comment{28}AcoMS                     &  \cite{P339}   & 2008 &  M                & M,G,On    & R,O & P & 5 &  \n    & $\checkmark$      & $\checkmark$      & A & H &  \n    & P &  \n    & \n  &  \n   &  \n    & $\checkmark$       \comment{28}\\ 
\comment{29}CROCO                     &  \cite{P334}   & 2008 &  M                & On           & R,O & Q & {    } & $\checkmark$      & $\checkmark$      &  \n    & D & H & $\checkmark$      & A & C,V &  \n    &  \n   &  \n    & $\checkmark$     \comment{29}\\ 
\comment{30}EmoCASN                   &  \cite{P274}   & 2008 &  S    & K & R & Q & {2,4} &  \n    &  \n    &  \n    & D & L &  \n    & P &  \n    &  \n    &  \n   &  \n    &  \n    \comment{30}\\ \\ 
\comment{31}Hydra                     &  \cite{P105}   & 2009 &  M                & K,On,Ob      & R,O & Q & 3 & $\checkmark$      & $\checkmark$      & $\checkmark$      &  \n    & H & $\checkmark$      & P & V &  \n    &  \n   &  \n    &  \n    \comment{31}\\ 
\comment{32}UPnP                      &  \cite{P300}   & 2009 &  M    & K,M         & R & Q & 4 & $\checkmark$      &  \n    & $\checkmark$      & D & H & $\checkmark$      & A &  \n    & A &  $\checkmark$    &   \n   &  $\checkmark$     \comment{32}\\ 
\comment{33}COSAR                     &  \cite{P187}   & 2009 &  M                & On           & S,O & Q & 5 &  \n    & $\checkmark$      & $\checkmark$      & A & H &  \n    & P &  \n    &  \n   &  \n   &  \n    &  \n    \comment{33}\\ 
\comment{34}SPBCA                     &  \cite{P420}   & 2009 &  M                & On           & R,O & Q & 2 &  \n    &  \n    & $\checkmark$      & A & H & $\checkmark$      & A &  \n    &  \n    &  \n   &  \n    &  \n    \comment{34}\\ 
\comment{35}C-CAST                    &  \cite{P280}   & 2009 &  M    & M           & R & P,Q & 5 & $\checkmark$      &  \n    & $\checkmark$      & D & H &  \n    & A &  \n    &  \n    &  \n   &  \n    & $\checkmark$     \comment{35}\\ 
\comment{36}---                       &  \cite{P312}   & 2009 &  M                & On           & O & P & 5 & $\checkmark$      &  \n    & $\checkmark$      & D & H &  \n    & A &  \n    & A &  \n   &  \n    &  \n    \comment{36}\\ 
\comment{37}CDA                       &  \cite{P341}   & 2009 &  M                & Ob          &  \n    & Q & {4,6} &  \n    &  \n    &  \n    &  \n    & H &  \n    & V &  \n    &  \n    &  \n   &  \n    &  $\checkmark$     \comment{37}\\ 
\comment{38}SALES                     &  \cite{P314}   & 2009 &  M                & M           & R & Q & {2,4} &  \n    &  \n    & $\checkmark$      & D & L &  \n    & P &  \n    & F &  \n   &  \n    &  $\checkmark$     \comment{38}\\ 
\comment{39}MidSen                    &  \cite{P275}   & 2009 &  M    & K           & R & P,Q & 5 &  \n    & $\checkmark$      & $\checkmark$      & D & H &  \n    & P &  \n    &  \n    &  \n   &  \n    & $\checkmark$    \comment{39}\\ \\ 
\comment{40}SCONSTREAM                &  \cite{P309}   & 2010 &  S                & G           & R & Q & 5 & $\checkmark$      &  \n    & $\checkmark$      &  \n    & H &  \n    & P &  \n    &  \n    &  \n   &  $\checkmark$     &  \n    \comment{40}\\ 
\comment{41}---                       &  \cite{P328}   & 2010 &  M                & M           & P & Q & {2,4} & $\checkmark$      &  \n    & $\checkmark$      &  \n    & H &  \n    & A &  \n    & F &  $\checkmark$    &  \n    &  \n    \comment{41}\\ 
\comment{42}Feel@Home                 &  \cite{P346}   & 2010 &  M                & G,On         & O & P,Q & {2,4} &  \n    & $\checkmark$      & $\checkmark$      &  \n    & H & $\checkmark$      & A &  \n    &  \n    &  \n   &  \n    &  $\checkmark$    \comment{42}\\ 
\comment{43}CoMiHoC                   &  \cite{P347}   & 2010 &  M                & Ob          & R,P & Q & 5 &  \n    & $\checkmark$      & $\checkmark$      & D & H &  \n    & A & V &  \n   &  \n   &  \n    &  \n    \comment{43}\\ 
\comment{44}Intelligibility    &  \cite{P384}   & 2010 &  T    &  \n    & R,S,P & Q & {1,5} &  \n    &  \n    & $\checkmark$      & D & H &  \n    & A & V &  \n   &  \n   &  \n    &  \n    \comment{44}\\ 
\comment{45}ezContext                 &  \cite{P294}   & 2010 &  M    & K,Ob       & R & Q & 5 & $\checkmark$      & $\checkmark$      & $\checkmark$      &  \n    & H &  \n    & A &  \n    & A &  \n   &  \n    &  $\checkmark$     \comment{45}\\ 
\comment{46}UbiQuSE                   &  \cite{P322}   & 2010 &  M    & M           & R & Q & 5 & $\checkmark$      &  \n    & $\checkmark$      & D,A & H &  \n    & A &  \n    &  \n    &  \n   &  $\checkmark$     &  \n    \comment{46}\\  
\comment{47}COPAL                      &  \cite{P571}   & 2010 & M & M & R & P,Q & {1,5} &  \n    &  \n    & $\checkmark$      & D & H & $\checkmark$      &  & V & A,F &  \n   &  $\checkmark$     &    $\checkmark$   \comment{47}\\ \\
\comment{48}Octopus                   &  \cite{P285}   & 2011 &  S                & $\checkmark$      & $\checkmark$      & P & {2,4} &  \n    &  \n    & $\checkmark$      & D & H &  \n    & A &  \n    & A &  $\checkmark$    &  \n    &    \n   \comment{48}\\ 
\comment{49}---                       &  \cite{P327}   & 2011 &  M                &  \n    & $\checkmark$      & P & 2 &  \n    &  \n    &  \n    & D & H &  \n    & P &  \n    & A &  \n   &  \n    &  $\checkmark$      \comment{49}\\  
\comment{50}---                       &  \cite{P289}   & 2011 &  S    & K,Ob        & S,P &      & {2,4} & $\checkmark$      & $\checkmark$      & $\checkmark$      & D,A & H &  \n    & M & V & A,F &  \n   &  \n    &  $\checkmark$  \comment{50}\\

\hline 

\multicolumn{20}{p{1\textwidth}}{\textit{Notes}: Refer Section \ref{chapter2:CDLC:Evaluation of Surveyed Research Efforts} for the meanings of the abbreviations and symbols used in the table}

\end{tabular}
\end{table*}

\subsubsection{\textbf{Real Time Processing}}
Most of the interactions are expected to be processed in real time in the IoT. This functionality has  been rarely addressed  by the research  community in the context-aware computing  domain.  The most important real time processing task is event detection as we explained  in Section  \ref{chapter2:CDLC:ESRF:Event_Detection}. However, context reasoning, and query processing can also be considered   as essential  real time processing  tasks. Real time processing  solutions are focused on processing faster than traditional methods, which allows sensor stream data processing  \cite{P309}.  The symbol ($\checkmark$) denoted the presence of real time processing functionality  in some form.

\subsubsection{\textbf{Registry Maintenance and Lookup Services}} We use the ($\checkmark$) symbol to denote the presence of registry maintenance and lookup services functionality  in the project. This functionality allows different components such as  context sources,  data fusion operators,  knowledge bases, and context  consumers to be registered. This functionality is also closely related to dynamic composition where it  needs  to select relevant and matching  components to be composed together. Registries need to be updated to reflect (dis)appearing components.


\subsection{Evaluation of Research Efforts}
\label{chapter2:CDLC:Evaluation of Surveyed Research Efforts2 }


\textbf{Context Toolkit }\cite{P143} aims to facilitating development and deployment of context-aware applications. This is one of the earliest efforts of providing framework support for context-aware application  development. Context Toolkit contains a combination  of features and abstractions to support context-aware application developers. It introduces three main abstractions: context widget (to retrieve data from sensors), context interpreter (to reason sensor data using different  reasoning techniques), and context aggregator. The research around Context Toolkit is still active and a number of extensions have been developed to enhance its context-aware capabilities. Enactor \cite{P393} provides a context decision modelling facility to the Context Toolkit. Further, the Intelligibility Toolkit  \cite{P384} extends the Enactor framework  by supporting more decision models for context reasoning. Context Toolkit identifies the common features  required by context-aware  applications   as capture and access of context, storage, distribution,  and independent execution from applications.


\textbf{Aura} \cite{P555} is a task oriented  system based on distributed architecture which focuses on different computational devices  used by human  users every  day. The objective is to run a set  of applications  called \textit{personal  aura} in all devices in order to manage  user tasks in a  context-aware   fashion across  all the devices smoothly.  Aura addresses  two major challenges. First, aura allows a user to preserve continuity  in his/her work when moving between different environments. Second, it is capable of adapting to the on-going computation of a particular  environment  in the presence of dynamic resource variability. Aura consists  of four major components: context observer  (collects context and send it to task and environment  managers), task manager (also called prism, four different kinds of changes: user moves to another environment, environment, task, and context), environment manager (handles context suppliers and related service),  and context suppliers (provides context information).  XML based markup schemes are used to describe services.
.


\textbf{CARISMA} \cite{P386} (Context-Aware Reflective middleware System for Mobile Applications) is focused on mobile systems where they are extremely dynamic. Adaptation  (also called reflection) is the main focus of  CARISMA. context is stored as application  profiles  (XML based), which allows each application  to maintain  meta-data under two categories: passive  and active. The passive category  defines  actions  that middleware  would take  when specific events  occur using rules, such as  shutting down if  battery is low. However, conflicts could arise when two profiles defines rules that conflict  each other. The active category allows relationships to be maintained between services used by the application, the policies, and context configurations.   This information tells how to behave under different environmental  and user conditions.  A conflict resolution mechanism is also introduced in CARISMA based on macroeconomic  techniques. An auction protocol is used to handle the resolution  as they support greater degrees of heterogeneity over other alternatives. In simple terms, rules are used in auctions with different constraints imposed on the bidding by different agents (also called applications).  Final decisions are made in order to maximise the social welfare among the agents.


\textbf{CoBrA} \cite{P419} (Context Broker Architecture) is a broker-centric agent architecture that provides knowledge sharing and context reasoning  for smart spaces.  It  is specially focused on smart meeting places. CoBrA addresses two major issues: supporting resource-limited mobile computing devices and addressing concerns over user privacy.  Context information is modelled using OWL ontologies. Context brokers are the main elements of CoBrA.  A context broker comprises the following four functional  components: context knowledge base (provides  persistent storage for context information),  context reasoning engine (performs reasoning over context information stored in storage), context acquisition module (retrieve context from context  sources), and policy management module (manages policies,  such as who has access to what data). Even though the architecture is centralised, several brokers can work together  through a broker federation.  Context knowledge  is represented in Resource Description Framework (RDF) triples using Jena.


\textbf{Gaia} \cite{P444} is a distributed  context  infrastructure  uncertainty based reasoning.  Ontologies  are used to represented context information.  Gaia has employed a Prolog based probabilistic reasoning framework.  The architecture  of Gaia consists of  six key  components:  context provider (data acquisition  from sensors or other data sources), context consumer (different  parties who are interest in context), context synthesiser  (generate  high-level context information using raw low-level context), context  provider lookup service (maintains  a detailed registry  of context providers   so  the appropriate  context providers  can be found based  on their capabilities  when required),  context history service (stores history of context), and ontology server (maintains different ontologies).


\textbf{SOCAM} \cite{P570} (Service Oriented Context-Aware Middleware) is an ontology based context-aware  middleware.  It separates  the ontologies  into two levels: upper level ontology for general concepts and lower level ontologies domain specific descriptions. SOCAM architecture comprises several key components: context provider (acquires data from sensors and other internal and external  data sources and converts the context in to OWL representation),  context interpreter (performs reasoning using reasoning engine and stores the processed  context information in the knowledge  base), context-aware  services (context  consumers), and services locating service (context providers and interpreter are allowed to register  so other components can search for appropriates providers  and interpreters based on their capabilities).


\textbf{e-SENSE} \cite{P266} enables ambient intelligence  using wireless  multi-sensor  networks  for making context-rich  information available  to applications  and services. e-SENSE combines body sensor networks (BSN),  object sensor networks (OSN), and environment  sensor networks  (ESN) to capture context in the IoT paradigm. The features required  by context-aware IoT middleware solutions are identified as sensor data capturing, data pre-filtering, context abstraction data source integration, context extraction, rule engine, and adaptation.


\textbf{HCoM} \cite{P336} (Hybrid Context Management) is  a  hybrid approach  which combines  semantic ontology and relational schemas. This approach claims that standard  database management  systems  alone cannot  be used to manage  context. In  contrast,  semantic  ontologies may not perform well in terms of efficiency and query processing with large volumes of data. So the hybrid approach is required. HCoM architecture  consists of five layers: acquisition layer, pre-processing layer, data modelling and storage layer, management modelling  layer, and utilising layer. HCoM has identified  several key requirements that a context management solution  should have that are encapsulated in several components: context  manager (aggregates the results and sends the data to reasoning engine), collaboration manager (if  context selector  decides  the existing  context information is not sufficient to perform reasoning, the collaboration manager attempts to gather more data from other possible context sources),  context filter (once the context is received, it  validates  and decide whether it needs to be stored  in RCDB),  context selector (based on the user request, it decides what context   should be used in reasoning  processing  based on the accuracy, time, and required computational  resources), context-onto  (manages the ontologies  and acts as a repository), rules and policy (users  are allowed to add rules to the system), RCDB (stores the captured context in a standard database management system), rule-mining (a data  base that consists of rules that tell what actions to perform when), and interfaces (provides interface to the context consumers).


\textbf{MoCA} \cite{P338} is a service  based distributed  middleware that employs ontologies to model and manage context. The primary conceptual component is context domain. The context  management node (CMN) is infrastructure that is responsible  for  managing the  context domain. Similar to most of the other context management solutions, the three key  components  in  MoCA are: context providers (responsible for generating or retrieving  context   from other sources available  to be used by the context management  system), context consumers  (consume  the context gathered  and processed  by the system),  and context service  (responsible for receiving, storing, and disseminating context information). MoCA uses  an object oriented model for context handling, instead of an ontology-based model due to the weaknesses posed by ontologies in terms of scalability  and performance. XML is used to model context. The XML files are fed into the context tool in order to check validation. Then the program  codes are generated automatically  to acquire data. These program codes will acquire context and insert the data into context repositories.


\textbf{CaSP} \cite{P317} (Context-aware Service Platform) is a  context gathering framework for mobile solutions based on middleware architecture. The platform provides six different functionalities: context sensing,  context modelling, context association, context storage, and retrieval. The paper also provides  a comprehensive evaluation of existing context sensing solutions.  CaSP consists of typical context   management components which handle the mentioned  functionalities.


\textbf{SIM} \cite{P349} (Sensor Information Management) is focused on the smart home domain  which  addresses location tracking. SIM uses  an agent based architecture  according  to the standard specifications provided in Foundation for Intelligent Physical Agents. Its emphasis is on collecting  sensor data from multiple sources and aggregating them together to analyse and derive more accurate information.  SIM  collects two types of information: node level and attribute level. In node level, node ID, location, and priority are collected. Attributes are stored  in attribute information  base comprising  attribute and the corresponding measurement. A location  tracking  algorithm  has been introduced using a  mobile positioning  device. A position  manager handles tracking. SIM has the capability  to resolve conflicts in sensor information based on sensor priority. Conflict resolution is handled by a context manager with the help of aggregation, classification, and decision components. Even though SIM is not focused on hardware level context management, the approach is closer to low-level instead of high-level  compared to other projects.


\textbf{COSMOS} \cite{P403} is middleware  that enables the processing of context information in ubiquitous environments. COSMOS consists of three layers: context collector (collects information from the sensors), context processing (derives high level information from raw sensor data), and context adaptation (provides access to the processed context   for the applications). In contrast  to the other context solutions,  the components  of COSMOS  are context nodes. In COSMOS,  each piece of context information is defined  as a context  node. COSMOS  can support any number of context nodes which are organised into hierarchies. Context node is an independently  operated module that consists of its own activity manager, context processor, context reasoner, context configurator,  and message managers. Therefore, COSMOS  follows distributed  architecture which increases the scalability of the middleware.


\textbf{DMS-CA} \cite{P308} (Data Management System-Context Architecture) is based on smart building domain. XML is used to define rules, contexts, and services. Further, an event driven rule checking technique is used to reason context.  Rules can be configured by mobile  devices and push them to the server to be used by the rule checking engine. Providing a mobile interface to build rules and queries is important in a dynamic and mobile environment such as the IoT.


\textbf{ACoMS} \cite{P339} (Autonomic Context Management System) can dynamically  configure and reconfigure its context information acquisition and pre-processing functionality to perform fault tolerant provisioning of context information. ACoMS architecture comprises application context subscription manager stores (manages context information  requests from the applications using a subscribe mechanism),  context source  manager (performs  actions such as low-level communication with context sources,  context source discovery,  registration, and configuration),  and reconfiguration  manager (performs monitoring  tasks such as mapping context  sources to context information).


\textbf{CROCO} \cite{P334} (CROss application COntext management) is an ontology based  context modelling and management service. CROCO  identifies  several requirements to be a cross application,  such as application  plug-in  capability.  CROCO has three responsibilities  where they are distributed  among three separate layers: data management (perform  operations such as storing inferred  data for historic  use, develop and maintain  fact database), consistency  checking  and reasoning  (consistency manager is responsible for checking the consistency,  such as data types, and cardinality  when sensor data arrives before it is feed in to reasoning or storage; reasoning manager performs reasoning based on the facts stored in the fact data base), and context update and provision  (allows context consumers to register themselves, retrieve context from context sources, and provide query interface to the consumers).


\textbf{EMoCASN} \cite{P274} (Environment Monitoring Oriented Context Aware Sensor Networks) proposes a context-aware  model for sensor networks (CASN).  This modelling  approach is narrowly focused on managing sensor networks using low level context such as node context,  task context,  and data context.  For example, CASN  uses low level context   such as remaining energy of a node, location  of the sensor, and orientation  of the sensor to decide  energy efficient routing.


\textbf{Hydra}\footnote{The name `Hydra' has changes its name due to name conflict between another project registered under same name in Germany. The new name of the middleware is the `LinkSmart' middleware.} \cite{P105} is an IoT middleware  that aims to integrate  wireless  devices and sensors  into ambient intelligence  systems. Hydra comprises   a Context Aware Framework (CAF). CAF provides the capabilities  of both high-level, powerful  reasoning, based on the use of ontologies and lower-level  semantic processing based on object-oriented/key-value approach. CAF consists of two main components: Data Acquisition Component  (DAqC) and the Context Manager (CM). DAqC is responsible for connecting and retrieving data from sensors.  CM is responsible  for context management, context  awareness, and context interpretation.  A rule engine called Drools platform \cite{P395} has  been  employed   as the core context reasoning  mechanism.  CAF models three distinct types of context: device contexts (e.g. data source), semantic contexts (e.g. location, environment, and entity), and application contexts (e.g. domain specific). Hydra identifies context reasoning rule engine, context storage, context querying, and event/action   management   as the key  components of a context-aware framework.


\textbf{C-Cast} \cite{P280} is middleware   that integrates WSN into context-aware  systems by addressing context acquisition, dissemination,  representation, recognising,  and reasoning about context and situations.  C-Cast lays its architecture  on four layers: sensor,  context detection, context acquisition, and application. In C-Cast, context  providers  (CP) are the main components.  Each context provider handles  one task. For example, WeatherCP collects weather information  and Address-bookCP collects  related addresses. Any amount of CPs can be added to the system to extend the system wide functionality. Each context provider independently handles data acquisition, context processing  (e.g. filter and aggregate  context), context provider management (e.g. handles subscriptions), and context access  and dissemination (e.g. handles queries).  C-Cast claims that complex reasoning and intuitive  reasoning can only be achieved by using rich representation models. In contrast, C-CAST avoids using ontologies to model context claiming  ontologies are too resource intensive.


\textbf{SALES} \cite{P314} (Scalable context-Aware middleware for mobiLe EnviromentS) is a context-aware  middleware  that achieves scalability in context dissemination. The main components of  this middleware  are nodes.  These  nodes  are not sensor nodes but servers, computers, laptops, PDAs, and mobile phones. SALES consists of four types of nodes. XML schemes are used to store and transfer context.


\textbf{MidSen} \cite{P275} is context-aware  middleware  for WSN. The system  is based  on Event-Condition-Action  (ECA) rules. It highlights the importance of efficient  event detection by processing two algorithms: event detection algorithm  (EDA) and context-aware  service discovery algorithm (CASDA). MidSen has proposed  a complete  architecture  to enable context awareness in WSN. It consists of the following key components: knowledge  manager, application notifiers, knowledge   base, inference engine, working memory, application interface, and network interface.


\textbf{Feel@Home} \cite{P346} is a context  management framework  that supports interaction  between different domains. The proposed approach  is  demonstrated  using three domains: smart home, smart office, and mobile.  The context information is stored  using OWL  \cite{P148}. Feel@Home supports two different interactions: intra-domain and cross domain. The cross domain interaction is essential in the IoT paradigm. Further, this is one of the major differences between sensor networks and the IoT. Sensor networks usually  only deal with one domain. However, IoT demands the capability of dealing with multiple domains. In addition, context management frameworks should not be limited to a  specific number  of domains.  Feel@Home  consists  of three parts: user queries, global administration  server (GAS), and domain  context  manager (DCM). User queries are first received  by GAS. It decides  what the relevant domain needs  to be contacted  to answer  the user query. Then, GAS redirects the user query to the relevant domain context managers.  Two components  reside  in GAS, context entry manager  (CEM) and context entry engine (CEE), which performs the above task. DCM consists of typical context management components  such as  context wrapper (gathers context from sensors and other sources), context  aggregator (triggers context reasoning), context reasoning, knowledge  base (stores context), and several other components to manage user queries, publish/subscribe mechanism. The answers to the user query will return by using the same path as when received.


\textbf{CoMiHoc} \cite{P347} (Context Middleware for ad-HoC network) is a  middleware   framework that supports  context management and situation  reasoning. CoMiHoc proposes  a CoMoS (Context Mobile Spaces), a context  modeling, and situation reasoning  mechanism  that extends  the context spaces   \cite{P195}. CoMiHoc uses Java Dempster-Shafer   library  \cite{P441}. CoMiHoc architecture  comprises six components:  context provisioner, request manager, situation reasoner, location  reasoner, communication manager, and On-Demand Multicast Routing Protocol (ODMRP).


\textbf{ezContext} \cite{P294} is a framework   that provides automatic context life  cycle management.  ezContext  comprises several  components:  context source (any source that provides  context,  either physical sensors,  databases or web service), context provider (retrieves context from various sources whether in push (passive) or pull (active) method, context manager (manages context modelling,  storage and producing high-level context using low-level  context), context wrapper (encapsulate retrieved context into correct format, in this approach, key-value pairs), and providers' registry (maintains list of context providers and their capabilities). JavaBeans are used as the main data format.


\textbf{Octopus} \cite{P285} is an open-source,  dynamically extensible system that supports data management and fusion for IoT applications. Octopus develops middleware abstractions and programming  models for the IoT. It  enables  non-specialised   developers  to deploy sensors  and applications without detailed knowledge  of the underlying  technologies and network.  Octopus is focused on the smart home/office domain and its main component is \textit{solver}. Solver is a module that performs sensor data fusion operations. Solvers can be added and removed from the system at any time based  on requirements. Further solvers  can be combined together dynamically  to build complex  operations.

\section{Lessons Learned}
\label{chapter2:LL}

\setcounter{subsubsection}{0}

\subsubsection{Development Aids and Practices}


\textcolor{orange}{ Toolkits in general are suitable for limited scale application.  Managing  context in the IoT paradigm requires middleware solutions that can provide more functionality towards managing data. Applications  should be able to be built on top of the middleware  so they can request context from the middleware.  Context Toolkit \cite{P143} has introduced   the notion of having common standard interfaces. For example, context widget component  encapsulate the communication  between context sources and the toolkit. Standardisation makes it easier to learn, use, and extend the toolkit. Standardisation is important in the IoT paradigm,  because it increases interoperability and extendibility. For example, standardising context modelling components will help to employ the different techniques we discussed  in Section \ref{chapter2:CAF:Context Modelling}  despite  the differences  in inner-workings. It also enables the addition of different  components when necessary.  In such a situation, standard  interfaces  and structures will  guarantee a smooth  interaction   between  new and old components. Further, Intelligibility Toolkit \cite{P384} provides explanations  to the users to improve the trust between users and the context-aware applications which helps in faster adaptation of the users towards IoT. }

\textcolor{orange}{Making correct design decisions is a critical task in IoT. For example, data modelling and communication can be done using different techniques as follows where each method has its own advantages and disadvantages \cite{P569}. 1) Binary is smaller in size than the other three formats and also portable due to its small size. In contrast, binary makes it difficult to extend and modify later. 2) Objects method allows complex data structures. 3) Attribute-value  pairs method provides more limited complexity than an object representation. In contrast, simpler representation allows language- and platform- independent applications. 4) XML  method provides more opportunities for complex data structures.  XML adds a substantial  overhead in term of network communications and processing.}


\textcolor{orange}{CoOL \cite{P190} shows how extensions (e.g. context modelling and reasoning)  can be developed to support general purpose service models.  CoOL allows context management functionality to be added to any model using context management access point, which is responsible for handle communication between CoOL and the rest of the general purpose architecture. Security and privacy  issues in context-aware computing are not researched and seriously  considered in many solutions. CoBrA \cite{P419} shows how an ontology based approach  can be used to manage user privacy  via policies which allow it to monitor  and access contextual  control context. As ontologies are getting popular and  adopted in web related developments, such practice will makes IoT development much easier.}

\textcolor{orange}{Octopus \cite{P285} highlights the significance of designing programming  models that enable non-technical  people to deploy sensors. As we mentioned earlier, the majority of the sensor deployments  are expected  to be carried out by non-technical users. Kim and Choi \cite{P333} models context meta-data from an operational perspective as discussed  in Section  \ref{chapter2:CAF:context Types}, which allows it to understand operational  parameters such as complexities,   quality, up-to-dateness, and cost of acquisition.}

\subsubsection{Mobility, Validity, and Sharing}

\textcolor{orange}{Monitoring continuity, which is also called mobility, is an important  task in the IoT. People move from one situation to another and IoT solutions need to track user movements and facilitate context-aware functionalities over different forms of devices.  Aura \cite{P555} shows the requirement of having IoT middleware running over many platforms and devices under different resource  limitations (i.e. from cloud server, computers, tablets, mobile phones to everyday objects) where different versions  (with different capabilities) would fit on different devices.}


\textcolor{orange}{CARISMA \cite{P386} shows how conflict resolution can be done using profiles and rules where it stress the importance of  making  decisions  to optimize the return for  every  party involved. In the IoT, there will  be many data sources  that will provide similar information that can be used to derive  the same knowledge where conflict resolution will help to make accurate actions. MoCA \cite{P338} also emphasizes validation  of context which has an impact on the accuracy of the reasoning. Further, it shows how context can be modelled in formats such as XML and then inserted into any programming  language via binding  techniques (e.g. data binding  in Java). In CROCO \cite{P334}, validation  (e.g. consistency), conflict resolution, and privacy concerns are given attention where they are rarely  addressed by many other solutions. Sharing context information allows mobility and smooth transition from device to device or situation to situation. Park et al. \cite{P289} highlight the importance of context sharing using mobile devices, which allows more comprehensive and accurate reasoning and high level context recognition.}

\subsubsection{On Demand Data Modelling}


\textcolor{orange}{Due to unpredictability and broadness of IoT, data models need to be extensible on demand. For example, IoT solutions may need to be expand its knowledge-base towards different domains. SOCAM \cite{P570} shows how knowledge  can be separated among different levels of  ontologies: upper ontology and domain specific ontology. In SOCAM,  upper ontology models general purpose data while domain specific ontologies model domain specific data, which is allowed to extend to both levels independently. As an IoT  solution will  be used  in  many different domains,  the ability to add ontologies (i.e. knowledge)  when necessary is critical for wider adaptation. SCK \cite{P332}, Zhan et al. \cite{P312}, and BIONETS \cite{P316} use  different ontologies for each context category.  As we discussed  in  Section  \ref{chapter2:CAF:context Types}, there are many different types of context categories which model context   in different perspectives. Therefore, in the IoT it is important to store different types of context   as they can help in a variety of situations. They also stresses the requirement of having domain specific and domain independent ontologies.}

\subsubsection{Hybrid Reasoning}


\textcolor{orange}{Gaia \cite{P444}, Ko and Sim \cite{P197}, CDMS \cite{P293}, and HCoM \cite{P336} highlight the importance of employing multiple reasoning  techniques  such as Bayesian networks, probabilistic and fuzzy logic, where each technique  performs  well in different situations. Incorporation of multiple modelling and reasoning techniques can mitigate individual weaknesses using each other's strengths. COSAR \cite{P187} combines statistical reasoning and ontological  reasoning techniques to achieve more accurate results. }

\subsubsection{Hardware Layer Support}


\textcolor{orange}{EMoCASN \cite{P274}, TRAILBLAZER \cite{P305} and  CASN \cite{P288}  shows the  importance of  embedding context-aware capabilities in low (hardware) layer communication. Context  awareness allows sensors to act more intelligently and save energy. In the IoT a majority of the communications are expected to happened between machines.  In such situations, context  awareness becomes critical for each individual object to optimize their actions. Further, in order to build a fully context-aware solution, we have to embed context-aware capabilities in both software and hardware layers. In an environment such as the IoT where billions of objects communicate with each other, significant amounts of energy can be saved by following fairly simple optimisation techniques as presented in PROCON \cite{P278}. SALES \cite{P314} shows how context   can be managed using distributed architecture with a variety of different devices with different  resource constraints in the hardware level.}

\subsubsection{Dynamic Configuration and Extensions}

\textcolor{orange}{Hydra \cite{P105} is one of  the early efforts at building IoT middleware which focuses on connecting  embedded devices to applications.  It  shows how the context modelling needs to be done in  order to model device information. Hydra also highlights  the importance of pluggable rules that allow insertions  when necessary  as  it  is a  major requirement  in IoT  middleware  applications,  where domains and required knowledge cannot be predicted during the development stage. A complementary technology has  proposed by ACoMS \cite{P339}. It  has proposed  a technique  that allows it to automatically connect sensors to an IoT solution using Transducer Electronic Data Sheet (TEDS) \cite{P258} and Sensor Markup Languages (SensorML) \cite{P256}.  UPnP FRAMEWORK \cite{P300} is strongly related to a vision of the IoT where machine-to-machine communication  play a significant role. This approach is applicable to devices such as cameras, web cams, and microwaves;  but, not for low end temperature or humidity sensors. UPnP approach is a key technology that enables automated configuration.}

\textcolor{orange}{Solar \cite{P569}, CMF (MAGNET) \cite{P344}, and COSMOS \cite{P403} promote the notion of dynamic composition which is critical in IoT solutions where possible interactions cannot be identified at the design and development stage.  ezContext \cite{P294} shares a common  notion of context providers similar to C-Cast \cite{P280}  uses them to decouple context sources from the system. Different  types of context providers, which are dedicated to communicating and retrieving data related to a specific domain, can be employed when necessary. In line with above solutions, COPAL \cite{P571} demonstrates the essential features IoT middleware should have,  such as loosely  coupled plug-in architecture  and automated  code generation  via abstracts  which stimulates extendibility and usability.}

\subsubsection{Distributed Processing}

\textcolor{orange}{This is a one of the most commons tasks need to be performed by IoT solutions.
UbiQuSE \cite{P322}  shows how real time query processing can be done incorporating live  streaming  data and historic context in repositories. Similarly, SCONSTREAM \cite{P309}  highlights the challenges in real-time context stream processing where real time processing is a significant component to be successful in the IoT. Most event detections need to be performed in real time. Further, Feel@Home \cite{P346} shows  how cross  domain context can be queried in order to answer complex user requirements. As we mentioned earlier, there is not a central point of management in the IoT paradigm. Therefore communicating, sharing, and querying  context   managed in a distributed fashion by different  managers is essential.}

\subsubsection{Other Aspects}


\textcolor{orange}{CARS \cite{P311} introduces   a  technique   that can be used to evaluate, test, and improve IoT solutions in social and user point of view. As we mentioned earlier, success of IoT depends on the user adaptation. CARS evaluates the process  of deriving high level information using low level sensor  data where users  will  appreciate  the work done by the software systems.}

\textcolor{orange}{Cloud computing offers significant  amounts of processing and storage capabilities.  With the three services models, Infrastructure-as-a-Service  (IaaS), Platform-as-a-Service (PaaS),  and Software-as-a-Service   (SaaS), context management can largely benefit from cloud computing in many ways in the IoT paradigm. In the IoT, sensors will be attached to almost every object around us. Further,  these sensors will be deployed by ordinary users, governments,  or business organisations.  Cloud computing allows all parties to share sensor data  based on a  financial model.  Sensor owners will advertise their sensors in the cloud. The consumers who want to access those  sensors  will  pay the owners and acquire the sensor readings. Therefore,  the cloud model perfectly matches with the IoT paradigm.  In addition, cloud resources can be used to reason and store large volumes of context where significant amounts of processing power and storage are required. The cloud brings added scalability  to context management in the IoT. Further, interoperability among different IoT solution can be achieved by following approaches such as CDA \cite{P341}. Data matching is the process to identify and matching records in diverse database that refer to the same real-world entities in situations where no entity identifiers are available, and therefore the available attributes have to be used to conduct the matching \cite{P595}. Context information plays a critical in data matching where sensors can be considered as entities in the sensing as a service model \cite{ZMP003}.}

\section{Challenges and Future Research Directions}
\label{chapter2:LLFRD}

As we mentioned earlier, one of our goal in this survey is to understand how context-aware computing can be applied in the IoT paradigm based on past experience. Specifically, we evaluated fifty context-aware projects and highlighted the lessons we can learn from them in the IoT perspective. In this section our objective is to discuss six unique  challenges in the IoT where novel techniques and solution  may  need to be employed.

\setcounter{subsubsection}{0}

\subsubsection{Automated configuration of sensors}
\label{chapter2:LLFRD:Automated_configuration_of_sensors}
In traditional pervasive/ubiquitous computing, we connect only a limited number of sensors to the applications (e.g. smart farm, smart home). In contrast, the IoT envisions billions of sensors to be connected together over the Internet. As a result,  a unique  challenge  would arise on connection and configuration  of sensors to applications. Due to the scale, it is not feasible to connect sensors manually  to an application or to a middleware \cite{ZMP005}.  There has to be an automated or at least semi-automated  process to connect sensors to applications.  In order to accomplish  this task, applications should be able to understand  the sensors  (e.g. sensors'  capabilities,  data structures they produce, hardware/driver  level configuration  details). Recent developments such as Transducer Electronic  Data Sheet (TEDS)  \cite{P258}, Open Geospatial Consortium (OGC) Sensor Web Enablement related standards such as Sensor Markup Languages (SensorML)   \cite{P256}, sensor ontologies \cite{P103}, and immature but promising  efforts such as Sensor Device Definitions \cite{ZMP002} show future directions to carry out the research work further, in order to tackle this challenge.

\subsubsection{Context discovery}

Once we connect sensors to a software  solution,  as mentioned  above, there has to be a method  to understand the sensor data produced by the sensors  and the related  context automatically.  We discussed context categorisation techniques comprehensively in Section  \ref{chapter2:CAF:context Types}. There are many types of context that can be used to enrich sensor data. However, understanding sensor data and appropriately annotating it automatically in a paradigm  such as the IoT, where application domains vary widely, is a challenging  task. Recent developments in semantic technologies \cite{P191, P103, P088}  and linked data \cite{P520, P068} show future directions to carry out further research  work. Semantic technology is popularly  used to encode domain knowledge.

\subsubsection{Acquisition, modelling, reasoning, and distribution}

After analysing acquisition, modelling, and reasoning in different perspectives, it is evident that no single technique would serve the requirements  of the IoT. Incorporating and integrating  multiple techniques  has shown  promising  success in the field. Some of the early work such as \cite{P216, P463} have  discussed  the process in detail. However, due to the immaturity of the field of IoT, it is difficult  to predict when and where to employ each technique. Therefore,  it is important to define and follow a standard  specification   so different techniques  can be added to the solutions without significant effort. Several  design principles  have been proposed by \cite{P143, P384} as a step towards standardisation of components and techniques. The inner-workings of each technique can be different from one solution to another. However, common standard interfaces will insure the interoperability among techniques.

\subsubsection{Selection of sensors in sensing-as-a-service model}

This is going to be one of the toughest challenges in the IoT. It is clear that we are going to have access to billions of sensors. In such an environment,  there could be many different  alternative sensors to be used. For example, let us consider a situation where an environmental  scientist  wants to measure  environmental pollution in New York city. There are two main problems: (1) `what sensors provide information  about pollution?' \cite{ZMP004} (2) when there are multiple sensors that can measure the same parameter (e.g. pH concentration in a lake), `what sensor should be used?' \cite{ZMP006} In order to answer question (1), domain  knowledge  needs to be incorporate with the IoT solution. Manually selecting the sensors that will provide information  about environmental pollution is not feasible in the IoT due to its scale. In order to answer question (2), quality frameworks  need to be defined and employed.  Such a  framework should be able to rank the sensors based on factors such as accuracy,  relevancy, user feedback,  reliability, cost, and completeness.  Similar challenges have been addressed in the web service domain during the last decade  \cite{P563, P564} where we can learn from those efforts.

\subsubsection{Security, privacy, and trust}
This has been a challenge for context-aware computing  since the beginning. The advantage of context is that it provides more meaningful information that will help us understand a situation  or data. At the same time, it increases the security threats due to possible misuse of the context (e.g. identity, location, activity,  and behaviour). However, the IoT will  increase this challenge significantly.  Even though security  and privacy  issues are addressed at the context-aware application  level, it is largely unattended at the context-aware middleware  level. In the IoT, security and privacy  need to be protected in several  layers: sensor hardware  layer, sensor data communication  (protocol)  layer, context annotation and context discovery layer, context modelling layer, and the context distribution layer. IoT is a community  based approach where  the acceptance of the users (e.g. general public) is essential. Therefore, security and privacy  protection  requirements need to be carefully addressed in order to win the trust of the users.

\subsubsection{Context Sharing}
\label{chapter2:LLFRD:Context_Data_Sharing}
This is largely neglected in the context-aware middleware  domain.  Most of the middleware solutions or architectures are designed to facilitate applications in isolated factions. Inter-middleware communication is not considered  to be a critical requirement. However, in the IoT, there would be no central point of control. Different middleware solutions developed by different parties will be employed to connect to sensors, collect, model, and reason  context. Therefore,  sharing context information between  different kinds of middleware solutions or different  instances of the same middleware  solution is important. Sensor data stream processing middleware solutions such as  GSN  \cite{P050} have  employed   this capability to share sensor  data among  different instances (e.g. installed and configured in different computers and locations)  where context is not the  focus. However,  in contrast to sensor  data,  context information has  strong relationships  between each other (e.g. context modelled using RDF). Therefore,  relationship  models also need to be transferred and shared among different  solutions, which enables the receiver to understand and model the context accurately at the receivers end.

\section{Conclusions}
\label{chapter2:Conclusions}
The IoT has  gained significant attention over the last few years.  With the advances  in sensor  hardware technology and cheap materials,  sensors are expected  to be attached to all the objects around us, so these can communicate with each other with minimum human intervention.  Understanding  sensor data is one of the main challenges that the IoT would face. This vision has been supported and heavily invested by governments,  interest groups, companies,  and research institutes. For example, context  awareness has been identified as an important  IoT research need by the Cluster of European Research Projects on the IoT (CERP-IoT)  \cite{P019} funded by the European Union. The EU has allocated a time frame for research and development into context-aware computing focused on the IoT to be carried out during 2015-2020.

In this survey paper, we analysed and evaluated context-aware computing research efforts to understand how the challenges in the field of context-aware computing  have been tackled in desktop, web, mobile, sensor networks,  and pervasive computing  paradigms. A large number  of solutions exist  in  terms of  systems, middleware,  applications,  techniques, and models proposed by researchers to solve different challenges in context-aware  computing.  We also  discussed some of the trends in the field that were identified during the survey. The results clearly show the importance of context awareness in the IoT paradigm. Our ultimate goal is to build a foundation that helps us to understand what has happened in the past so we can plan for the future more efficiently and effectively.


%



\section*{Acknowledgment}


Authors acknowledge support from SSN TCP, CSIRO, Australia and ICT OpenIoT Project, which is co-funded by the European Commission under seventh framework program, contract number FP7-ICT-2011-7-287305-OpenIoT. The Author(s) acknowledge help and contributions from The Australian National University.

\vspace{-3pt}
\ifCLASSOPTIONcaptionsoff
  \newpage
\fi




\bibliography{IEEEabrv,Bibliography}

\begin{thebibliography}{100}
\providecommand{\url}[1]{#1}
\csname url@samestyle\endcsname
\providecommand{\newblock}{\relax}
\providecommand{\bibinfo}[2]{#2}
\providecommand{\BIBentrySTDinterwordspacing}{\spaceskip=0pt\relax}
\providecommand{\BIBentryALTinterwordstretchfactor}{4}
\providecommand{\BIBentryALTinterwordspacing}{\spaceskip=\fontdimen2\font plus
\BIBentryALTinterwordstretchfactor\fontdimen3\font minus
  \fontdimen4\font\relax}
\providecommand{\BIBforeignlanguage}[2]{{%
\expandafter\ifx\csname l@#1\endcsname\relax
\typeout{** WARNING: IEEEtran.bst: No hyphenation pattern has been}%
\typeout{** loaded for the language `#1'. Using the pattern for}%
\typeout{** the default language instead.}%
\else
\language=\csname l@#1\endcsname
\fi
#2}}
\providecommand{\BIBdecl}{\relax}
\BIBdecl

\bibitem{P506}
\BIBentryALTinterwordspacing
M.~Weiser, ``The computer for the 21st century,'' \emph{Scientific American},
  vol. 265, no.~3, pp. 66--75, July 1991. [Online]. Available:
  \url{http://doi.acm.org/10.1145/329124.329126}
\BIBentrySTDinterwordspacing

\bibitem{P173}
\BIBentryALTinterwordspacing
B.~Schilit and M.~Theimer, ``Disseminating active map information to mobile
  hosts,'' \emph{Network, IEEE}, vol.~8, no.~5, pp. 22 --32, sep/oct 1994.
  [Online]. Available: \url{http://dx.doi.org/10.1109/65.313011}
\BIBentrySTDinterwordspacing

\bibitem{P104}
\BIBentryALTinterwordspacing
G.~D. Abowd, A.~K. Dey, P.~J. Brown, N.~Davies, M.~Smith, and P.~Steggles,
  ``Towards a better understanding of context and context-awareness,'' in
  \emph{Proceedings of the 1st international symposium on Handheld and
  Ubiquitous Computing}, ser. HUC '99.\hskip 1em plus 0.5em minus 0.4em\relax
  London, UK: Springer-Verlag, 1999, pp. 304--307. [Online]. Available:
  \url{http://dl.acm.org/citation.cfm?id=647985.743843}
\BIBentrySTDinterwordspacing

\bibitem{P029}
H.~Sundmaeker, P.~Guillemin, P.~Friess, and S.~Woelffle, ``Vision and
  challenges for realising the internet of things,'' European Commission
  Information Society and Media, Tech. Rep., March 2010,
  \url{http://www.internet-of-things-research.eu/pdf/IoT_Clusterbook_March_2010.pdf}
  [Accessed on: 2011-10-10].

\bibitem{ZMP003}
A.~Zaslavsky, C.~Perera, and D.~Georgakopoulos, ``Sensing as a service and big
  data,'' in \emph{International Conference on Advances in Cloud Computing
  (ACC-2012)}, Bangalore, India, July 2012, pp. 21--29.

\bibitem{P431}
G.~Chen and D.~Kotz, ``A survey of context-aware mobile computing research,''
  Department of Computer Science, Dartmouth College, Hanover, NH, USA, Tech.
  Rep., 2000, \url{http://www.cs.dartmouth.edu/reports/TR2000-381.pdf}
  [Accessed on: 2011-12-05].

\bibitem{P184}
\BIBentryALTinterwordspacing
T.~Strang and C.~Linnhoff-Popien, ``A context modeling survey,'' in \emph{In:
  Workshop on Advanced Context Modelling, Reasoning and Management, UbiComp
  2004 - The Sixth International Conference on Ubiquitous Computing,
  Nottingham/England}, 2004. [Online]. Available:
  \url{http://elib.dlr.de/7444/1/Ubicomp2004ContextWSCameraReadyVersion.pdf}
\BIBentrySTDinterwordspacing

\bibitem{P417}
\BIBentryALTinterwordspacing
M.~M. Molla and S.~I. Ahamed, ``A survey of middleware for sensor network and
  challenges,'' in \emph{Proceedings of the 2006 International Conference
  Workshops on Parallel Processing}, ser. ICPPW '06.\hskip 1em plus 0.5em minus
  0.4em\relax Washington, DC, USA: IEEE Computer Society, 2006, pp. 223--228.
  [Online]. Available: \url{http://dx.doi.org/10.1109/ICPPW.2006.18}
\BIBentrySTDinterwordspacing

\bibitem{P035}
\BIBentryALTinterwordspacing
K.~E. Kjaer, ``A survey of context-aware middleware,'' in \emph{Proceedings of
  the 25th conference on IASTED International Multi-Conference: Software
  Engineering}.\hskip 1em plus 0.5em minus 0.4em\relax ACTA Press, 2007, pp.
  148--155. [Online]. Available:
  \url{http://dl.acm.org/citation.cfm?id=1332044.1332069}
\BIBentrySTDinterwordspacing

\bibitem{P402}
\BIBentryALTinterwordspacing
M.~Baldauf, S.~Dustdar, and F.~Rosenberg, ``A survey on context aware
  systems,'' \emph{Int. J. Ad Hoc Ubiquitous Comput.}, vol.~2, no.~4, pp.
  263--277, Jun. 2007. [Online]. Available:
  \url{http://dx.doi.org/10.1504/IJAHUC.2007.014070}
\BIBentrySTDinterwordspacing

\bibitem{P185}
\BIBentryALTinterwordspacing
M.~Perttunen, J.~Riekki, and O.~Lassila, ``Context representation and reasoning
  in pervasive computing: a review,'' \emph{International Journal of Multimedia
  and Ubiquitous Engineering}, vol.~4, no.~4, pp. 1--28, 2009. [Online].
  Available: \url{http://www.sersc.org/journals/IJMUE/vol4_no4_2009/1.pdf}
\BIBentrySTDinterwordspacing

\bibitem{P216}
\BIBentryALTinterwordspacing
C.~Bettini, O.~Brdiczka, K.~Henricksen, J.~Indulska, D.~Nicklas,
  A.~Ranganathan, and D.~Riboni, ``A survey of context modelling and reasoning
  techniques,'' \emph{Pervasive Mob. Comput.}, vol.~6, pp. 161--180, April
  2010. [Online]. Available: \url{http://dx.doi.org/10.1016/j.pmcj.2009.06.002}
\BIBentrySTDinterwordspacing

\bibitem{P359}
\BIBentryALTinterwordspacing
A.~Saeed and T.~Waheed, ``An extensive survey of context-aware middleware
  architectures,'' in \emph{Electro/Information Technology (EIT), 2010 IEEE
  International Conference on}, may 2010, pp. 1 --6. [Online]. Available:
  \url{http://dx.doi.org/10.1109/EIT.2010.5612118}
\BIBentrySTDinterwordspacing

\bibitem{P118}
\BIBentryALTinterwordspacing
S.~Bandyopadhyay, M.~Sengupta, S.~Maiti, and S.~Dutta, ``Role of middleware for
  internet of things,'' \emph{International Journal of Computer Science and
  Engineering Survey}, vol.~2, pp. 94--105, 2011. [Online]. Available:
  \url{http://airccse.org/journal/ijcses/papers/0811cses07.pdf}
\BIBentrySTDinterwordspacing

\bibitem{P594}
\BIBentryALTinterwordspacing
P.~Makris, D.~Skoutas, and C.~Skianis, ``A survey on context-aware mobile and
  wireless networking: On networking and computing environments' integration,''
  \emph{Communications Surveys Tutorials, IEEE}, vol.~PP, no.~99, pp. 1 --25,
  2012. [Online]. Available:
  \url{http://dx.doi.org/10.1109/SURV.2012.040912.00180}
\BIBentrySTDinterwordspacing

\bibitem{P291}
\BIBentryALTinterwordspacing
P.~Bellavista, A.~Corradi, M.~Fanelli, and L.~Foschini, ``A survey of context
  data distribution for mobile ubiquitous systems,'' \emph{ACM Computing
  Surveys}, vol.~xx, no.~xx, p.~49, 2013. [Online]. Available:
  \url{http://www-lia.deis.unibo.it/Staff/LucaFoschini/pdfDocs/context_survey_CSUR.pdf}
\BIBentrySTDinterwordspacing

\bibitem{P260}
\BIBentryALTinterwordspacing
N.~Olifer and V.~Olifer, \emph{Computer Networks: Principles, Technologies and
  Protocols for Network Design}.\hskip 1em plus 0.5em minus 0.4em\relax John
  Wiley \& Sons, 2005. [Online]. Available:
  \url{http://au.wiley.com/WileyCDA/WileyTitle/productCd-EHEP000983.html}
\BIBentrySTDinterwordspacing

\bibitem{P575}
D.~Guinard, ``Towards the web of things: Web mashups for embedded devices,'' in
  \emph{In MEM 2009 in Proceedings of WWW 2009. ACM}, 2009.

\bibitem{P018}
{Casaleggio Associati}, ``The evolution of internet of things,'' {Casaleggio
  Associati}, Tech. Rep., February 2011,
  \url{http://www.casaleggio.it/pubblicazioni/Focus_internet_of_things_v1.81%20-%20eng.pdf}
  [Accessed on: 2011-06-08].

\bibitem{P006}
{European Commission}, ``Internet of things in 2020 road map for the future,''
  Working Group RFID of the ETP EPOSS, Tech. Rep., May 2008,
  \url{http://ec.europa.eu/information_society/policy/rfid/documents/iotprague2009.pdf}
  [Accessed on: 2011-06-12].

\bibitem{P019}
P.~Guillemin and P.~Friess, ``Internet of things strategic research roadmap,''
  The Cluster of European Research Projects, Tech. Rep., September 2009,
  \url{http://www.internet-of-things-research.eu/pdf/IoT_Cluster_Strategic_Research_Agenda_2009.pdf}
  [Accessed on: 2011-08-15].

\bibitem{P007}
{Carnot Institutes}, ``Smart networked objects and internet of things,'' Carnot
  Institutes' Information Communication Technologies and Micro Nano
  Technologies alliance, White Paper, January 2011,
  \url{http://www.internet-of-things-research.eu/pdf/IoT_Clusterbook_March_2010.pdf}
  [Accessed on:2011-11-28].

\bibitem{P003}
\BIBentryALTinterwordspacing
L.~Atzori, A.~Iera, and G.~Morabito, ``The internet of things: A survey,''
  \emph{Comput. Netw.}, vol.~54, no.~15, pp. 2787--2805, Oct. 2010. [Online].
  Available: \url{http://dx.doi.org/10.1016/j.comnet.2010.05.010}
\BIBentrySTDinterwordspacing

\bibitem{P041}
\BIBentryALTinterwordspacing
G.~Kortuem, F.~Kawsar, D.~Fitton, and V.~Sundramoorthy, ``Smart objects as
  building blocks for the internet of things,'' \emph{Internet Computing,
  IEEE}, vol.~14, no.~1, pp. 44 --51, jan.-feb. 2010. [Online]. Available:
  \url{http://dx.doi.org/10.1109/MIC.2009.143}
\BIBentrySTDinterwordspacing

\bibitem{P026}
\BIBentryALTinterwordspacing
D.~Le-Phuoc, A.~Polleres, M.~Hauswirth, G.~Tummarello, and C.~Morbidoni,
  ``Rapid prototyping of semantic mash-ups through semantic web pipes,'' in
  \emph{Proceedings of the 18th international conference on World wide web},
  ser. WWW 2009.\hskip 1em plus 0.5em minus 0.4em\relax ACM, 2009, pp.
  581--590. [Online]. Available:
  \url{http://dx.doi.org/10.1145/1526709.1526788}
\BIBentrySTDinterwordspacing

\bibitem{P040}
\BIBentryALTinterwordspacing
A.~Dohr, R.~Modre-Opsrian, M.~Drobics, D.~Hayn, and G.~Schreier, ``The internet
  of things for ambient assisted living,'' in \emph{Information Technology: New
  Generations (ITNG), 2010 Seventh International Conference on}, 2010, pp.
  804--809. [Online]. Available: \url{http://dx.doi.org/10.1109/ITNG.2010.104}
\BIBentrySTDinterwordspacing

\bibitem{P065}
K.~Ashton, ``That 'internet of things' thing in the real world, things matter
  more than ideas,'' \emph{RFID Journal}, June 2009,
  \url{http://www.rfidjournal.com/article/print/4986} [Accessed on:
  2012-07-30].

\bibitem{P361}
D.~L. Brock, ``The electronic product code (epc) a naming scheme for physical
  objects,'' Auto-ID Center, White Paper, January 2001,
  \url{http://www.autoidlabs.org/uploads/media/MIT-AUTOID-WH-002.pdf} [Accessed
  on: 2011-08-25].

\bibitem{P020}
{International Telecommunication Union}, ``Itu internet reports 2005: The
  internet of things,'' International Telecommunication Union, Workshop Report,
  November 2005,
  \url{http://www.itu.int/dms_pub/itu-s/opb/pol/S-POL-IR.IT-2005-SUM-PDF-E.pdf}
  [Accessed on: 2011-12-12].

\bibitem{P002}
\BIBentryALTinterwordspacing
T.~Lu and W.~Neng, ``Future internet: The internet of things,'' in \emph{3rd
  International Conference on Advanced Computer Theory and
  Engineering(ICACTE)}, vol.~5, August 2010, pp. V5--376--V5--380. [Online].
  Available: \url{http://dx.doi.org/10.1109/ICACTE.2010.5579543}
\BIBentrySTDinterwordspacing

\bibitem{P017}
\BIBentryALTinterwordspacing
L.~W.~F. Chaves and C.~Decker, ``A survey on organic smart labels for the
  internet-of-things,'' in \emph{Networked Sensing Systems (INSS), 2010 Seventh
  International Conference on}, 2010, pp. 161--164. [Online]. Available:
  \url{http://dx.doi.org/10.1109/INSS.2010.5573467}
\BIBentrySTDinterwordspacing

\bibitem{P005}
\BIBentryALTinterwordspacing
Y.~Chen, J.~Guo, and X.~Hu, ``The research of internet of things' supporting
  technologies which face the logistics industry,'' in \emph{Computational
  Intelligence and Security (CIS), 2010 International Conference on}, 2010, pp.
  659--663. [Online]. Available: \url{http://dx.doi.org/10.1109/CIS.2010.148}
\BIBentrySTDinterwordspacing

\bibitem{P362}
\BIBentryALTinterwordspacing
Y.-W. Wang, H.-L. Yu, and Y.~Li, ``Internet of things technology applied in
  medical information,'' in \emph{Consumer Electronics, Communications and
  Networks (CECNet), 2011 International Conference on}, april 2011, pp. 430
  --433. [Online]. Available:
  \url{http://dx.doi.org/10.1109/CECNET.2011.5768647}
\BIBentrySTDinterwordspacing

\bibitem{P363}
\BIBentryALTinterwordspacing
G.~Chong, L.~Zhihao, and Y.~Yifeng, ``The research and implement of smart home
  system based on internet of things,'' in \emph{Electronics, Communications
  and Control (ICECC), 2011 International Conference on}, sept. 2011, pp. 2944
  --2947. [Online]. Available:
  \url{http://dx.doi.org/10.1109/ICECC.2011.6066672}
\BIBentrySTDinterwordspacing

\bibitem{P244}
\BIBentryALTinterwordspacing
J.~Burrell, T.~Brooke, and R.~Beckwith, ``Vineyard computing: sensor networks
  in agricultural production,'' \emph{Pervasive Computing, IEEE}, vol.~3,
  no.~1, pp. 38 -- 45, jan.-march 2004. [Online]. Available:
  \url{http://dx.doi.org/10.1109/MPRV.2004.1269130}
\BIBentrySTDinterwordspacing

\bibitem{P013}
\BIBentryALTinterwordspacing
L.~Lin, ``Application of the internet of thing in green agricultural products
  supply chain management,'' in \emph{Intelligent Computation Technology and
  Automation (ICICTA), 2011 International Conference on}, vol.~1, 2011, pp.
  1022--1025. [Online]. Available:
  \url{http://dx.doi.org/10.1109/ICICTA.2011.256}
\BIBentrySTDinterwordspacing

\bibitem{P416}
A.~Asin and D.~Gascon, ``50 sensor applications for a smarter world,''
  {Libelium Comunicaciones Distribuidas}, Tech. Rep., 2012,
  \url{http://www.libelium.com/top_50_iot_sensor_applications_ranking/pdf}
  [Accessed on: 2012-05-02].

\bibitem{P255}
{BCC Research}, ``Sensors: Technologies and global markets,'' {BCC Research},
  Market Forecasting, March 2011,
  \url{http://www.bccresearch.com/report/sensors-technologies-markets-ias006d.html}
  [Accessed on: 2012-01-05].

\bibitem{P009}
\BIBentryALTinterwordspacing
I.~Akyildiz, W.~Su, Y.~Sankarasubramaniam, and E.~Cayirci, ``A survey on sensor
  networks,'' \emph{Communications Magazine, IEEE}, vol.~40, no.~8, pp. 102 --
  114, aug 2002. [Online]. Available:
  \url{http://dx.doi.org/10.1109/MCOM.2002.1024422}
\BIBentrySTDinterwordspacing

\bibitem{P193}
\BIBentryALTinterwordspacing
A.~Mainwaring, D.~Culler, J.~Polastre, R.~Szewczyk, and J.~Anderson, ``Wireless
  sensor networks for habitat monitoring,'' in \emph{Proceedings of the 1st ACM
  international workshop on Wireless sensor networks and applications}, ser.
  WSNA '02.\hskip 1em plus 0.5em minus 0.4em\relax New York, NY, USA: ACM,
  2002, pp. 88--97. [Online]. Available:
  \url{http://doi.acm.org/10.1145/570738.570751}
\BIBentrySTDinterwordspacing

\bibitem{P158}
\BIBentryALTinterwordspacing
D.~Malan, T.~Fulford-jones, M.~Welsh, and S.~Moulton, ``Codeblue: An ad hoc
  sensor network infrastructure for emergency medical care,'' in \emph{In
  International Workshop on Wearable and Implantable Body Sensor Networks},
  2004. [Online]. Available:
  \url{http://www.eecs.harvard.edu/~mdw/papers/codeblue-bsn04.pdf}
\BIBentrySTDinterwordspacing

\bibitem{P113}
\BIBentryALTinterwordspacing
S.~Rooney, D.~Bauer, and P.~Scotton, ``Techniques for integrating sensors into
  the enterprise network,'' \emph{Network and Service Management, IEEE
  Transactions on}, vol.~3, no.~1, pp. 43 --52, jan. 2006. [Online]. Available:
  \url{http://dx.doi.org/10.1109/TNSM.2006.4798306}
\BIBentrySTDinterwordspacing

\bibitem{P067}
\BIBentryALTinterwordspacing
A.~R. Da~Rocha, F.~C. Delicato, J.~N. de~Souza, D.~G. Gomes, and L.~Pirmez, ``A
  semantic middleware for autonomic wireless sensor networks,'' in
  \emph{Proceedings of the 2009 Workshop on Middleware for Ubiquitous and
  Pervasive Systems}, ser. WMUPS '09.\hskip 1em plus 0.5em minus 0.4em\relax
  New York, NY, USA: ACM, 2009, pp. 19--25. [Online]. Available:
  \url{http://doi.acm.org/10.1145/1551693.1551697}
\BIBentrySTDinterwordspacing

\bibitem{P266}
\BIBentryALTinterwordspacing
A.~Gluhak and W.~Schott, ``A wsn system architecture to capture context
  information for beyond 3g communication systems,'' in \emph{Intelligent
  Sensors, Sensor Networks and Information, 2007. ISSNIP 2007. 3rd
  International Conference on}, dec. 2007, pp. 49 --54. [Online]. Available:
  \url{http://dx.doi.org/10.1109/ISSNIP.2007.4496818}
\BIBentrySTDinterwordspacing

\bibitem{P086}
\BIBentryALTinterwordspacing
F.~Neto and C.~Ribeiro, ``Dynamic change of services in wireless sensor network
  middleware based on semantic technologies,'' in \emph{Autonomic and
  Autonomous Systems (ICAS), 2010 Sixth International Conference on}, march
  2010, pp. 58 --63. [Online]. Available:
  \url{http://dx.doi.org/10.1109/ICAS.2010.17}
\BIBentrySTDinterwordspacing

\bibitem{P475}
\BIBentryALTinterwordspacing
K.~Al~Nuaimi, M.~Al~Nuaimi, N.~Mohamed, I.~Jawhar, and K.~Shuaib, ``Web-based
  wireless sensor networks: a survey of architectures and applications,'' in
  \emph{Proceedings of the 6th International Conference on Ubiquitous
  Information Management and Communication}, ser. ICUIMC '12.\hskip 1em plus
  0.5em minus 0.4em\relax New York, NY, USA: ACM, 2012, pp. 113:1--113:9.
  [Online]. Available: \url{http://doi.acm.org/10.1145/2184751.2184881}
\BIBentrySTDinterwordspacing

\bibitem{P318}
\BIBentryALTinterwordspacing
S.-H. Sho, K.-S. Kim, W.-r. Jun, J.-S. Kim, S.-H. Kim, and J.-D. Lee, ``Ttcg:
  three-tier context gathering technique for mobile devices,'' in
  \emph{Proceedings of the 5th international conference on Pervasive services},
  ser. ICPS '08.\hskip 1em plus 0.5em minus 0.4em\relax New York, NY, USA: ACM,
  2008, pp. 157--162. [Online]. Available:
  \url{http://doi.acm.org/10.1145/1387269.1387296}
\BIBentrySTDinterwordspacing

\bibitem{P351}
C.~Alcaraz, P.~Najera, J.~Lopez, and R.~Roman, ``Wireless sensor networks and
  the internet of things: Do we need a complete integration?'' in \emph{1st
  International Workshop on the Security of the Internet of Things
  (SecIoT'10)}, 2010.

\bibitem{P277}
\BIBentryALTinterwordspacing
E.~Elnahrawy and B.~Nath, ``Context-aware sensors,'' in \emph{Wireless Sensor
  Networks}, ser. Lecture Notes in Computer Science, H.~Karl, A.~Wolisz, and
  A.~Willig, Eds.\hskip 1em plus 0.5em minus 0.4em\relax Springer Berlin /
  Heidelberg, 2004, vol. 2920, pp. 77--93. [Online]. Available:
  \url{http://dx.doi.org/10.1007/978-3-540-24606-0_6}
\BIBentrySTDinterwordspacing

\bibitem{P285}
\BIBentryALTinterwordspacing
B.~Firner, R.~S. Moore, R.~Howard, R.~P. Martin, and Y.~Zhang, ``Poster: Smart
  buildings, sensor networks, and the internet of things,'' in
  \emph{Proceedings of the 9th ACM Conference on Embedded Networked Sensor
  Systems}, ser. SenSys '11.\hskip 1em plus 0.5em minus 0.4em\relax New York,
  NY, USA: ACM, 2011, pp. 337--338. [Online]. Available:
  \url{http://doi.acm.org/10.1145/2070942.2070978}
\BIBentrySTDinterwordspacing

\bibitem{P038}
\BIBentryALTinterwordspacing
S.~Alam, M.~M.~R. Chowdhury, and J.~Noll, ``Senaas: An event-driven sensor
  virtualization approach for internet of things cloud,'' in \emph{Networked
  Embedded Systems for Enterprise Applications (NESEA), 2010 IEEE International
  Conference on}, November 2010, pp. 1--6. [Online]. Available:
  \url{http://dx.doi.org/10.1109/NESEA.2010.5678060}
\BIBentrySTDinterwordspacing

\bibitem{P275}
\BIBentryALTinterwordspacing
P.~Patel, S.~Jardosh, S.~Chaudhary, and P.~Ranjan, ``Context aware middleware
  architecture for wireless sensor network,'' in \emph{Services Computing,
  2009. SCC '09. IEEE International Conference on}, sept. 2009, pp. 532 --535.
  [Online]. Available: \url{http://dx.doi.org/10.1109/SCC.2009.49}
\BIBentrySTDinterwordspacing

\bibitem{P069}
\BIBentryALTinterwordspacing
D.~Le-Phuoc, J.~X. Parreira, M.~Hausenblas, Y.~Han, and M.~Hauswirth, ``Live
  linked open sensor database,'' in \emph{Proceedings of the 6th International
  Conference on Semantic Systems}, ser. I-SEMANTICS '10.\hskip 1em plus 0.5em
  minus 0.4em\relax New York, NY, USA: ACM, 2010, pp. 46:1--46:4. [Online].
  Available: \url{http://doi.acm.org/10.1145/1839707.1839763}
\BIBentrySTDinterwordspacing

\bibitem{P083}
\BIBentryALTinterwordspacing
M.~Botts, G.~Percivall, C.~Reed, and J.~Davidson, ``Ogc sensor web enablement:
  Overview and high level architecture,'' in \emph{Geosensor Networks Lecture
  Notes In Computer Science}, S.~Nittel, A.~Labrinidis, and A.~Stefanidis,
  Eds.\hskip 1em plus 0.5em minus 0.4em\relax Berlin, Heidelberg:
  Springer-Verlag, 2008, ch. OGC Sensor Web Enablement: Overview and High Level
  Architecture, pp. 175--190. [Online]. Available:
  \url{http://dx.doi.org/10.1007/978-3-540-79996-2_10}
\BIBentrySTDinterwordspacing

\bibitem{P498}
\BIBentryALTinterwordspacing
S.~Patidar, D.~Rane, and P.~Jain, ``A survey paper on cloud computing,'' in
  \emph{Advanced Computing Communication Technologies (ACCT), 2012 Second
  International Conference on}, jan. 2012, pp. 394 --398. [Online]. Available:
  \url{http://dx.doi.org/10.1109/ACCT.2012.15}
\BIBentrySTDinterwordspacing

\bibitem{P502}
\BIBentryALTinterwordspacing
M.~Zhou, R.~Zhang, D.~Zeng, and W.~Qian, ``Services in the cloud computing era:
  A survey,'' in \emph{Universal Communication Symposium (IUCS), 2010 4th
  International}, oct. 2010, pp. 40 --46. [Online]. Available:
  \url{http://dx.doi.org/10.1109/IUCS.2010.5666772}
\BIBentrySTDinterwordspacing

\bibitem{P533}
\BIBentryALTinterwordspacing
P.~Banerjee, R.~Friedrich, C.~Bash, P.~Goldsack, B.~Huberman, J.~Manley,
  C.~Patel, P.~Ranganathan, and A.~Veitch, ``Everything as a service: Powering
  the new information economy,'' \emph{Computer}, vol.~44, no.~3, pp. 36 --43,
  march 2011. [Online]. Available: \url{http://dx.doi.org/10.1109/MC.2011.67}
\BIBentrySTDinterwordspacing

\bibitem{P064}
\BIBentryALTinterwordspacing
V.~Issarny, M.~Caporuscio, and N.~Georgantas, ``A perspective on the future of
  middleware-based software engineering,'' in \emph{2007 Future of Software
  Engineering}, ser. FOSE '07.\hskip 1em plus 0.5em minus 0.4em\relax
  Washington, DC, USA: IEEE Computer Society, 2007, pp. 244--258. [Online].
  Available: \url{http://dx.doi.org/10.1109/FOSE.2007.2}
\BIBentrySTDinterwordspacing

\bibitem{P028}
M.~Nagy, A.~Katasonov, O.~Khriyenko, S.~Nikitin, M.~Szydlowski, and
  V.~Terziyan, ``Challenges of middleware for the internet of things,''
  University of Jyvaskyla, Tech. Rep., 2009,
  \url{http://cdn.intechopen.com/pdfs/8786/InTech-Challenges_of_middleware_for_the_internet_of_things.pdf}
  [Accessed on: 2011-12-20].

\bibitem{P377}
{OpenIoT Consortium}, ``Open source solution for the internet of things into
  the cloud,'' January 2012, \url{http://www.openiot.eu} [Accessed on:
  2012-04-08].

\bibitem{P105}
\BIBentryALTinterwordspacing
A.~Badii, M.~Crouch, and C.~Lallah, ``A context-awareness framework for
  intelligent networked embedded systems,'' in \emph{Advances in Human-Oriented
  and Personalized Mechanisms, Technologies and Services (CENTRIC), 2010 Third
  International Conference on}, aug. 2010, pp. 105 --110. [Online]. Available:
  \url{http://dx.doi.org/10.1109/CENTRIC.2010.29}
\BIBentrySTDinterwordspacing

\bibitem{P375}
M.~Simonov, ``Ismb middleware for iot (rfid),'' Presentation Slides, March
  2010,
  \url{http://ec.europa.eu/information-society/activities/foi/events/fippp/docs/mikhail-simonov.pdf}
  [Accessed on: 2012-04-04].

\bibitem{P366}
\BIBentryALTinterwordspacing
N.~Kefalakis, N.~Leontiadis, J.~Soldatos, and D.~Donsez, ``Middleware building
  blocks for architecting rfid systems,'' in \emph{Mobile Lightweight Wireless
  Systems}, ser. Lecture Notes of the Institute for Computer Sciences, Social
  Informatics and Telecommunications Engineering, F.~Granelli, C.~Skianis,
  P.~Chatzimisios, Y.~Xiao, S.~Redana, O.~Akan, P.~Bellavista, J.~Cao,
  F.~Dressler, D.~Ferrari, M.~Gerla, H.~Kobayashi, S.~Palazzo, S.~Sahni, X.~S.
  Shen, M.~Stan, J.~Xiaohua, A.~Zomaya, and G.~Coulson, Eds.\hskip 1em plus
  0.5em minus 0.4em\relax Springer Berlin Heidelberg, 2009, vol.~13, pp.
  325--336. [Online]. Available:
  \url{http://dx.doi.org/10.1007/978-3-642-03819-8_31}
\BIBentrySTDinterwordspacing

\bibitem{P146}
\BIBentryALTinterwordspacing
A.~Katasonov, O.~Kaykova, O.~Khriyenko, S.~Nikitin, and V.~Y. Terziyan, ``Smart
  semantic middleware for the internet of things.'' in \emph{ICINCO-ICSO'08},
  2008, pp. 169--178. [Online]. Available:
  \url{http://www.mit.jyu.fi/ai/papers/ICINCO-2008.pdf}
\BIBentrySTDinterwordspacing

\bibitem{P367}
\BIBentryALTinterwordspacing
M.~Caporuscio, P.-G. Raverdy, and V.~Issarny, ``ubisoap: A service-oriented
  middleware for ubiquitous networking,'' \emph{Services Computing, IEEE
  Transactions on}, vol.~5, no.~1, pp. 86 --98, jan.-march 2012. [Online].
  Available: \url{http://dx.doi.org/10.1109/TSC.2010.60}
\BIBentrySTDinterwordspacing

\bibitem{P119}
\BIBentryALTinterwordspacing
V.~Terziyan, O.~Kaykova, and D.~Zhovtobryukh, ``Ubiroad: Semantic middleware
  for context-aware smart road environments,'' in \emph{Internet and Web
  Applications and Services (ICIW), 2010 Fifth International Conference on},
  may 2010, pp. 295 --302. [Online]. Available:
  \url{http://dx.doi.org/10.1109/ICIW.2010.50}
\BIBentrySTDinterwordspacing

\bibitem{P050}
A.~Salehi, ``Design and implementation of an efficient data stream processing
  system.'' Ph.D. dissertation, Ecole Polytechnique Federale de Lausanne
  (EPFL), 2010,
  \url{http://biblion.epfl.ch/EPFL/theses/2010/4611/EPFL_TH4611.pdf} [Accessed
  on: 2011-10-05].

\bibitem{P371}
\BIBentryALTinterwordspacing
M.~Albano, A.~Brogi, R.~Popescu, M.~Diaz, and J.~A. Dianes, ``Towards secure
  middleware for embedded peer-to-peer systems: Objectives and requirements,''
  in \emph{RSPSI '07: Workshop on Requirements and Solutions for Pervasive
  Software Infrastructures}, 2007, pp. 1--6. [Online]. Available:
  \url{http://citeseerx.ist.psu.edu/viewdoc/download?doi=10.1.1.90.5982&rep=rep1&type=pdf}
\BIBentrySTDinterwordspacing

\bibitem{P373}
\BIBentryALTinterwordspacing
A.~Cannata, M.~Gerosa, and M.~Taisch, ``Socrades: A framework for developing
  intelligent systems in manufacturing,'' in \emph{Industrial Engineering and
  Engineering Management, 2008. IEEM 2008. IEEE International Conference on},
  dec. 2008, pp. 1904 --1908. [Online]. Available:
  \url{http://dx.doi.org/10.1109/IEEM.2008.4738203}
\BIBentrySTDinterwordspacing

\bibitem{P368}
\BIBentryALTinterwordspacing
H.~Bohn, A.~Bobek, and F.~Golatowski, ``Sirena - service infrastructure for
  real-time embedded networked devices: A service oriented framework for
  different domains,'' in \emph{Networking, International Conference on Systems
  and International Conference on Mobile Communications and Learning
  Technologies, 2006. ICN/ICONS/MCL 2006. International Conference on}, april
  2006, p.~43. [Online]. Available:
  \url{http://dx.doi.org/10.1109/ICNICONSMCL.2006.196}
\BIBentrySTDinterwordspacing

\bibitem{P370}
\BIBentryALTinterwordspacing
D.~Giusto, A.~Iera, G.~Morabito, L.~Atzori, A.~Puliafito, A.~Cucinotta, A.~L.
  Minnolo, and A.~Zaia, ``Making the internet of things a reality: The wherex
  solution,'' in \emph{The Internet of Things}.\hskip 1em plus 0.5em minus
  0.4em\relax Springer New York, 2010, pp. 99--108. [Online]. Available:
  \url{http://dx.doi.org/10.1007/978-1-4419-1674-7_10}
\BIBentrySTDinterwordspacing

\bibitem{P143}
\BIBentryALTinterwordspacing
A.~K. Dey, G.~D. Abowd, and D.~Salber, ``A conceptual framework and a toolkit
  for supporting the rapid prototyping of context-aware applications,''
  \emph{Hum.-Comput. Interact.}, vol.~16, pp. 97--166, December 2001. [Online].
  Available: \url{http://dx.doi.org/10.1207/S15327051HCI16234_02}
\BIBentrySTDinterwordspacing

\bibitem{P175}
\BIBentryALTinterwordspacing
P.~J. Brown, ``The stick-e document: a framework for creating context-aware
  applications,'' \emph{Electronic Publishing}, vol.~8, no. 2 \& 3, pp.
  259--272, 1996. [Online]. Available:
  \url{http://citeseerx.ist.psu.edu/viewdoc/download?doi=10.1.1.8.7472&rep=rep1&type=pdf}
\BIBentrySTDinterwordspacing

\bibitem{P178}
\BIBentryALTinterwordspacing
D.~Franklin and J.~Flachsbart, ``All gadget and no representation makes jack a
  dull environment,'' \emph{Proceedings of the AAAI 1998 Spring Symposium on
  Intelligent Environments}, vol. Technical Report SS-98-02, pp. 155--160,
  1998. [Online]. Available:
  \url{http://infolab.northwestern.edu/media/papers/paper10072.pdf}
\BIBentrySTDinterwordspacing

\bibitem{P181}
\BIBentryALTinterwordspacing
T.~Rodden, K.~Chervest, N.~Davies, and A.~Dix, ``Exploiting context in hci
  design for mobile systems,'' in \emph{in Workshop on Human Computer
  Interaction with Mobile Devices}, 1998. [Online]. Available:
  \url{http://eprints.lancs.ac.uk/11619/}
\BIBentrySTDinterwordspacing

\bibitem{P179}
\BIBentryALTinterwordspacing
R.~Hull, P.~Neaves, and J.~Bedford-Roberts, ``Towards situated computing,'' in
  \emph{Wearable Computers, 1997. Digest of Papers., First International
  Symposium on}, oct 1997, pp. 146 --153. [Online]. Available:
  \url{http://dx.doi.org/10.1109/ISWC.1997.629931}
\BIBentrySTDinterwordspacing

\bibitem{P183}
\BIBentryALTinterwordspacing
A.~Ward, A.~Jones, and A.~Hopper, ``A new location technique for the active
  office,'' \emph{Personal Communications, IEEE}, vol.~4, no.~5, pp. 42 --47,
  oct 1997. [Online]. Available: \url{http://dx.doi.org/10.1109/98.626982}
\BIBentrySTDinterwordspacing

\bibitem{P115}
\BIBentryALTinterwordspacing
G.~D. Abowd and E.~D. Mynatt, ``Charting past, present, and future research in
  ubiquitous computing,'' \emph{ACM Trans. Comput.-Hum. Interact.}, vol.~7, pp.
  29--58, March 2000. [Online]. Available:
  \url{http://doi.acm.org/10.1145/344949.344988}
\BIBentrySTDinterwordspacing

\bibitem{P116}
\BIBentryALTinterwordspacing
B.~Schilit, N.~Adams, and R.~Want, ``Context-aware computing applications,'' in
  \emph{Mobile Computing Systems and Applications, 1994. Proceedings., Workshop
  on}, dec 1994, pp. 85 --90. [Online]. Available:
  \url{http://dx.doi.org/10.1109/MCSA.1994.512740}
\BIBentrySTDinterwordspacing

\bibitem{P180}
\BIBentryALTinterwordspacing
J.~Pascoe, ``Adding generic contextual capabilities to wearable computers,'' in
  \emph{Wearable Computers, 1998. Digest of Papers. Second International
  Symposium on}, oct 1998, pp. 92 --99. [Online]. Available:
  \url{http://dx.doi.org/10.1109/ISWC.1998.729534}
\BIBentrySTDinterwordspacing

\bibitem{P437}
{Dictionary.com LLC}, ``Thesaurus.com,'' 1995, \url{http://thesaurus.com/}
  [accessed on: 2012-05-15].

\bibitem{P438}
{Foldoc.org}, ``Free on-line dictionary of computing,'' 2010,
  \url{http://foldoc.org/context} [Accessed on: 2012-05-21].

\bibitem{P439}
{princeton.edu}, ``Wordnet: Lexical database for the english language,'' 2005,
  \url{http://wordnetweb.princeton.edu/} [Accessed on: 2012-05-10].

\bibitem{P440}
{Pearson Education Limited}, ``Longman dictionary of contemporary english
  advanced learner's dictionary.'' 2012,
  \url{http://www.ldoceonline.com/dictionary/context} [Accessed on:
  2012-05-10].

\bibitem{P344}
\BIBentryALTinterwordspacing
L.~Sanchez, J.~Lanza, R.~Olsen, M.~Bauer, and M.~Girod-Genet, ``A generic
  context management framework for personal networking environments,'' in
  \emph{Mobile and Ubiquitous Systems - Workshops, 2006. 3rd Annual
  International Conference on}, july 2006, pp. 1 --8. [Online]. Available:
  \url{http://dx.doi.org/10.1109/MOBIQW.2006.361743}
\BIBentrySTDinterwordspacing

\bibitem{P278}
\BIBentryALTinterwordspacing
S.~Ahn and D.~Kim, ``Proactive context-aware sensor networks,'' in
  \emph{Wireless Sensor Networks}, ser. Lecture Notes in Computer Science,
  K.~Römer, H.~Karl, and F.~Mattern, Eds.\hskip 1em plus 0.5em minus
  0.4em\relax Springer Berlin / Heidelberg, 2006, vol. 3868, pp. 38--53.
  [Online]. Available: \url{http://dx.doi.org/10.1007/11669463_6}
\BIBentrySTDinterwordspacing

\bibitem{P182}
\BIBentryALTinterwordspacing
N.~S. Ryan, J.~Pascoe, and D.~R. Morse, ``Enhanced reality fieldwork: the
  context-aware archaeological assistant,'' in \emph{Computer Applications in
  Archaeology 1997}, ser. British Archaeological Reports, V.~Gaffney, M.~van
  Leusen, and S.~Exxon, Eds.\hskip 1em plus 0.5em minus 0.4em\relax Oxford:
  Tempus Reparatum, October 1998. [Online]. Available:
  \url{http://www.cs.kent.ac.uk/pubs/1998/616}
\BIBentrySTDinterwordspacing

\bibitem{P339}
\BIBentryALTinterwordspacing
P.~Hu, J.~Indulska, and R.~Robinson, ``An autonomic context management system
  for pervasive computing,'' in \emph{Pervasive Computing and Communications,
  2008. PerCom 2008. Sixth Annual IEEE International Conference on}, march
  2008, pp. 213 --223. [Online]. Available:
  \url{http://dx.doi.org/10.1109/PERCOM.2008.56}
\BIBentrySTDinterwordspacing

\bibitem{P389}
K.~Henricksen, ``A framework for context-aware pervasive computing
  applications,'' Computer Science, School of Information Technology and
  Electrical Engineering, The University of Queensland, September 2003,
  \url{http://henricksen.id.au/publications/phd-thesis.pdf} [Accessed
  on:2012-01-05].

\bibitem{P352}
A.~Moses, ``Lg smart fridge tells you what to buy, cook and eat,'' The Sydney
  Morning Herald, January 2012,
  \url{http://www.smh.com.au/digital-life/hometech/lg-smart-fridge-tells-you-what-to-buy-cook-and-eat-20120110-1ps9z.html}
  [Accessed on: 2012-04-04].

\bibitem{P109}
M.~Raskino, J.~Fenn, and A.~Linden, ``Extracting value from the massively
  connected world of 2015,'' Gartner Research, Tech. Rep., April 2005,
  \url{http://www.gartner.com/resources/125900/125949/extracting_valu.pdf}
  [Accessed on: 2011-08-20].

\bibitem{P540}
\BIBentryALTinterwordspacing
P.~Prekop and M.~Burnett, ``Activities, context and ubiquitous computing,''
  \emph{Special Issue on Ubiquitous Computing Computer Communications},
  vol.~26, no.~11, p. 1168–1176, 2003. [Online]. Available:
  \url{http://arxiv.org/ftp/cs/papers/0209/0209021.pdf}
\BIBentrySTDinterwordspacing

\bibitem{P541}
\BIBentryALTinterwordspacing
R.~M. Gustavsen, ``Condor - an application framework for mobility-based
  context-aware applications,'' in \emph{Proceedings of the Workshop on
  Concepts and Models for Ubiquitous Computing}, Goeteborg, Sweden, 2002.
  [Online]. Available:
  \url{http://www.alandix.com/academic/conf/ubicomp2002-models/pdf/Gustavsen-goteborg%20sept-02.pdf}
\BIBentrySTDinterwordspacing

\bibitem{P542}
\BIBentryALTinterwordspacing
T.~Hofer, W.~Schwinger, M.~Pichler, G.~Leonhartsberger, J.~Altmann, and
  W.~Retschitzegger, ``Context-awareness on mobile devices - the hydrogen
  approach,'' in \emph{Proceedings of the 36th Annual Hawaii International
  Conference on System Sciences}, ser. HICSS '03, 2003, p. 292–302. [Online].
  Available: \url{http://dx.doi.org/10.1109/HICSS.2003.1174831}
\BIBentrySTDinterwordspacing

\bibitem{P304}
\BIBentryALTinterwordspacing
A.~van Bunningen, L.~Feng, and P.~Apers, ``Context for ubiquitous data
  management,'' in \emph{Ubiquitous Data Management, 2005. UDM 2005.
  International Workshop on}, april 2005, pp. 17 -- 24. [Online]. Available:
  \url{http://dx.doi.org/10.1109/UDM.2005.7}
\BIBentrySTDinterwordspacing

\bibitem{P281}
\BIBentryALTinterwordspacing
Z.~Miao and B.~Yuan, ``Spontaneous sensor networks for context-aware
  computing,'' in \emph{Wireless, Mobile and Multimedia Networks, 2006 IET
  International Conference on}, nov. 2006, pp. 1 --4. [Online]. Available:
  \url{http://ieeexplore.ieee.org/xpls/abs_all.jsp?arnumber=5195749}
\BIBentrySTDinterwordspacing

\bibitem{P331}
\BIBentryALTinterwordspacing
D.~Guan, W.~Yuan, S.~Lee, and Y.-K. Lee, ``Context selection and reasoning in
  ubiquitous computing,'' in \emph{Intelligent Pervasive Computing, 2007. IPC.
  The 2007 International Conference on}, oct. 2007, pp. 184 --187. [Online].
  Available: \url{http://dx.doi.org/10.1109/IPC.2007.102}
\BIBentrySTDinterwordspacing

\bibitem{P284}
\BIBentryALTinterwordspacing
S.~K. Chong, I.~McCauley, S.~W. Loke, and S.~Krishnaswamy, ``Context-aware
  sensors and data muling,'' in \emph{Context awareness for self-managing
  systems (devices, applications and networks) proceeding}.\hskip 1em plus
  0.5em minus 0.4em\relax Berlin : VDE-Verlag, 2007, pp. 103--117. [Online].
  Available:
  \url{http://arrow.latrobe.edu.au:8080/vital/access/HandleResolver/1959.9/122313}
\BIBentrySTDinterwordspacing

\bibitem{P432}
\BIBentryALTinterwordspacing
G.~Jun-zhong, ``Context aware computing,'' \emph{Journal of East China Normal
  University (Natural Science)}, vol.~5, pp. 1--20, 2009. [Online]. Available:
  \url{http://en.cnki.com.cn/Article_en/CJFDTOTAL-HDSZ200905002.htm}
\BIBentrySTDinterwordspacing

\bibitem{P297}
\BIBentryALTinterwordspacing
L.~Mei and S.~Easterbrook, ``Capturing and modeling human cognition for
  context-aware software,'' in \emph{International conference for research on
  computational models and computation-based theories of human behavior}, 2009.
  [Online]. Available:
  \url{http://sideshow.psyc.bbk.ac.uk/rcooper/iccm2009/proceedings/papers/0090/paper0090.pdf}
\BIBentrySTDinterwordspacing

\bibitem{P328}
\BIBentryALTinterwordspacing
S.~Rizou, K.~Haussermann, F.~Durr, N.~Cipriani, and K.~Rothermel, ``A system
  for distributed context reasoning,'' in \emph{Autonomic and Autonomous
  Systems (ICAS), 2010 Sixth International Conference on}, march 2010, pp. 84
  --89. [Online]. Available: \url{http://dx.doi.org/10.1109/ICAS.2010.21}
\BIBentrySTDinterwordspacing

\bibitem{P211}
\BIBentryALTinterwordspacing
W.~Liu, X.~Li, and D.~Huang, ``A survey on context awareness,'' in
  \emph{Computer Science and Service System (CSSS), 2011 International
  Conference on}, june 2011, pp. 144 --147. [Online]. Available:
  \url{http://dx.doi.org/10.1109/CSSS.2011.5972040}
\BIBentrySTDinterwordspacing

\bibitem{P271}
\BIBentryALTinterwordspacing
S.~Yanwei, Z.~Guangzhou, and P.~Haitao, ``Research on the context model of
  intelligent interaction system in the internet of things,'' in \emph{IT in
  Medicine and Education (ITME), 2011 International Symposium on}, vol.~2, dec.
  2011, pp. 379 --382. [Online]. Available:
  \url{http://dx.doi.org/10.1109/ITiME.2011.6132129}
\BIBentrySTDinterwordspacing

\bibitem{P430}
\BIBentryALTinterwordspacing
L.~Barkhuus, L.~Barkhuus, and A.~Dey, ``Is context-aware computing taking
  control away from the user? three levels of interactivity examined,'' in
  \emph{In Proceedings of Ubicomp 2003}.\hskip 1em plus 0.5em minus 0.4em\relax
  Springer, 2003, pp. 149--156. [Online]. Available:
  \url{http://www.itu.dk/people/barkhuus/barkhuus_ubicomp.pdf}
\BIBentrySTDinterwordspacing

\bibitem{P294}
\BIBentryALTinterwordspacing
D.~Martin, C.~Lamsfus, and A.~Alzua, ``Automatic context data life cycle
  management framework,'' in \emph{Pervasive Computing and Applications
  (ICPCA), 2010 5th International Conference on}, dec. 2010, pp. 330 --335.
  [Online]. Available: \url{http://dx.doi.org/10.1109/ICPCA.2010.5704122}
\BIBentrySTDinterwordspacing

\bibitem{P340}
\BIBentryALTinterwordspacing
F.~Ramparany, R.~Poortinga, M.~Stikic, J.~Schmalenstroer, and T.~Prante, ``An
  open context information management infrastructure the ist-amigo project,''
  in \emph{Intelligent Environments, 2007. IE 07. 3rd IET International
  Conference on}, sept. 2007, pp. 398 --403. [Online]. Available:
  \url{http://dx.doi.org/10.1049/cp:20070398}
\BIBentrySTDinterwordspacing

\bibitem{P302}
\BIBentryALTinterwordspacing
A.~Bernardos, P.~Tarrio, and J.~Casar, ``A data fusion framework for
  context-aware mobile services,'' in \emph{Multisensor Fusion and Integration
  for Intelligent Systems, 2008. MFI 2008. IEEE International Conference on},
  aug. 2008, pp. 606 --613. [Online]. Available:
  \url{http://dx.doi.org/10.1109/MFI.2008.4648011}
\BIBentrySTDinterwordspacing

\bibitem{P384}
\BIBentryALTinterwordspacing
B.~Y. Lim and A.~K. Dey, ``Toolkit to support intelligibility in context-aware
  applications,'' in \emph{Proceedings of the 12th ACM international conference
  on Ubiquitous computing}, ser. Ubicomp '10.\hskip 1em plus 0.5em minus
  0.4em\relax New York, NY, USA: ACM, 2010, pp. 13--22. [Online]. Available:
  \url{http://doi.acm.org/10.1145/1864349.1864353}
\BIBentrySTDinterwordspacing

\bibitem{P024}
\BIBentryALTinterwordspacing
G.~Hynes, V.~Reynolds, and M.~Hauswirth, ``A context lifecycle for web-based
  context management services,'' in \emph{Smart Sensing and Context}, ser.
  Lecture Notes in Computer Science, P.~Barnaghi, K.~Moessner, M.~Presser, and
  S.~Meissner, Eds.\hskip 1em plus 0.5em minus 0.4em\relax Springer Berlin /
  Heidelberg, 2009, vol. 5741, pp. 51--65. [Online]. Available:
  \url{http://dx.doi.org/10.1007/978-3-642-04471-7_5}
\BIBentrySTDinterwordspacing

\bibitem{P516}
M.~Peterson and E.~Pierre, ``Snias vision for information life cycle management
  (ilm),'' in \emph{Storage Networking World}.\hskip 1em plus 0.5em minus
  0.4em\relax Computer World, 2004.

\bibitem{P517}
AIIM, ``What is enterprise content management (ecm)?'' February 2009,
  \url{http://www.aiim.org/What-is-ECM-Enterprise-Content-Management.aspx}
  [Accessed on: 2012-06-20].

\bibitem{P515}
E.~Hayden, ``Data lifecycle management model shows risks and integrated data
  flow,'' in \emph{Information Security Magazine}, July 2008.

\bibitem{P170}
A.~N. Shulsky and G.~J. Schmitt, \emph{Silent Warfare: Understanding the World
  of Intelligence}, 3rd~ed.\hskip 1em plus 0.5em minus 0.4em\relax Potomac
  Books Inc, May 2002.

\bibitem{P171}
J.~R. Boyd, ``A discourse on winning and losing.'' Unpublished set of briefing
  slides available at Air University Library, Maxwell AFB, Alabama, 1987,
  \url{http://www.ausairpower.net/JRB/intro.pdf} [Accessed: 2011-12-18].

\bibitem{P114}
\BIBentryALTinterwordspacing
M.~Chantzara and M.~Anagnostou, ``Evaluation and selection of context
  information,'' in \emph{In: Second International Workshop on Modeling and
  Retrieval of Context, Edinburgh}, 2005. [Online]. Available:
  \url{http://ceur-ws.org/Vol-146/paper7.pdf}
\BIBentrySTDinterwordspacing

\bibitem{P518}
\BIBentryALTinterwordspacing
A.~Ferscha, S.~Vogl, and W.~Beer, ``Context sensing, aggregation,
  representation and exploitation in wireless networks,'' \emph{Scalable
  Computing: Practice and Experience}, vol.~6, no.~2, p. 71–81, 2005. [Online].
  Available: \url{http://www.scpe.org/index.php/scpe/article/view/327/17}
\BIBentrySTDinterwordspacing

\bibitem{P519}
\BIBentryALTinterwordspacing
K.~Wrona and L.~Gomez, ``Context-aware security and secure context-awareness in
  ubiquitous computing environments,'' in \emph{XXI Autumn Meeting of Polish
  Information Processing Society}, 2005. [Online]. Available:
  \url{http://proceedings2005.imcsit.org/docs/75.pdf}
\BIBentrySTDinterwordspacing

\bibitem{P334}
\BIBentryALTinterwordspacing
S.~Pietschmann, A.~Mitschick, R.~Winkler, and K.~Meissner, ``Croco:
  Ontology-based, cross-application context management,'' in \emph{Semantic
  Media Adaptation and Personalization, 2008. SMAP '08. Third International
  Workshop on}, dec. 2008, pp. 88 --93. [Online]. Available:
  \url{http://dx.doi.org/10.1109/SMAP.2008.10}
\BIBentrySTDinterwordspacing

\bibitem{P419}
\BIBentryALTinterwordspacing
H.~Chen, T.~Finin, A.~Joshi, L.~Kagal, F.~Perich, and D.~Chakraborty,
  ``Intelligent agents meet the semantic web in smart spaces,'' \emph{Internet
  Computing, IEEE}, vol.~8, no.~6, pp. 69 -- 79, nov.-dec. 2004. [Online].
  Available: \url{http://dx.doi.org/10.1109/MIC.2004.66}
\BIBentrySTDinterwordspacing

\bibitem{P543}
\BIBentryALTinterwordspacing
J.~Indulska and P.~Sutton, ``Location management in pervasive systems,'' in
  \emph{Proceedings of the Australasian information security workshop
  conference on ACSW frontiers 2003 - Volume 21}, ser. ACSW Frontiers
  '03.\hskip 1em plus 0.5em minus 0.4em\relax Darlinghurst, Australia,
  Australia: Australian Computer Society, Inc., 2003, pp. 143--151. [Online].
  Available: \url{http://dl.acm.org/citation.cfm?id=827987.828003}
\BIBentrySTDinterwordspacing

\bibitem{P544}
\BIBentryALTinterwordspacing
A.~Schmidt and K.~van Laerhoven, ``How to build smart appliances?''
  \emph{Personal Communications, IEEE}, vol.~8, no.~4, pp. 66 --71, aug. 2001.
  [Online]. Available: \url{http://dx.doi.org/10.1109/98.944006}
\BIBentrySTDinterwordspacing

\bibitem{P269}
\BIBentryALTinterwordspacing
K.~Balavalad, S.~Manvi, and A.~Sutagundar, ``Context aware computing in
  wireless sensor networks,'' in \emph{Advances in Recent Technologies in
  Communication and Computing, 2009. ARTCom '09. International Conference on},
  oct. 2009, pp. 514 --516. [Online]. Available:
  \url{http://dx.doi.org/10.1109/ARTCom.2009.85}
\BIBentrySTDinterwordspacing

\bibitem{P287}
\BIBentryALTinterwordspacing
G.~Castelli, M.~Mamei, A.~Rosi, and F.~Zambonelli, ``Extracting high-level
  information from location data: the w4 diary example,'' \emph{Mob. Netw.
  Appl.}, vol.~14, no.~1, pp. 107--119, Feb. 2009. [Online]. Available:
  \url{http://dx.doi.org/10.1007/s11036-008-0104-y}
\BIBentrySTDinterwordspacing

\bibitem{P335}
\BIBentryALTinterwordspacing
H.~Chang, S.~Shin, and C.~Chung, ``Context life cycle management scheme in
  ubiquitous computing environments,'' in \emph{Mobile Data Management, 2007
  International Conference on}, may 2007, pp. 315 --319. [Online]. Available:
  \url{http://dx.doi.org/10.1109/MDM.2007.66}
\BIBentrySTDinterwordspacing

\bibitem{P529}
{W3C Ubiquitos Web Domain}, ``Composite capabilities/preference profiles:
  Structure and vocabularies 2.0,'' June 2007,
  \url{http://www.w3.org/Mobile/CCPP/} [Accessed on: 2012-05-26].

\bibitem{P423}
\BIBentryALTinterwordspacing
M.~Knappmeyer, S.~L. Kiani, C.~Fra, B.~Moltchanov, and N.~Baker, ``Contextml: A
  light-weight context representation and context management schema,'' in
  \emph{Wireless Pervasive Computing (ISWPC), 2010 5th IEEE International
  Symposium on}, may 2010, pp. 367 --372. [Online]. Available:
  \url{http://dx.doi.org/10.1109/ISWPC.2010.5483753}
\BIBentrySTDinterwordspacing

\bibitem{P530}
{uml.org}, ``Unified modeling language (uml),'' 2012, \url{http://www.uml.org/}
  [Addressed on: 2012-05-29].

\bibitem{P531}
{ormfoundation.org}, ``The orm foundation,'' 1989,
  \url{http://www.ormfoundation.org} [Addressed on: 2012-05-29].

\bibitem{P556}
\BIBentryALTinterwordspacing
J.~Han, E.~Haihong, G.~Le, and J.~Du, ``Survey on nosql database,'' in
  \emph{Pervasive Computing and Applications (ICPCA), 2011 6th International
  Conference on}, oct. 2011, pp. 363 --366. [Online]. Available:
  \url{http://dx.doi.org/10.1109/ICPCA.2011.6106531}
\BIBentrySTDinterwordspacing

\bibitem{P557}
\BIBentryALTinterwordspacing
D.~Allemang and J.~Hendler, \emph{Semantic Web for the Working Ontologist,
  Second Edition: Effective Modeling in RDFS and OWL}, 2nd~ed.\hskip 1em plus
  0.5em minus 0.4em\relax Morgan Kaufmann, 2011. [Online]. Available:
  \url{http://mkp.com/news/semantic-web-for-the-working-ontologist-2nd-edition-effective-modeling-in-rdfs-and-owl-by-dean-allemang-james-hendler}
\BIBentrySTDinterwordspacing

\bibitem{P558}
\BIBentryALTinterwordspacing
L.~Yu, \emph{A Developer's Guide to the Semantic Web}, 1st~ed.\hskip 1em plus
  0.5em minus 0.4em\relax Springer, 2011. [Online]. Available:
  \url{http://www.springer.com/computer/database+management+%26+information+retrieval/book/978-3-642-15969-5}
\BIBentrySTDinterwordspacing

\bibitem{P378}
\BIBentryALTinterwordspacing
P.~Hitzler, M.~Krötzsch, and S.~Rudolph, \emph{Foundations of Semantic Web
  Technologies}.\hskip 1em plus 0.5em minus 0.4em\relax Chapman \& Hall/CRC,
  2009. [Online]. Available:
  \url{http://www.semantic-web-book.org/page/Foundations_of_Semantic_Web_Technologies}
\BIBentrySTDinterwordspacing

\bibitem{P256}
M.~Botts and A.~Robin, ``Opengis sensor model language (sensorml)
  implementation specification,'' Open Geospatial Consortium Inc, Tech. Rep.,
  2007,
  \url{https://portal.opengeospatial.org/modules/admin/license_agreement.php?suppressHeaders=0&access_license_id=3&target=http://portal.opengeospatial.org/files/%3fartifact_id=12606}
  [Accessed on: 2011-12-15].

\bibitem{P057}
\BIBentryALTinterwordspacing
B.~Khoo, ``Rfid- from tracking to the internet of things: A review of
  developments,'' in \emph{Proceedings of the 2010 IEEE/ACM Int'l Conference on
  Green Computing and Communications \& Int'l Conference on Cyber, Physical and
  Social Computing}, ser. GREENCOM-CPSCOM '10.\hskip 1em plus 0.5em minus
  0.4em\relax Washington, DC, USA: IEEE Computer Society, 2010, pp. 533--538.
  [Online]. Available: \url{http://dx.doi.org/10.1109/GreenCom-CPSCom.2010.22}
\BIBentrySTDinterwordspacing

\bibitem{P191}
\BIBentryALTinterwordspacing
X.~H. Wang, T.~Gu, D.~Q. Zhang, and H.~K. Pung, ``Ontology based context
  modeling and reasoning using owl,'' in \emph{Pervasive Computing and
  Communications Workshops, 2004. Proceedings of the Second IEEE Annual
  Conference on}, march 2004, pp. 18 -- 22. [Online]. Available:
  \url{http://dx.doi.org/10.1109/PERCOMW.2004.1276898}
\BIBentrySTDinterwordspacing

\bibitem{P447}
N.~F. Noy and D.~L. McGuinness, ``Ontology development 101: A guide to creating
  your first ontology,'' Stanford University, Stanford, CA, Tech. Rep., March
  2001,
  \url{http://protege.stanford.edu/publications/ontology_development/ontology101.pdf}
  [Accessed on: 2011-12-15].

\bibitem{P546}
\BIBentryALTinterwordspacing
R.~Studer, V.~Benjamins, and D.~Fensel, ``Knowledge engineering: Principles and
  methods,'' \emph{Data \& Knowledge Engineering}, vol.~25, no. 1–2, pp.
  161 -- 197, 1998. [Online]. Available:
  \url{http://dx.doi.org/10.1016/S0169-023X(97)00056-6}
\BIBentrySTDinterwordspacing

\bibitem{P445}
\BIBentryALTinterwordspacing
M.~Uschold and M.~Gruninger, ``Ontologies: Principles, methods and
  applications,'' \emph{The Knowledge Engineering Review}, vol.~11, no.~02, pp.
  93--136, 1996. [Online]. Available:
  \url{http://dx.doi.org/10.1017/S0269888900007797}
\BIBentrySTDinterwordspacing

\bibitem{P545}
\BIBentryALTinterwordspacing
P.~Korpipaa, J.~Mantyjarvi, J.~Kela, H.~Keranen, and E.-J. Malm, ``Managing
  context information in mobile devices,'' \emph{IEEE Pervasive Computing},
  vol.~2, no.~3, pp. 42--51, Jul. 2003. [Online]. Available:
  \url{http://dx.doi.org/10.1109/MPRV.2003.1228526}
\BIBentrySTDinterwordspacing

\bibitem{P034}
\BIBentryALTinterwordspacing
P.~Korpipaa and J.~Mantyjarvi, ``An ontology for mobile device sensor-based
  context awareness,'' in \emph{Proceedings of the 4th international and
  interdisciplinary conference on Modeling and using context}, ser.
  CONTEXT'03.\hskip 1em plus 0.5em minus 0.4em\relax Berlin, Heidelberg:
  Springer-Verlag, 2003, pp. 451--458. [Online]. Available:
  \url{http://dl.acm.org/citation.cfm?id=1763142.1763181}
\BIBentrySTDinterwordspacing

\bibitem{P197}
\BIBentryALTinterwordspacing
K.-E. Ko and K.-B. Sim, ``Development of context aware system based on bayesian
  network driven context reasoning method and ontology context modeling,'' in
  \emph{Control, Automation and Systems, 2008. ICCAS 2008. International
  Conference on}, oct. 2008, pp. 2309 --2313. [Online]. Available:
  \url{http://dx.doi.org/10.1109/ICCAS.2008.4694191}
\BIBentrySTDinterwordspacing

\bibitem{P332}
\BIBentryALTinterwordspacing
R.~de~Freitas Bulcao~Neto and M.~da~Graca Campos~Pimentel, ``Toward a
  domain-independent semantic model for context-aware computing,'' in \emph{Web
  Congress, 2005. LA-WEB 2005. Third Latin American}, oct.-2 nov. 2005, p. 10
  pp. [Online]. Available: \url{http://dx.doi.org/10.1109/LAWEB.2005.43}
\BIBentrySTDinterwordspacing

\bibitem{P103}
\BIBentryALTinterwordspacing
M.~Compton, C.~Henson, H.~Neuhaus, L.~Lefort, and A.~Sheth, ``A survey of the
  semantic specification of sensors,'' in \emph{2nd International Workshop on
  Semantic Sensor Networks, at 8th International Semantic Web Conference,},
  Oct. 2009. [Online]. Available:
  \url{http://sunsite.informatik.rwth-aachen.de/Publications/CEUR-WS/Vol-522/p6.pdf}
\BIBentrySTDinterwordspacing

\bibitem{P252}
{w3.org}, ``Resource description framework (rdf),'' 2004,
  \url{http://www.w3.org/RDF/} [accesed on: 2012-01-21].

\bibitem{P559}
------, ``Rdf vocabulary description language 1.0: Rdf schema (rdfs),'' 2004,
  \url{http://www.w3.org/2001/sw/wiki/RDFS} [accesed on: 2012-01-21].

\bibitem{P148}
{W3.org}, ``Web ontology language (owl),'' Webpage, October 2007,
  \url{www.w3.org/2004/OWL/} [Accessed: 2011-12-18].

\bibitem{P215}
\BIBentryALTinterwordspacing
A.~Bikakis, T.~Patkos, G.~Antoniou, , and D.~Plexousaki, ``A survey of
  semantics-based approaches for context reasoning in ambient intelligence,''
  in \emph{Ambient Intelligence 2007 Workshops}, M.~M, F.~A, and A.~E, Eds.,
  vol.~11.\hskip 1em plus 0.5em minus 0.4em\relax SPRINGER-VERLAG BERLIN, 2008.
  [Online]. Available:
  \url{http://www.csd.uoc.gr/~bikakis/pubs/survey-ami07.pdf}
\BIBentrySTDinterwordspacing

\bibitem{P214}
P.~Nurmi and P.~Floree. (2004) Reasoning in context-aware systems. Position
  Paper. Department of Computer Science, University of Helsinki.

\bibitem{P248}
\BIBentryALTinterwordspacing
D.~Hall and J.~Llinas, ``An introduction to multisensor data fusion,''
  \emph{Proceedings of the IEEE}, vol.~85, no.~1, pp. 6 --23, jan 1997.
  [Online]. Available: \url{http://dx.doi.org/10.1109/5.554205}
\BIBentrySTDinterwordspacing

\bibitem{P217}
\BIBentryALTinterwordspacing
N.~Lane, E.~Miluzzo, H.~Lu, D.~Peebles, T.~Choudhury, and A.~Campbell, ``A
  survey of mobile phone sensing,'' \emph{Communications Magazine, IEEE},
  vol.~48, no.~9, pp. 140 --150, sept. 2010. [Online]. Available:
  \url{http://dx.doi.org/10.1109/MCOM.2010.5560598}
\BIBentrySTDinterwordspacing

\bibitem{P187}
\BIBentryALTinterwordspacing
D.~Riboni and C.~Bettini, ``Context-aware activity recognition through a
  combination of ontological and statistical reasoning,'' in \emph{Proceedings
  of the 6th International Conference on Ubiquitous Intelligence and
  Computing}, ser. UIC '09.\hskip 1em plus 0.5em minus 0.4em\relax Berlin,
  Heidelberg: Springer-Verlag, 2009, pp. 39--53. [Online]. Available:
  \url{http://dx.doi.org/10.1007/978-3-642-02830-4_5}
\BIBentrySTDinterwordspacing

\bibitem{P561}
\BIBentryALTinterwordspacing
S.-H. Huang, T.-T. Wu, H.-C. Chu, and G.-J. Hwang, ``A decision tree approach
  to conducting dynamic assessment in a context-aware ubiquitous learning
  environment,'' in \emph{Wireless, Mobile, and Ubiquitous Technology in
  Education, 2008. WMUTE 2008. Fifth IEEE International Conference on}, march
  2008, pp. 89 --94. [Online]. Available:
  \url{http://dx.doi.org/10.1109/WMUTE.2008.10}
\BIBentrySTDinterwordspacing

\bibitem{P289}
\BIBentryALTinterwordspacing
H.-S. Park, K.~Oh, and S.-B. Cho, ``Bayesian network-based high-level context
  recognition for mobile context sharing in cyber-physical system,''
  \emph{International Journal of Distributed Sensor Networks}, vol. 2011,
  p.~10, 2011. [Online]. Available:
  \url{http://downloads.hindawi.com/journals/ijdsn/2011/650387.pdf}
\BIBentrySTDinterwordspacing

\bibitem{P267}
\BIBentryALTinterwordspacing
B.~Korel and S.~Koo, ``A survey on context-aware sensing for body sensor
  networks,'' \emph{Scientific Research Publishing: Wireless Sensor Network},
  vol. 2 (8), pp. 571--583, 2010. [Online]. Available:
  \url{http://www.scirp.org/journal/PaperDownload.aspx?paperID=2345}
\BIBentrySTDinterwordspacing

\bibitem{P562}
\BIBentryALTinterwordspacing
C.~Doukas, I.~Maglogiannis, P.~Tragas, D.~Liapis, and G.~Yovanof, ``Patient
  fall detection using support vector machines,'' in \emph{Artificial
  Intelligence and Innovations 2007: from Theory to Applications}, ser. IFIP
  International Federation for Information Processing, C.~Boukis,
  A.~Pnevmatikakis, and L.~Polymenakos, Eds.\hskip 1em plus 0.5em minus
  0.4em\relax Springer Boston, 2007, vol. 247, pp. 147--156. [Online].
  Available: \url{http://dx.doi.org/10.1007/978-0-387-74161-1_16}
\BIBentrySTDinterwordspacing

\bibitem{P209}
\BIBentryALTinterwordspacing
O.~Brdiczka, J.~Crowley, and P.~Reignier, ``Learning situation models in a
  smart home,'' \emph{Systems, Man, and Cybernetics, Part B: Cybernetics, IEEE
  Transactions on}, vol.~39, no.~1, pp. 56 --63, feb. 2009. [Online].
  Available: \url{http://dx.doi.org/10.1109/TSMCB.2008.923526}
\BIBentrySTDinterwordspacing

\bibitem{P565}
\BIBentryALTinterwordspacing
T.-N. Lin and P.-C. Lin, ``Performance comparison of indoor positioning
  techniques based on location fingerprinting in wireless networks,'' in
  \emph{Wireless Networks, Communications and Mobile Computing, 2005
  International Conference on}, vol.~2, june 2005, pp. 1569 -- 1574 vol.2.
  [Online]. Available: \url{http://dx.doi.org/10.1109/WIRLES.2005.1549647}
\BIBentrySTDinterwordspacing

\bibitem{P566}
\BIBentryALTinterwordspacing
K.~V. Laerhoven, ``Combining the self-organizing map and k-means clustering for
  on-line classification of sensor data,'' in \emph{Proceedings of the
  International Conference on Artificial Neural Networks}, ser. ICANN
  '01.\hskip 1em plus 0.5em minus 0.4em\relax London, UK, UK: Springer-Verlag,
  2001, pp. 464--469. [Online]. Available:
  \url{http://dl.acm.org/citation.cfm?id=646258.683987}
\BIBentrySTDinterwordspacing

\bibitem{P268}
\BIBentryALTinterwordspacing
R.~Shtykh and Q.~Jin, ``Capturing user contexts: Dynamic profiling for
  information seeking tasks,'' in \emph{Systems and Networks Communications,
  2008. ICSNC '08. 3rd International Conference on}, oct. 2008, pp. 365 --370.
  [Online]. Available: \url{http://dx.doi.org/10.1109/ICSNC.2008.55}
\BIBentrySTDinterwordspacing

\bibitem{P243}
{w3.org}, ``Swrl: A semantic web rule language combining owl and ruleml,'' May
  2004, \url{http://www.w3.org/Submission/SWRL/} [Accessed on:2012-01-03].

\bibitem{P420}
\BIBentryALTinterwordspacing
X.~Zhou, X.~Tang, X.~Yuan, and D.~Chen, ``Spbca: Semantic pattern-based
  context-aware middleware,'' in \emph{Parallel and Distributed Systems
  (ICPADS), 2009 15th International Conference on}, dec. 2009, pp. 891 --895.
  [Online]. Available: \url{http://dx.doi.org/10.1109/ICPADS.2009.146}
\BIBentrySTDinterwordspacing

\bibitem{P421}
\BIBentryALTinterwordspacing
C.~Kessler, M.~Raubal, and C.~Wosniok, ``Semantic rules for context-aware
  geographical information retrieval,'' in \emph{Proceedings of the 4th
  European conference on Smart sensing and context}, ser. EuroSSC'09.\hskip 1em
  plus 0.5em minus 0.4em\relax Berlin, Heidelberg: Springer-Verlag, 2009, pp.
  77--92. [Online]. Available:
  \url{http://dx.doi.org/10.1007/978-3-642-04471-7_7}
\BIBentrySTDinterwordspacing

\bibitem{P298}
\BIBentryALTinterwordspacing
C.~Choi, I.~Park, S.~Hyun, D.~Lee, and D.~Sim, ``Mire: A minimal rule engine
  for context-aware mobile devices,'' in \emph{Digital Information Management,
  2008. ICDIM 2008. Third International Conference on}, nov. 2008, pp. 172
  --177. [Online]. Available:
  \url{http://dx.doi.org/10.1109/ICDIM.2008.4746772}
\BIBentrySTDinterwordspacing

\bibitem{P136}
\BIBentryALTinterwordspacing
C.~Barbero, P.~D. Zovo, and B.~Gobbi, ``A flexible context aware reasoning
  approach for iot applications,'' in \emph{Proceedings of the 2011 IEEE 12th
  International Conference on Mobile Data Management - Volume 01}, ser. MDM
  '11.\hskip 1em plus 0.5em minus 0.4em\relax Washington, DC, USA: IEEE
  Computer Society, 2011, pp. 266--275. [Online]. Available:
  \url{http://dx.doi.org/10.1109/MDM.2011.55}
\BIBentrySTDinterwordspacing

\bibitem{P128}
\BIBentryALTinterwordspacing
K.~Teymourian, O.~Streibel, A.~Paschke, R.~Alnemr, and C.~Meinel, ``Towards
  semantic event-driven systems,'' in \emph{Proceedings of the 3rd
  international conference on New technologies, mobility and security}, ser.
  NTMS'09.\hskip 1em plus 0.5em minus 0.4em\relax Piscataway, NJ, USA: IEEE
  Press, 2009, pp. 347--352. [Online]. Available:
  \url{http://dx.doi.org/10.1109/NTMS.2009.5384713}
\BIBentrySTDinterwordspacing

\bibitem{P139}
\BIBentryALTinterwordspacing
N.~Konstantinou, E.~Solidakis, S.~Zoi, A.~Zafeiropoulos, P.~Stathopoulos, and
  N.~Mitrou, ``Priamos: a middleware architecture for real-time semantic
  annotation of context features,'' in \emph{Intelligent Environments, 2007. IE
  07. 3rd IET International Conference on}, sept. 2007, pp. 96 --103. [Online].
  Available: \url{http://ieeexplore.ieee.org/xpls/abs_all.jsp?arnumber=4449917}
\BIBentrySTDinterwordspacing

\bibitem{P567}
\BIBentryALTinterwordspacing
T.~Terada, M.~Tsukamoto, K.~Hayakawa, T.~Yoshihisa, Y.~Kishino, A.~Kashitani,
  and S.~Nishio, ``Ubiquitous chip: A rule-based i/o control device for
  ubiquitous computing,'' in \emph{Pervasive Computing}, ser. Lecture Notes in
  Computer Science, A.~Ferscha and F.~Mattern, Eds.\hskip 1em plus 0.5em minus
  0.4em\relax Springer Berlin Heidelberg, 2004, vol. 3001, pp. 238--253.
  [Online]. Available: \url{http://dx.doi.org/10.1007/978-3-540-24646-6_18}
\BIBentrySTDinterwordspacing

\bibitem{P444}
\BIBentryALTinterwordspacing
M.~Roman, C.~Hess, R.~Cerqueira, A.~Ranganathan, R.~H. Campbell, and
  K.~Nahrstedt, ``A middleware infrastructure for active spaces,'' \emph{IEEE
  Pervasive Computing}, vol.~1, no.~4, pp. 74--83, Oct. 2002. [Online].
  Available: \url{http://dx.doi.org/10.1109/MPRV.2002.1158281}
\BIBentrySTDinterwordspacing

\bibitem{P568}
\BIBentryALTinterwordspacing
A.~Ranganathan and R.~H. Campbell, ``A middleware for context-aware agents in
  ubiquitous computing environments,'' in \emph{Proceedings of the
  ACM/IFIP/USENIX 2003 International Conference on Middleware}, ser. Middleware
  '03.\hskip 1em plus 0.5em minus 0.4em\relax New York, NY, USA:
  Springer-Verlag New York, Inc., 2003, pp. 143--161. [Online]. Available:
  \url{http://dl.acm.org/citation.cfm?id=1515915.1515926}
\BIBentrySTDinterwordspacing

\bibitem{P547}
\BIBentryALTinterwordspacing
J.~Mantyjarvi and T.~Seppanen, ``Adapting applications in mobile terminals
  using fuzzy context information,'' in \emph{Proceedings of the 4th
  International Symposium on Mobile Human-Computer Interaction}, ser. Mobile
  HCI '02.\hskip 1em plus 0.5em minus 0.4em\relax London, UK, UK:
  Springer-Verlag, 2002, pp. 95--107. [Online]. Available:
  \url{http://dx.doi.org/10.1007/3-540-45756-9_9}
\BIBentrySTDinterwordspacing

\bibitem{P548}
\BIBentryALTinterwordspacing
A.~Padovitz, S.~W. Loke, and A.~Zaslavsky, ``The ecora framework: A hybrid
  architecture for context-oriented pervasive computing,'' \emph{Pervasive Mob.
  Comput.}, vol.~4, no.~2, pp. 182--215, Apr. 2008. [Online]. Available:
  \url{http://dx.doi.org/10.1016/j.pmcj.2007.10.002}
\BIBentrySTDinterwordspacing

\bibitem{P253}
D.~Tsarkov, ``Fact++,'' Software, 2007,
  \url{http://owl.man.ac.uk/factplusplus/} [Accessed on: 2012-01-21].

\bibitem{P150}
{Clark andParsia}, ``Pellet: Owl 2 reasoner for java,'' Software, 2004,
  \url{http://clarkparsia.com/pellet/} [Accessed: 2011-12-18].

\bibitem{P072}
\BIBentryALTinterwordspacing
A.~Zafeiropoulos, N.~Konstantinou, S.~Arkoulis, D.-E. Spanos, and N.~Mitrou,
  ``A semantic-based architecture for sensor data fusion,'' in \emph{Mobile
  Ubiquitous Computing, Systems, Services and Technologies, 2008. UBICOMM '08.
  The Second International Conference on}, 29 2008-oct. 4 2008, pp. 116 --121.
  [Online]. Available: \url{http://dx.doi.org/10.1109/UBICOMM.2008.67}
\BIBentrySTDinterwordspacing

\bibitem{P073}
\BIBentryALTinterwordspacing
A.~Zafeiropoulos, D.-E. Spano, S.~Arkoulis, N.~Konstantinou, and N.~Mitrou,
  \emph{Data Management in the Semantic Web}, ser. Distributed, Cluster and
  Grid Computing - Yi Pan (Georgia State University), Series Edito, H.~Jin,
  Ed.\hskip 1em plus 0.5em minus 0.4em\relax NOVA Publishers, 2011. [Online].
  Available:
  \url{https://www.novapublishers.com/catalog/product_info.php?products_id=20094}
\BIBentrySTDinterwordspacing

\bibitem{P071}
\BIBentryALTinterwordspacing
Z.~Song, A.~C{\'a}~andrdenas, and R.~Masuoka, ``Semantic middleware for the
  internet of things,'' in \emph{Internet of Things (IOT), 2010}, 29 2010-dec.
  1 2010, pp. 1 --8. [Online]. Available:
  \url{http://dx.doi.org/10.1109/IOT.2010.5678448}
\BIBentrySTDinterwordspacing

\bibitem{P238}
\BIBentryALTinterwordspacing
L.~Peizhi and Z.~Jian, ``A context-aware application infrastructure with
  reasoning mechanism based on dempster-shafer evidence theory,'' in
  \emph{Vehicular Technology Conference, 2008. VTC Spring 2008. IEEE}, may
  2008, pp. 2834 --2838. [Online]. Available:
  \url{http://dx.doi.org/10.1109/VETECS.2008.618}
\BIBentrySTDinterwordspacing

\bibitem{P236}
\BIBentryALTinterwordspacing
C.~H. Lyu, M.~S. Choi, Z.~Y. Li, and H.~Y. Youn, ``Reasoning with imprecise
  context using improved dempster-shafer theory,'' in \emph{Web Intelligence
  and Intelligent Agent Technology (WI-IAT), 2010 IEEE/WIC/ACM International
  Conference on}, vol.~2, 31 2010-sept. 3 2010, pp. 475 --478. [Online].
  Available: \url{http://dx.doi.org/10.1109/WI-IAT.2010.190}
\BIBentrySTDinterwordspacing

\bibitem{P235}
\BIBentryALTinterwordspacing
D.~Zhang, J.~Cao, J.~Zhou, and M.~Guo, ``Extended dempster-shafer theory in
  context reasoning for ubiquitous computing environments,'' in
  \emph{Computational Science and Engineering, 2009. CSE '09. International
  Conference on}, vol.~2, aug. 2009, pp. 205 --212. [Online]. Available:
  \url{http://dx.doi.org/10.1109/CSE.2009.201}
\BIBentrySTDinterwordspacing

\bibitem{P553}
P.~Blunsom, ``Hidden markov models,'' University of Melbourne, tutorial, August
  2004, \url{http://digital.cs.usu.edu/~cyan/CS7960/hmm-tutorial.pdf} [Accessed
  on: 2012-07-10].

\bibitem{P326}
\BIBentryALTinterwordspacing
M.~Krause, C.~Linnhoff-Popien, and M.~Strassberger, ``Concurrent inference on
  high level context using alternative context construction trees,'' in
  \emph{Autonomic and Autonomous Systems, 2007. ICAS07. Third International
  Conference on}, june 2007, p.~7. [Online]. Available:
  \url{http://dx.doi.org/10.1109/CONIELECOMP.2007.67}
\BIBentrySTDinterwordspacing

\bibitem{P560}
\BIBentryALTinterwordspacing
T.~Buchholz, M.~Krause, C.~Linnhoff-Popien, and M.~Schiffers, ``Coco: dynamic
  composition of context information,'' in \emph{Mobile and Ubiquitous Systems:
  Networking and Services, 2004. MOBIQUITOUS 2004. The First Annual
  International Conference on}, aug. 2004, pp. 335 -- 343. [Online]. Available:
  \url{http://dx.doi.org/10.1109/MOBIQ.2004.1331740}
\BIBentrySTDinterwordspacing

\bibitem{P463}
\BIBentryALTinterwordspacing
W.~Dargie, ``The role of probabilistic schemes in multisensor
  context-awareness,'' in \emph{Proceedings of the Fifth IEEE International
  Conference on Pervasive Computing and Communications Workshops}, ser. PERCOMW
  '07.\hskip 1em plus 0.5em minus 0.4em\relax Washington, DC, USA: IEEE
  Computer Society, 2007, pp. 27--32. [Online]. Available:
  \url{http://dx.doi.org/10.1109/PERCOMW.2007.115}
\BIBentrySTDinterwordspacing

\bibitem{P569}
\BIBentryALTinterwordspacing
G.~Chen, M.~Li, and D.~Kotz, ``Data-centric middleware for context-aware
  pervasive computing,'' \emph{Pervasive Mob. Comput.}, vol.~4, no.~2, pp.
  216--253, Apr. 2008. [Online]. Available:
  \url{http://dx.doi.org/10.1016/j.pmcj.2007.10.001}
\BIBentrySTDinterwordspacing

\bibitem{P317}
\BIBentryALTinterwordspacing
A.~Devaraju, S.~Hoh, and M.~Hartley, ``A context gathering framework for
  context-aware mobile solutions,'' in \emph{Proceedings of the 4th
  international conference on mobile technology, applications, and systems and
  the 1st international symposium on Computer human interaction in mobile
  technology}, ser. Mobility '07.\hskip 1em plus 0.5em minus 0.4em\relax New
  York, NY, USA: ACM, 2007, pp. 39--46. [Online]. Available:
  \url{http://doi.acm.org/10.1145/1378063.1378070}
\BIBentrySTDinterwordspacing

\bibitem{P290}
A.~K. Dey, G.~D. Abowd, and D.~Salber, ``A context-based infrastructure for
  smart environments,'' Georgia Institute of Technology, Tech. Rep., 1999,
  \url{http://www.cc.gatech.edu/fce/contexttoolkit/pubs/MANSE99.pdf} [Accessed
  on: 2011-12-05].

\bibitem{P031}
\BIBentryALTinterwordspacing
O.~Corcho and R.~Garcia-Castro, ``Five challenges for the semantic sensor
  web,'' \emph{Semant. web}, vol.~1, no. 1,2, pp. 121--125, Apr. 2010.
  [Online]. Available: \url{http://dl.acm.org/citation.cfm?id=2019445.2019450}
\BIBentrySTDinterwordspacing

\bibitem{P288}
\BIBentryALTinterwordspacing
Q.~Huaifeng and Z.~Xingshe, ``Context aware sensornet,'' in \emph{Proceedings
  of the 3rd international workshop on Middleware for pervasive and ad-hoc
  computing}, ser. MPAC '05.\hskip 1em plus 0.5em minus 0.4em\relax New York,
  NY, USA: ACM, 2005, pp. 1--7. [Online]. Available:
  \url{http://doi.acm.org/10.1145/1101480.1101489}
\BIBentrySTDinterwordspacing

\bibitem{P310}
\BIBentryALTinterwordspacing
J.~Filho and N.~Agoulmine, ``A quality-aware approach for resolving context
  conflicts in context-aware systems,'' in \emph{Embedded and Ubiquitous
  Computing (EUC), 2011 IFIP 9th International Conference on}, oct. 2011, pp.
  229 --236. [Online]. Available: \url{http://dx.doi.org/10.1109/EUC.2011.9}
\BIBentrySTDinterwordspacing

\bibitem{P130}
\BIBentryALTinterwordspacing
E.~F. Nakamura, A.~A.~F. Loureiro, and A.~C. Frery, ``Information fusion for
  wireless sensor networks: Methods, models, and classifications,'' \emph{ACM
  Comput. Surv.}, vol.~39, no.~3, pp. 9/1--9/55, Sep. 2007. [Online].
  Available: \url{http://doi.acm.org/10.1145/1267070.1267073}
\BIBentrySTDinterwordspacing

\bibitem{P555}
\BIBentryALTinterwordspacing
D.~Garlan, D.~Siewiorek, A.~Smailagic, and P.~Steenkiste, ``Project aura:
  Toward distraction-free pervasive computing,'' \emph{IEEE Pervasive
  Computing}, vol.~1, no.~2, pp. 22--31, Apr. 2002. [Online]. Available:
  \url{http://dx.doi.org/10.1109/MPRV.2002.1012334}
\BIBentrySTDinterwordspacing

\bibitem{P190}
\BIBentryALTinterwordspacing
T.~Strang, C.~Linnhoff-Popien, and K.~Frank, ``Cool: A context ontology
  language to enable contextual interoperability,'' \emph{Ifip International
  Federation For Information Processing}, vol. 2893, pp. 236--247, 2003.
  [Online]. Available:
  \url{http://www.springerlink.com/index/DJNHU2GVPV7CQLTV.pdf}
\BIBentrySTDinterwordspacing

\bibitem{P386}
\BIBentryALTinterwordspacing
L.~Capra, W.~Emmerich, and C.~Mascolo, ``Carisma: context-aware reflective
  middleware system for mobile applications,'' \emph{Software Engineering, IEEE
  Transactions on}, vol.~29, no.~10, pp. 929 -- 945, oct. 2003. [Online].
  Available: \url{http://dx.doi.org/10.1109/TSE.2003.1237173}
\BIBentrySTDinterwordspacing

\bibitem{P570}
\BIBentryALTinterwordspacing
T.~Gu, H.~K. Pung, and D.~Q. Zhang, ``A service-oriented middleware for
  building context-aware services,'' \emph{J. Netw. Comput. Appl.}, vol.~28,
  no.~1, pp. 1--18, Jan. 2005. [Online]. Available:
  \url{http://dx.doi.org/10.1016/j.jnca.2004.06.002}
\BIBentrySTDinterwordspacing

\bibitem{P311}
\BIBentryALTinterwordspacing
D.~H. Wilson, A.~C. Long, and C.~Atkeson, ``A context-aware recognition survey
  for data collection using ubiquitous sensors in the home,'' in \emph{CHI '05
  extended abstracts on Human factors in computing systems}, ser. CHI EA
  '05.\hskip 1em plus 0.5em minus 0.4em\relax New York, NY, USA: ACM, 2005, pp.
  1865--1868. [Online]. Available:
  \url{http://doi.acm.org/10.1145/1056808.1057042}
\BIBentrySTDinterwordspacing

\bibitem{P305}
\BIBentryALTinterwordspacing
C.-H. Hou, H.-C. Hsiao, C.-T. King, and C.-N. Lu, ``Context discovery in sensor
  networks,'' in \emph{Information Technology: Research and Education, 2005.
  ITRE 2005. 3rd International Conference on}, june 2005, pp. 2 -- 6. [Online].
  Available: \url{http://dx.doi.org/10.1109/ITRE.2005.1503053}
\BIBentrySTDinterwordspacing

\bibitem{P316}
\BIBentryALTinterwordspacing
C.~Jacob, D.~Linner, S.~Steglich, and I.~Radusch, ``Bio-inspired context
  gathering in loosely coupled computing environments,'' in \emph{Bio-Inspired
  Models of Network, Information and Computing Systems, 2006. 1st}, dec. 2006,
  pp. 1 --6. [Online]. Available:
  \url{http://dx.doi.org/10.1109/BIMNICS.2006.361803}
\BIBentrySTDinterwordspacing

\bibitem{P336}
\BIBentryALTinterwordspacing
D.~Ejigu, M.~Scuturici, and L.~Brunie, ``Semantic approach to context
  management and reasoning in ubiquitous context-aware systems,'' in
  \emph{Digital Information Management, 2007. ICDIM '07. 2nd International
  Conference on}, vol.~1, oct. 2007, pp. 500 --505. [Online]. Available:
  \url{http://dx.doi.org/10.1109/ICDIM.2007.4444272}
\BIBentrySTDinterwordspacing

\bibitem{P338}
\BIBentryALTinterwordspacing
R.~de~Rocha and M.~Endler, ``Middleware: Context management in heterogeneous,
  evolving ubiquitous environments,'' \emph{Distributed Systems Online, IEEE},
  vol.~7, no.~4, p.~1, april 2006. [Online]. Available:
  \url{http://dx.doi.org/10.1109/MDSO.2006.28}
\BIBentrySTDinterwordspacing

\bibitem{P349}
\BIBentryALTinterwordspacing
S.-H. Baek, E.-C. Choi, J.-D. Huh, and K.-R. Park, ``Sensor information
  management mechanism for context-aware service in ubiquitous home,''
  \emph{Consumer Electronics, IEEE Transactions on}, vol.~53, no.~4, pp. 1393
  --1400, nov. 2007. [Online]. Available:
  \url{http://dx.doi.org/10.1109/TCE.2007.4429229}
\BIBentrySTDinterwordspacing

\bibitem{P403}
\BIBentryALTinterwordspacing
D.~Conan, R.~Rouvoy, and L.~Seinturier, ``Scalable processing of context
  information with cosmos,'' in \emph{Proceedings of the 7th IFIP WG 6.1
  international conference on Distributed applications and interoperable
  systems}, ser. DAIS'07.\hskip 1em plus 0.5em minus 0.4em\relax Berlin,
  Heidelberg: Springer-Verlag, 2007, pp. 210--224. [Online]. Available:
  \url{http://hal.inria.fr/docs/00/15/50/45/PDF/article.pdf}
\BIBentrySTDinterwordspacing

\bibitem{P308}
\BIBentryALTinterwordspacing
J.~Herbert, J.~O'Donoghue, and X.~Chen, ``A context-sensitive rule-based
  architecture for a smart building environment,'' in \emph{Future Generation
  Communication and Networking, 2008. FGCN '08. Second International Conference
  on}, vol.~2, dec. 2008, pp. 437 --440. [Online]. Available:
  \url{http://dx.doi.org/10.1109/FGCN.2008.169}
\BIBentrySTDinterwordspacing

\bibitem{P293}
\BIBentryALTinterwordspacing
W.~Xue, H.~Pung, W.~Ng, and T.~Gu, ``Data management for context-aware
  computing,'' in \emph{Embedded and Ubiquitous Computing, 2008. EUC '08.
  IEEE/IFIP International Conference on}, vol.~1, dec. 2008, pp. 492 --498.
  [Online]. Available: \url{http://dx.doi.org/10.1109/EUC.2008.27}
\BIBentrySTDinterwordspacing

\bibitem{P333}
\BIBentryALTinterwordspacing
E.~Kim and J.~Choi, ``A context management system for supporting context-aware
  applications,'' in \emph{Embedded and Ubiquitous Computing, 2008. EUC '08.
  IEEE/IFIP International Conference on}, vol.~2, dec. 2008, pp. 577 --582.
  [Online]. Available: \url{http://dx.doi.org/10.1109/EUC.2008.168}
\BIBentrySTDinterwordspacing

\bibitem{P274}
\BIBentryALTinterwordspacing
Z.~Li, X.~Zhou, H.~Qing, and S.~Li, ``Model and implementation of context-aware
  sensor networks,'' in \emph{Information Science and Engineering, 2008. ISISE
  '08. International Symposium on}, vol.~2, dec. 2008, pp. 16 --19. [Online].
  Available: \url{http://dx.doi.org/10.1109/ISISE.2008.202}
\BIBentrySTDinterwordspacing

\bibitem{P300}
\BIBentryALTinterwordspacing
S.~Gashti, G.~Pujolle, and J.~Rotrou, ``An upnp-based context-aware framework
  for ubiquitous mesh home networks,'' in \emph{Personal, Indoor and Mobile
  Radio Communications, 2009 IEEE 20th International Symposium on}, sept. 2009,
  pp. 400 --404. [Online]. Available:
  \url{http://dx.doi.org/10.1109/PIMRC.2009.5449966}
\BIBentrySTDinterwordspacing

\bibitem{P280}
\BIBentryALTinterwordspacing
E.~S. Reetz, R.~Tonjes, and N.~Baker, ``Towards global smart spaces: Merge
  wireless sensor networks into context-aware systems,'' in \emph{Wireless
  Pervasive Computing (ISWPC), 2010 5th IEEE International Symposium on}, may
  2010, pp. 337 --342. [Online]. Available:
  \url{http://dx.doi.org/10.1109/ISWPC.2010.5483728}
\BIBentrySTDinterwordspacing

\bibitem{P312}
\BIBentryALTinterwordspacing
Y.~Zhan, S.~Wang, Z.~Zhao, C.~Chen, and J.~Ma, ``A mobile device oriented
  framework for context information management,'' in \emph{Information,
  Computing and Telecommunication, 2009. YC-ICT '09. IEEE Youth Conference on},
  sept. 2009, pp. 150 --153. [Online]. Available:
  \url{http://dx.doi.org/10.1109/YCICT.2009.5382404}
\BIBentrySTDinterwordspacing

\bibitem{P341}
\BIBentryALTinterwordspacing
P.~Pawar, H.~Boros, F.~Liu, G.~Heijenk, and B.-J. van Beijnum, ``Bridging
  context management systems in the ad hoc and mobile environments,'' in
  \emph{Computers and Communications, 2009. ISCC 2009. IEEE Symposium on}, july
  2009, pp. 882 --888. [Online]. Available:
  \url{http://dx.doi.org/10.1109/ISCC.2009.5202323}
\BIBentrySTDinterwordspacing

\bibitem{P314}
\BIBentryALTinterwordspacing
A.~Corradi, M.~Fanelli, and L.~Foschini, ``Implementing a scalable
  context-aware middleware,'' in \emph{Computers and Communications, 2009. ISCC
  2009. IEEE Symposium on}, july 2009, pp. 868 --874. [Online]. Available:
  \url{http://dx.doi.org/10.1109/ISCC.2009.5202318}
\BIBentrySTDinterwordspacing

\bibitem{P309}
\BIBentryALTinterwordspacing
O.~Kwon, Y.-S. Song, J.-H. Kim, and K.-J. Li, ``Sconstream: A spatial context
  stream processing system,'' in \emph{International Conference on
  Computational Science and Its Applications}, march 2010, pp. 165 --170.
  [Online]. Available: \url{http://dx.doi.org/10.1109/ICCSA.2010.48}
\BIBentrySTDinterwordspacing

\bibitem{P346}
\BIBentryALTinterwordspacing
B.~Guo, L.~Sun, and D.~Zhang, ``The architecture design of a cross-domain
  context management system,'' in \emph{8th IEEE International Conference on
  Pervasive Computing and Communications Workshops}, 29 2010-april 2 2010, pp.
  499 --504. [Online]. Available:
  \url{http://dx.doi.org/10.1109/PERCOMW.2010.5470618}
\BIBentrySTDinterwordspacing

\bibitem{P347}
\BIBentryALTinterwordspacing
W.~Wibisono, A.~Zaslavsky, and S.~Ling, ``Comihoc: A middleware framework for
  context management in manet environment,'' in \emph{Advanced Information
  Networking and Applications (AINA), 2010 24th IEEE International Conference
  on}, april 2010, pp. 620 --627. [Online]. Available:
  \url{http://dx.doi.org/10.1109/AINA.2010.153}
\BIBentrySTDinterwordspacing

\bibitem{P322}
\BIBentryALTinterwordspacing
A.~Shaeib, P.~Cappellari, and M.~Roantree, ``A framework for real-time context
  provision in ubiquitous sensing environments,'' in \emph{Computers and
  Communications (ISCC), 2010 IEEE Symposium on}, june 2010, pp. 1083 --1085.
  [Online]. Available: \url{http://dx.doi.org/10.1109/ISCC.2010.5546645}
\BIBentrySTDinterwordspacing

\bibitem{P571}
\BIBentryALTinterwordspacing
F.~Li, S.~Sehic, and S.~Dustdar, ``Copal: An adaptive approach to context
  provisioning.'' in \emph{WiMob}.\hskip 1em plus 0.5em minus 0.4em\relax IEEE,
  2010, pp. 286--293. [Online]. Available:
  \url{http://dx.doi.org/10.1109/WIMOB.2010.5645051}
\BIBentrySTDinterwordspacing

\bibitem{P327}
\BIBentryALTinterwordspacing
E.~Badidi and I.~Taleb, ``Towards a cloud-based framework for context
  management,'' in \emph{Innovations in Information Technology (IIT), 2011
  International Conference on}, april 2011, pp. 35 --40. [Online]. Available:
  \url{http://dx.doi.org/10.1109/INNOVATIONS.2011.5893849}
\BIBentrySTDinterwordspacing

\bibitem{P393}
\BIBentryALTinterwordspacing
A.~K. Dey and A.~Newberger, ``Support for context-aware intelligibility and
  control,'' in \emph{Proceedings of the 27th international conference on Human
  factors in computing systems}, ser. CHI '09.\hskip 1em plus 0.5em minus
  0.4em\relax New York, NY, USA: ACM, 2009, pp. 859--868. [Online]. Available:
  \url{http://doi.acm.org/10.1145/1518701.1518832}
\BIBentrySTDinterwordspacing

\bibitem{P395}
{jboss.org}, ``Drools - the business logic integration platform,'' 2001,
  \url{http://www.jboss.org/drools} [Accessed on: 2012-04-23].

\bibitem{P195}
\BIBentryALTinterwordspacing
A.~Padovitz, S.~Loke, and A.~Zaslavsky, ``Towards a theory of context spaces,''
  in \emph{Pervasive Computing and Communications Workshops, Proceedings of the
  Second IEEE Annual Conference on}, march 2004, pp. 38 -- 42. [Online].
  Available: \url{http://dx.doi.org/10.1109/PERCOMW.2004.1276902}
\BIBentrySTDinterwordspacing

\bibitem{P441}
T.~Reineking, ``Java dempster shafer library,'' July 2010,
  \url{https://sourceforge.net/projects/jds/} [Accessed on: 2012-05-12].

\bibitem{P258}
\BIBentryALTinterwordspacing
{IEEE Instrumentation and Measurement Society}, ``Ieee standard for a smart
  transducer interface for sensors and actuators wireless communication
  protocols and transducer electronic data sheet (teds) formats,'' \emph{IEEE
  Std 1451.5-2007}, pp. C1 --236, 5 2007. [Online]. Available:
  \url{http://dx.doi.org/10.1109/IEEESTD.2007.4346346}
\BIBentrySTDinterwordspacing

\bibitem{P595}
P.~Christen, \emph{Data Matching}, ser. Data-Centric Systems and
  Applications.\hskip 1em plus 0.5em minus 0.4em\relax Springer, 2012.

\bibitem{ZMP005}
C.~Perera, P.~Jayaraman, A.~Zaslavsky, P.~Christen, and D.~Georgakopoulos,
  ``Dynamic configuration of sensors using mobile sensor hub in internet of
  things paradigm,'' in \emph{IEEE 8th International Conference on Intelligent
  Sensors, Sensor Networks, and Information Processing (ISSNIP)}, Melbourne,
  Australia, April 2013, pp. 473--478.

\bibitem{ZMP002}
C.~Perera, A.~Zaslavsky, P.~Christen, A.~Salehi, and D.~Georgakopoulos,
  ``Connecting mobile things to global sensor network middleware using
  system-generated wrappers,'' in \emph{International ACM Workshop on Data
  Engineering for Wireless and Mobile Access 2012 (ACM SIGMOD/PODS
  2012-Workshop-MobiDE)}, Scottsdale, Arizona, USA, May 2012, pp. 23--30.

\bibitem{P088}
D.-E. Spanos, P.~Stavrou, N.~Konstantinou, and N.~Mitrou, ``Sensorstream: A
  semantic real-time stream management system,'' in \emph{International Journal
  of Ad Hoc and Ubiquitous Computing}, 2011.

\bibitem{P520}
\BIBentryALTinterwordspacing
T.~Heath and C.~Bizer, \emph{Linked Data (Synthesis Lectures on the Semantic
  Web: Theory and Technology)}, 1st~ed., J.~Hendler, Ed.\hskip 1em plus 0.5em
  minus 0.4em\relax Morgan \& Claypool Publishers, February 2011. [Online].
  Available: \url{http://linkeddatabook.com/editions/1.0/}
\BIBentrySTDinterwordspacing

\bibitem{P068}
\BIBentryALTinterwordspacing
D.~L. Phuoc and M.~Hauswirth, ``Linked open data in sensor data mashups,'' in
  \emph{In Proceedings of the 2nd International Workshop on Semantic Sensor
  Networks (SSN09)}, vol. 522.\hskip 1em plus 0.5em minus 0.4em\relax CEUR
  Workshop at ISWC 2009, Washington DC, USA, 2009, pp. 1--16. [Online].
  Available: \url{http://ceur-ws.org/Vol-522/p3.pdf}
\BIBentrySTDinterwordspacing

\bibitem{ZMP004}
\BIBentryALTinterwordspacing
C.~Perera, A.~Zaslavsky, P.~Christen, and D.~Georgakopoulos, ``Ca4iot: Context
  awareness for internet of things,'' in \emph{IEEE International Conference on
  Conference on Internet of Things (iThing)}, Besançon, France, November 2012,
  pp. 775--782. [Online]. Available:
  \url{http://dx.doi.org/10.1109/GreenCom.2012.128}
\BIBentrySTDinterwordspacing

\bibitem{ZMP006}
C.~Perera, A.~Zaslavsky, P.~Christen, M.~Compton, and D.~Georgakopoulos,
  ``Context-aware sensor search, selection and ranking model for internet of
  things middleware,'' in \emph{IEEE 14th International Conference on Mobile
  Data Management (MDM)}, Milan, Italy, June 2013.

\bibitem{P563}
\BIBentryALTinterwordspacing
E.~Maximilien and M.~Singh, ``A framework and ontology for dynamic web services
  selection,'' \emph{Internet Computing, IEEE}, vol.~8, no.~5, pp. 84 -- 93,
  sept.-oct. 2004. [Online]. Available:
  \url{http://dx.doi.org/10.1109/MIC.2004.27}
\BIBentrySTDinterwordspacing

\bibitem{P564}
\BIBentryALTinterwordspacing
S.~Ran, ``A model for web services discovery with qos,'' \emph{SIGecom Exch.},
  vol.~4, no.~1, pp. 1--10, Mar. 2003. [Online]. Available:
  \url{http://doi.acm.org/10.1145/844357.844360}
\BIBentrySTDinterwordspacing

\end{thebibliography}
\bibliographystyle{IEEEtran}

%
%

%

\vspace{-30pt}

\begin{IEEEbiography}[{\includegraphics[width=1in,height=1.25in,clip,keepaspectratio]{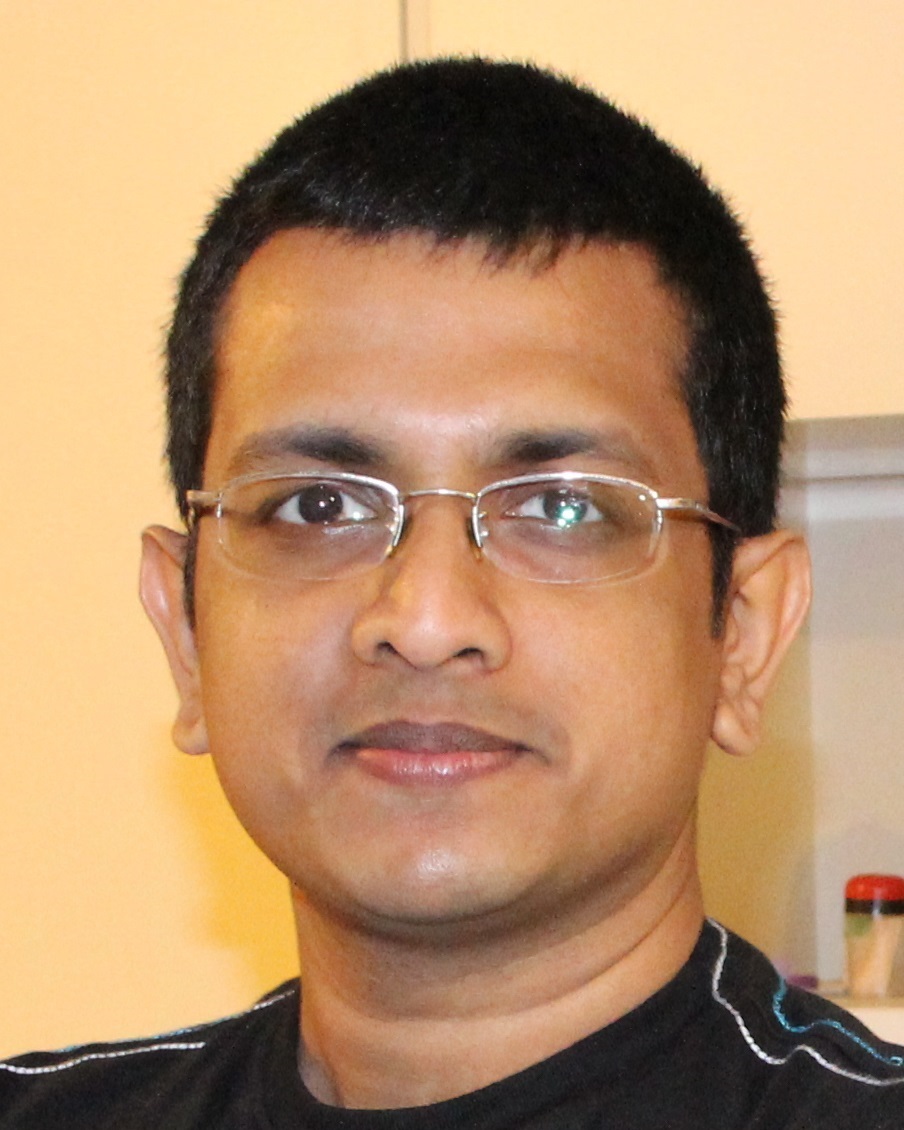}}]{Charith Perera}
received his BSc (Hons) in Computer Science in 2009 from Staffordshire University, Stoke-on-Trent, United Kingdom and MBA in Business Administration in 2012 from University of Wales, Cardiff, United Kingdom. He is currently pursing his PhD in Computer Science at The Australian National University, Canberra, Australia. He is also working at Information Engineering Laboratory, ICT Centre, CSIRO and involved in OpenIoT Project (Open source blueprint for large scale self organizing cloud environments for IoT applications), which is co-funded by the European Commission under seventh framework program. His research interests include Internet of Things, pervasive and ubiquitous computing with a focus on sensor networks, middleware, context aware computing, mobile computing and semantic technologies. He is a member of the Association for Computing Machinery (ACM) and the Institute of Electrical and Electronics Engineers (IEEE).

\end{IEEEbiography}

\vspace{-45pt}

\begin{IEEEbiography}[{\includegraphics[width=1in,height=1.25in,clip,keepaspectratio]{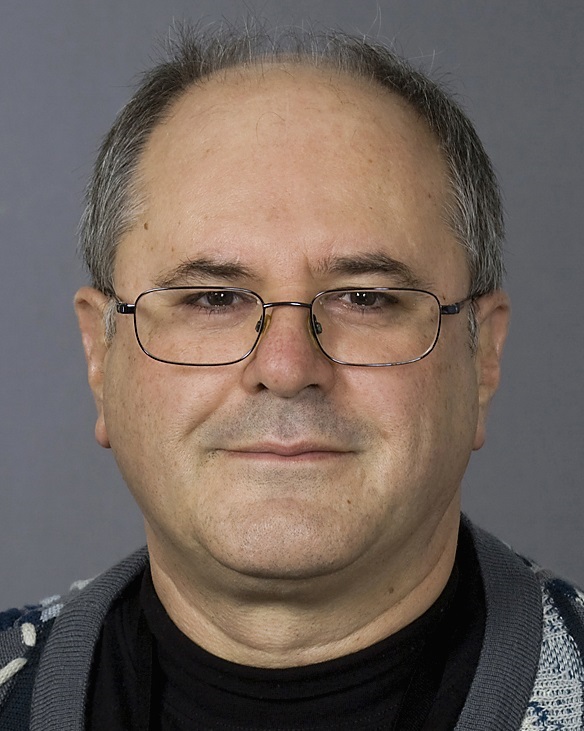}}]{Arkady Zaslavsky}
is the Science Leader of the Semantic Data Management science area at Information Engineering Laboratory, ICT Centre, CSIRO. He is also holding positions of Adjunct Professor at ANU, Research Professor at LTU and Adjunct Professor at UNSW. He is currently involved and is leading a number of European and national research projects. Before coming to CSIRO in July 2011, he held a position of a Chaired Professor in Pervasive and Mobile Computing at Luleå University of Technology, Sweden where he was involved in a number of European research projects, collaborative projects with Ericsson Research, PhD supervision and postgraduate education. Between 1992 and 2008 Arkady was a full-time academic staff member at Monash University, Australia. Arkady made internationally recognised contribution in the area of disconnected transaction management and replication in mobile computing environments, context-awareness as well as in mobile agents. He made significant internationally recognised contributions in the areas of data stream mining on mobile devices, adaptive mobile computing systems, ad-hoc mobile networks, efficiency and reliability of mobile computing systems, mobile agents and mobile file systems. Arkady received MSc in Applied Mathematics majoring in Computer Science from Tbilisi State University (Georgia, USSR) in 1976 and PhD in Computer Science from the Moscow Institute for Control Sciences (IPU-IAT), USSR Academy of Sciences in 1987. Before coming to Australia in 1991, Arkady worked in various research positions at industrial R\&D labs as well as at the Institute for Computational Mathematics of Georgian Academy of Sciences where he lead a systems software research laboratory. Arkady Zaslavsky has published more than 300 research publications throughout his professional career and supervised to completion more than 30 PhD students. Arkady Zaslavsky is a Senior Member of ACM, a member of IEEE Computer and Communication Societies.
\end{IEEEbiography}

\begin{IEEEbiography}[{\includegraphics[width=1in,height=1.25in,clip,keepaspectratio]{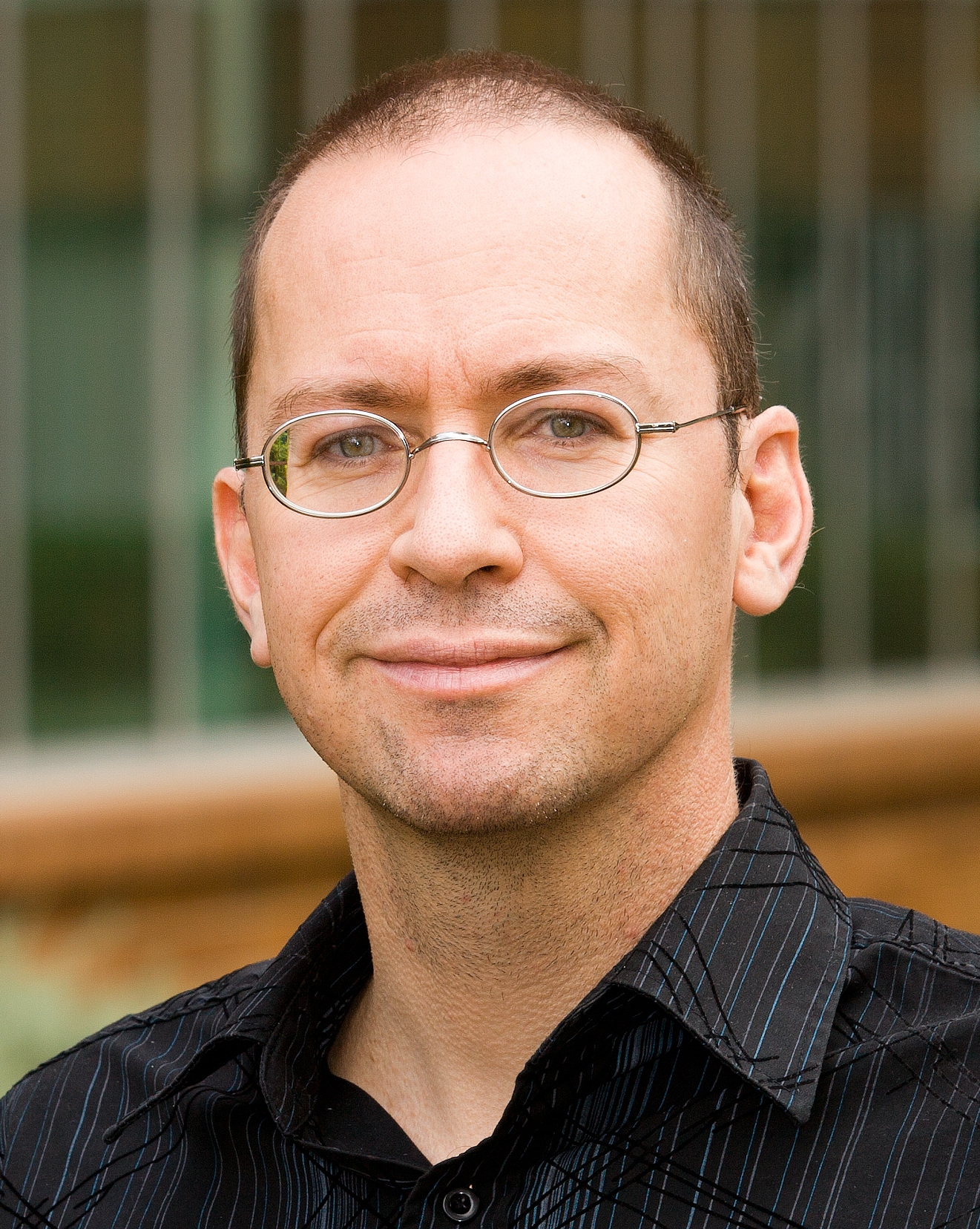}}]{Peter Christen}
is an Associate Professor in the Research School of Computer Science
at the Australian National University. He received his Diploma in Computer Science Engineering from ETH Z\"urich in 1995 and his PhD in Computer Science from the University of Basel in 1999 (both in Switzerland). His research interests are in data mining and data matching (entity resolution). He is especially interested in the development of scalable and real-time algorithms for data matching, and privacy and confidentiality aspects of data matching and data mining. He has published over 80 papers in these areas, including in 2012 the book `Data Matching' (by Springer), and he is the principle developer of the \emph{Febrl} (Freely Extensible Biomedical Record Linkage) open source data cleaning, deduplication and record linkage system.
\end{IEEEbiography}

\vspace{-220pt}

\begin{IEEEbiography}[{\includegraphics[width=1in,height=1.25in,clip,keepaspectratio]{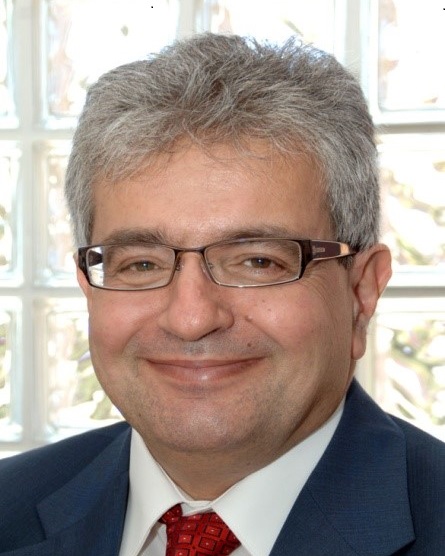}}]{Dimitrios Georgakopoulos}
is a Research Director at the CSIRO ICT Centre where he heads the Information Engineering Laboratory that is based in Canberra and Sydney. The laboratory has 70 researchers and more than 40 visiting scientists, students, and interns specializing in the areas of Service/Cloud Computing, Human Computer Interaction, Machine Learning, and Semantic Data Management. Dimitrios is also an Adjunct Professor at the Australian National University. Before coming to CSIRO in October 2008, Dimitrios held research and management positions in several industrial laboratories in the US. From 2000 to 2008, he was a Senior Scientist with Telcordia, where he helped found Telcordia’s Research Centers in Austin, Texas, and Poznan, Poland. From 1997 to 2000, Dimitrios was a Technical Manager in the Information Technology organization of Microelectronics and Computer Corporation (MCC), and the Chief Architect of MCC’s Collaboration Management Infrastructure (CMI) consortial project. From 1990-1997, Dimitrios was a Principal Scientist at GTE (currently Verizon) Laboratories Inc. Dimitrios has received a GTE (Verizon) Excellence Award, two IEEE Computer Society Outstanding Paper Awards, and was nominated for the Computerworld Smithsonian Award in Science. He has published more than one hundred journal and conference papers. Dimitrios is the Vice-Chair of the 12th International Semantic Web Conference (ISWC 2013) in Sydney, Australia, 2013, and the General Co-Chair of the 9th IEEE International Conference on Collaborative Computing (CollaborateCom 2013) in Austin, Texas, USA, 2013. In 2011, Dimitrios was the General chair of the 12th International Conference on Web Information System Engineering (WISE), Sydney, Australia, and the 7th CollaborateCom, Orlando, Florida, October 2011. In 2007, he was the Program Chair of the 8th WISE in Nancy France, and the 3rd CollaborateCom in New York, USA. In 2005, he was the General chair of the 6th WISE in New York. In 2002, and he served as the General Chair of the 18th International Conference on Data Engineering (ICDE) in San Jose, California. In 2001, he was the Program Chair of the 17th ICDE in Heidelberg, Germany. Before that he was the Program Chair of 1st International Conference on Work Activity Coordination (WACC) in San Francisco, California, 1999, and has served as Program Chair in a dozen smaller conferences and workshops.

\end{IEEEbiography}

%


%








\end{document}